\documentclass[runningheads,dvipsnames]{lmcs}
\pdfoutput=1
\usepackage[utf8]{inputenc}

\usepackage{lastpage}
\lmcsdoi{21}{3}{1}
\lmcsheading{}{\pageref{LastPage}}{}{}%
{Jun.~21,~2024}{Jul.~01,~2025}{}

\usepackage{mathpartir}
\usepackage{amstext,amsmath,amssymb,amsfonts,mathbbol}
\usepackage{mathtools}
\usepackage{thmtools}
\usepackage{thm-restate}
\usepackage{glossaries}
\usepackage{tikz}
\usepackage{dsfont}
\usepackage{xcolor}
\usepackage{listings}
\usepackage{stmaryrd}
\usepackage{hyperref}
\usepackage{booktabs}
\usepackage{scalerel}

\usetikzlibrary{arrows,automata}
\usetikzlibrary{decorations.pathmorphing}
\tikzset{snake it/.style={decorate, decoration=snake}}

\newcommand{\mc}[1]{\mathcal{#1}}
\newcommand{\ms}[1]{\mathsf{#1}}
\newcommand{\tms}[1]{$\mathsf{#1}$}

\newcommand{\red}[1]{\textcolor{red}{#1}}
\newcommand{\green}[1]{{\color{Green}#1}}
\newcommand{\purple}[1]{\textcolor{purple}{#1}}

\newcommand{\cset}[1]{\mathbb{#1}}

\newcommand{\rel}[1]{\mathcal{#1}}
\newcommand{\relation}[1]{\mathcal{#1}}

\newcommand{\nil}{{\bf{0}}}
\newcommand{\res}[1]{\nu\, #1.}

\newcommand{\sub}[2]{\{#1/#2\}}

\newcommand{\set}[1]{\{ #1\}}

\newcommand{\sys}[2]{#1 \triangleright #2}
\newcommand{\net}{\Delta}

\newcommand{\step}{ \longrightarrow}
\newcommand{\wstep}{\Longrightarrow}
\newcommand{\ltrans}[1]{\llparenthesis #1 \rrparenthesis}
\newcommand{\lstep}[1]{\;\xrightarrow{\;\;#1\;\;}\;}
\newcommand{\lwstep}[1]{\;\xRightarrow{\;\;#1\;\;}\;}
\newcommand{\barb}[1]{{\downarrow_{#1}}}
\newcommand{\nbarb}[1]{\hspace*{-0.3em}\not \downarrow_{#1}}
\newcommand{\wbarb}[1]{{\Downarrow_{#1}}}

\newcommand{\vect}[1]{\tilde{#1}}
\newcommand{\vvect}[1]{\widetilde{#1}}

\newcommand{\tuple}[1]{\langle #1 \rangle}
\newcommand{\parop}{\mid}
\newcommand{\fv}[1]{\mathsf{fv}(#1)}
\newcommand{\fn}[1]{\mathsf{fn}(#1)}

\newcommand{\scal}{\mathcal{S}}

\newcommand{\connected}{\leftrightsquigarrow}
\newcommand{\connection}{\leftrightarrow}

\makeatletter
\newcommand{\ctx}[1]{\mathds{#1}}
\newcommand{\lrl}[2]{\textnormal{\textsc{\LeftTirNameStyle \textsc{L}-}\rl[\text{\uppercase{#2}}]{#1}}}
\newcommand{\rl}[2][]{\textnormal{\textsc{\LeftTirNameStyle{#2\textsubscript{\uppercase{#1}}}}}}

\newcommand{\eout}[1]{\overline{#1}}
\newcommand{\out}[2]{\overline{#1}{\langle #2 \rangle}}

\newcommand{\node}[1]{\ms{node}(#1)}

\newcommand{\proc}[3]{\textnormal{\textbf{[}}\,#1\,\textnormal{\textbf{]}}^{#2}_{#3}}

\newcommand{\sbreak}[1]{\ms{unlink}\: #1}
\newcommand{\link}[1]{\ms{link}\: #1}
\newcommand{\skill}{\ms{kill}}

\newcommand{\create}[2]{\ms{create}\:#1.#2}

\newcommand{\spawn}[2]{\ms{spawn}\: #1. #2}
\newcommand{\spawnc}[3]{\langle #1 \rangle^{#2}_{#3}}
\newcommand{\spwn}[2]{\ms{go}\: #1. #2}

\newcommand{\views}{\succ}
\newcommand{\ift}[2]{\ms{if}~#1~\ms{then} \,#2}
\newcommand{\ifte}[3]{\ms{if}~#1~\ms{then}\,#2~\ms{else}~#3}
\newcommand{\remove}[1]{\ms{forget} \: #1}

\newcommand{\godloc}{\odot}
\DeclareMathOperator{\bcong}{\dot\approx}

\newcommand{\bsim}{\sim}
\newcommand{\wbsim}{\approx}
\newcommand*{\defeq}{\stackrel{\text{def}}{=}}
\makeatother

\lstdefinestyle{DOS}
{
  backgroundcolor=\color{white},
  basicstyle=\scriptsize\color{black}\ttfamily
}

\newif\ifwithproof

\withprooftrue

\newcommand{\paral}{\parallel}

\newcommand{\fnull}{\mathbf{\hat{0}}}

\begin{document}

\title[A Behavioral Theory For Distributed Systems With Weak Recovery]{A Behavioral Theory For Distributed Systems\texorpdfstring{\\}{} With Weak Recovery}

\author[G.~Fabbretti]{Giovanni Fabbretti\lmcsorcid{0000-0003-3002-0697}}[a]
\author[I.~Lanese]{Ivan Lanese\lmcsorcid{0000-0003-2527-9995}}[b]
\author[J.B.~Stefani]{Jean-Bernard Stefani\lmcsorcid{0000-0003-1373-7602}}[a]

\address{Univ. Grenoble Alpes, INRIA, CNRS, Grenoble INP, LIG, 38000
  Grenoble, France}
\email{giovanni.fabbretti@inria.fr, jean-bernard.stefani@inria.fr}

\address{Olas Team, Univ. of Bologna/INRIA, 40137 Bologna, Italy}
\email{ivan.lanese@gmail.com}

\thanks{Partially supported by the EU Marie
    Sk\l{}odowska-Curie action ReGraDe-CS (No: 101106046), by French
    ANR project SmartCloud ANR-23-CE25-0012, by French ANR project
    DCore ANR-18-CE25-0007, and by INdAM – GNCS 2024 project MARVEL, code CUP
    E53C23001670001.}

\begin{abstract} Distributed systems can be subject to various kinds of partial
  failures, therefore building fault-tolerance or failure mitigation mechanisms
  for distributed systems remains an important domain of research. In this
  paper, we present a calculus to formally model distributed systems subject to
  crash failures with recovery. The recovery model considered in the paper is
  \emph{weak}, in the sense that it makes no assumption on the exact state in
  which a failed node resumes its execution, only its identity has to be
  distinguishable from past incarnations of itself. Our calculus is inspired in
  part by the Erlang programming language and in part by the distributed
  $\pi$-calculus with nodes and link failures (D$\pi$F) introduced by
  Francalanza and Hennessy. In order to reason about distributed systems with
  failures and recovery we develop a behavioral theory for our calculus, in the
  form of a contextual equivalence, and of a fully abstract coinductive
  characterization of this equivalence by means of a labelled transition system
  semantics and its associated weak bisimilarity. This result is valuable for it
  provides a compositional proof technique for proving or disproving contextual
  equivalence between systems.
\end{abstract}

\maketitle

\section{Introduction}\label{sec:intro}

\textbf{Goal and motivations}. A key characteristic of distributed computer
systems is the occurrence of partial failures, which can affect part of a
system, e.g.\ failures of the system nodes (computers and the processes they
support) or failures of the connections between them. The model of crash
failures with recovery, where a node can fail by ceasing to operate entirely,
and later on recover its operation (typically, after some administrator
or management system intervention) is an important model of failure to consider because of its
relevance in practice. For instance, it is the failure model assumed by the Paxos protocol used in many 
cloud services around the world~\cite{PaxosWikipedia}, and the basic failure model considered 
by the Kubernetes cloud configuration system~\cite{Kubernetes} for its self-healing facility. 
However, the correct design of systems based on this model is
far from trivial, as a recent study on bugs affecting crash recovery in
distributed systems demonstrates~\cite{GaoDQGW0HZW18}.
Developing a process calculus analysis can help in this respect for it can reveal subtle phenomena 
in the behavior of such systems, and it can be leveraged for the development of
analysis tools, such as verifiers, testers and debuggers.

Unfortunately, the literature contains precious few examples of process calculi accounting for crash failures with recovery,
and those that do \cite{FournetGLMR96,Amadio97,BergerH00,BocchiLTV23} are not really satisfactory: they fail
to support what we call a weak recovery model as well as two other key features of actual distributed systems, namely
dynamic nodes and links, and imperfect knowledge of their environment.
The goal of this paper, then, is to introduce a process calculus exhibiting the
following key features:
\begin{enumerate}

	\item \label{dynamic} \emph{Dynamic nodes and links}: the number of nodes in a system, and of
	communication links between them, is not fixed and may vary during execution.

	\item \label{crash} \emph{Crash failures}: when a node or a link fails it does so silently.
	A failed link simply ceases to suppport communication between nodes. 
	A failed node simply ceases to execute the processes it hosts.

	\item \label{imperfect} 
    \emph{Imperfect knowledge}: in general, in a distributed system,
	a node has only a partial and possibly outdated knowledge of the
	overall state of the system.
  
	\item \emph{Weak recovery model}\label{recovery}: when a link fails it can be re-established. 
	When a node fails it can be restarted with an arbitrary, possibly empty, set of hosted processes. 
	There is no assumption that the state of hosted processes prior to the node crash has been preserved,
	only that it is possible to distinguish previous failed instances
	of a node from the current running one.
\end{enumerate}

Our calculus is positioned so that it stays close to the D$\pi$F process
calculus by Francalanza and Hennessy~\cite{FrancalanzaH08} while being faithful
to the failure behavior of systems built with the Erlang~\cite{Armstrong07Book}
programming language. We do so for the following reasons. On the one hand,
D$\pi$F constitutes a good benchmark for comparison. While not tackling
recovery, it exhibits several of the features mentioned above, and develops a
behavioral theory with node and link failures. Staying close to it allows us to
compare our behavioral theory with an established one, when restricting our
attention to systems without recovery. On the other hand, Erlang and its
environment are representative of modern distributed programming facilities.
Erlang is a functional, concurrent and distributed language based on the actor
model, and it is used in several large distributed
projects~\cite{ErlangProjects}. It is well known for its ``let it fail'' policy,
whereby a service that is not working as expected is killed as soon as the
faulty behavior is detected, and restarted by its supervisor. Staying close to
Erlang ensures our modeling remains faithful to actual distributed systems. 
% This
% mechanism of recovery appears also in Kubernetes~\cite{Kubernetes} system under
% the name of \emph{self-healing} and is well in line with our approach.

The table below summarizes our analysis of the state of the art with respect to the key features mentioned above.
\begin{table}[h!]
  \centering
  \begin{tabular}{|l|c|c|c|c|c|}
    \hline & \cite{FrancalanzaH08} & \cite{FournetGLMR96} & \cite{Amadio97} & \cite{BergerH00} & \cite{BocchiLTV23}\\
    \hline Dynamic nodes & ~~yes~~ &~~ yes~~ & ~~yes~~ & no & no \\
	\hline Dynamic links & partially & no & no & no & no \\
    \hline Crash failures & yes & yes & yes & ~~yes~~ & ~~yes~~ \\
    \hline Imperfect knowledge & ~~partially~~ & no & no & no & no \\
    \hline Weak recovery model~~ & no & no & no & no & no\\ \hline
  \end{tabular}
  % \label{table:related-work}
  % \caption{Related Work Comparison}
\end{table}

In D$\pi$F~\cite{FrancalanzaH08} fresh links can be established only when a new
location is created and only between the new location and other locations
reachable from the creating one. Moreover, once a link between two locations is broken it cannot be
restored. Hence, the partial meeting of the dynamic link criterion. Also, in
D$\pi$F, locations can have a partial view of the system but cannot hold a
(wrong) belief on the status of a remote location, hence the partial meeting of
the imperfect knowledge criterion.

\textbf{Approach and contributions}.
Coming up with a small calculus, meeting all the above criteria, is far from
trivial because of the interplay between the different features we target. 
Highlighting key elements of our approach can give a sense of the difficulties.

The starting point for our process
calculus, which we call D$\pi$FR for Distributed $\pi$-calculus with Failures and Recovery, 
is the D$\pi$F calculus presented in \cite{FrancalanzaH08}, however our technical
developments are markedly different.

First, D$\pi$F is a dynamically-typed calculus: whole systems are typed by the
public part of the network of locations and links on which they run, and new
location names are typed with their status (alive or dead) and the links
connecting new locations to other existing ones in the network (locations are
abstractions of nodes in D$\pi$F and D$\pi$FR). 
Types in D$\pi$F are annotations that encode information about the network structure.
As such they do not constrain D$\pi$F systems, but they constrain their equivalence:
two equivalent systems in
D$\pi$F are by definition equally typed, and thus their networks have the same public part.
But this is precisely a result which we
would expect to derive from a behavioral theory
(cf.~Proposition~\ref{prop:net-eq-bsim}). So we opt instead for an untyped
approach, where we can indeed prove such a result for (strongly) equivalent
systems.

Second, the dynamicity of networks in D$\pi$F is limited by the fact
that when creating a new location, links can only be established with existing
locations which are reachable by the creating location. Hence, if a location is
or becomes disconnected from all the other locations in a network, it will
remain so forever. But if we want to allow for link recovery this should clearly not be
the case.
D$\pi$FR thus allows for links to be (re)established freely, with no constraint
on reachability.
As a result, as we will see later in Section~\ref{sec:examples} (cf.~Example~\ref{ex:network}), our behavioral theory is more
discriminating than that of D$\pi$F even in absence of location recovery, which is an unexpected result. 

Third, the types in D$\pi$F induce complexity in the handling of
their restriction operator and in the definition of a fully abstract labelled transition system
semantics and a weak bisimilarity equivalence. Francalanza and Hennessy argue that this
leads to gains in the form of reduced sizes of the bisimulation relations they
need to consider in proofs of equivalence between systems. However, they can
obtain this only because their network (modulo the creation of fresh nodes) can
only diminish. In our setting, since nodes and links can be reestablished, their
approach would not work anyway. We thus introduce in D$\pi$FR a novel
handling of restriction, eschewing in particular traditional close
rule for scope extrusion from the $\pi$-calculus, and gaining in the process
simpler developments for our behavioral theory (a simpler labelled transition
system semantics, a simpler notion of weak bisimulation, and a much simpler proof of
completeness for our weak bisimilarity equivalence).

A key construct for supporting recovery in D$\pi$FR is the notion of
\emph{incarnation number}. Briefly, each location in D$\pi$FR is identified by
a name and an incarnation number. When a failed location recovers, 
its incarnation number is incremented, to distinguish the recovered location 
from its previous instances. 
We introduce this notion in D$\pi$FR for two main reasons. 
First, we do so to faithfully reflect the behavior of Erlang
systems in presence of failure. Incarnation numbers are called creation numbers
in Erlang~\cite{CreationErlang} and allow the Erlang environment to safely drop
messages issued by a previous incarnation of a node, avoiding message confusion
across different incarnations of the same node. Second, incarnation numbers are
key elements used in several failure recovery schemes and distributed
algorithms. Under various names (epoch numbers, incarnation numbers) they are
required in several rollback-recovery schemes surveyed in \cite{ElnozahyAWJ02},
including optimistic recovery schemes \cite{StromY85,DamaniTG99} and causal
logging schemes \cite{ElnozahyZ92}. They are also a key ingredient in several
distributed algorithms such as, e.g., algorithms for scalable distributed failure
detection \cite{GuptaCG01}, membership management \cite{DasGM02}, and diskless
crash-recovery \cite{MichaelPSS17}. With this notion, in D$\pi$FR we have basic
support in place for encoding these recovery schemes and algorithms.

To summarize, the key contributions of our work are as follows:
\begin{enumerate}
\item We formalize in D$\pi$FR support for a weak recovery model with incarnation numbers, 
  commonly found
  in actual systems such as Erlang ones, and at the basis of several recovery
  schemes which have been proposed in the literature.
\item To reason about failures and recoveries in distributed systems we develop
  a behavioral theory for D$\pi$FR including a contextual equivalence in the
  form of a weak barbed congruence, and its coinductive fully abstract
  characterization in the form of a labelled transition system semantics and its
  weak bisimilarity equivalence. Our behavioral theory agrees with the one for
  D$\pi$F on key examples without recovery.
\end{enumerate}

\textbf{Organization}.
The rest of this paper is organized as follows. Section~\ref{sec:running-example}
introduces a motivating example, which serves also as an introduction to our calculus. 
Section~\ref{sec:calculus} presents the calculus and its
reduction semantics.
Section~\ref{sec:behavioral-theory}
equips our calculus with a notion of weak barbed congruence and characterizes it by a
notion of weak bisimilarity to obtain a proof technique for checking system
equivalence.
Section~\ref{sec:examples} shows our behavioral theory in action on the
motivating example from Section~\ref{sec:running-example} and on other examples from~\cite{FrancalanzaH08}.
Section~\ref{sec:discussion} discusses key design choices in the definition of our calculus.
Section~\ref{sec:experiment} discusses the experiments we did with Erlang to validate our semantics.
Finally, Section~\ref{sec:related} discusses related work and concludes.
Details of proofs and complementary material are available in the Appendices.

%%% Local Variables:
%%% mode: latex
%%% TeX-master: "main-distributed"
%%% End:

\section{Crash and Recovery: Motivating Example}\label{sec:running-example}

Before presenting our calculus, we discuss a motivating
example which serves also as an informal introduction to the D$\pi$FR calculus
(for the time being, we omit some details).
An Erlang implementation of the example is available in the companion repository~\cite{RepositoryErlang}.

Consider the following system:
\[
  \textbf{servD} \defeq
  \sys{\res{\vect{u}}\net}{
    \proc{ I  }{n_i}{}
    \paral \proc{ R }{n_r}{}
    \paral \proc{ B }{n_b}{}
  }
\]
where:
\begin{align*}
  & \vect{u} = n_r,n_b,r_1,r_2,b \\
	& I = req(y,z).\spawn{n_r}{\out{r_1}{y,z}} \qquad
    B = b(y,z).\spawn{n_r}{\out{r_2}{z,w_y}}  \\
	& 	 R = (r_1(y,z).\spawn{n_b}{\out{b}{y,z}}) \parop (r_2(z,w).\spawn{n_i}{\out{z}{w}})
\end{align*}

System \textbf{servD} depicts a distributed server running on a network $\net$.
The network, $\net$, whose formal definition we omit for the moment, can be
graphically represented as follows,
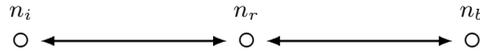
{
  \[
  \begin{tikzpicture}

    \node[label={[label distance=-1mm]\footnotesize $n_i$}] (l1) at (-3,0)
    {\Large $\circ$}; \node[label={[label distance=-1mm]\footnotesize $n_r$}]
    (l2) at (0,0) {\Large $\circ$}; \node[label={[label
      distance=-1mm]\footnotesize $n_b$}] (l3) at (3,0) {\Large $\circ$};

    \draw[latex-latex, thick] (l1) -- (l2); \draw[latex-latex, thick] (l2) --
    (l3);

  \end{tikzpicture}
  \]
}
where {\Large $\circ$} represents an alive location and the arrow {
  \hspace{-0.3cm}
  \tikz{
    \node[] at (0,0.05) {}; %required to shift coordinates
    \draw[latex-latex, thick] (0,0) -- (1,0);
  }
} represents a live bidirectional communication link between locations.

A location in our calculus represents a locus of computation. 
It can represent a hardware node in a distributed system, 
or a virtual machine or a container running on a hardware node. A location
constitutes also a unit of crash failure: when a location fails, all the processes inside the location
cease to function. Located processes take the form $\proc{P}{n}{}$, where $P$ is a process 
and $n$ is the location name. There can be several processes
located at the same location: for instance, $\proc{P}{n}{} \paral \proc{Q}{n}{}$ denotes two processes
$P$ and $Q$ running in parallel inside the same location $n$.

The (admittedly simplistic) system \textbf{servD} behaves as follows. 
The \emph{interface} process $I$, running on $n_i$, awaits a single request from the environment on channel $req$. 
Once received, the elements of the request (a parameter $y$, and a return channel $z$) 
are routed to location $n_b$, which runs the \emph{backend} process $B$,
through location $n_r$, which hosts the \emph{router} process $R$, as there is no direct link between $n_i$ and $n_b$. 
The router awaits for the elements of the request on a private channel $r_1$ and forwards them 
by spawning a message $\out{b}{y,z}$ on $n_b$, where $b$ is a private channel.
Message sending in our calculus can only occur locally.
For two remote locations to communicate, like for $n_i$ and $n_r$, it is
necessary for one to asynchronously spawn the message on the other one.
The backend handles the information $y$ and returns the answer $w_y$ to the
interface by routing it through $n_r$. Finally, the interface emits the answer on
$z$ for the client to consume it. 

The following is a possible client for \textbf{servD}
\[
  \proc{\spawn{n_i}{(\out{req}{h,z} \parop z(w).Q})}{n_k}{}
\]
It sends a request to the interface $n_i$ by spawning it on $n_i$. It also spawns a process on $n_i$
to handle the response, which will take the form of a message on channel $z$ located on $n_i$.

% Note that the backend cannot directly release the response on return channel $z$: 
% because of the restriction $\nu\, n_r, n_b$, locations
% $n_r$ and $n_b$ are not known from  the environment, only the interface location $n_i$ is; 
% hence the need to pass the response from $n_b$ to $n_i$ via the router location $n_r$.

Now, consider the following system:
\[
  \textbf{servDF} \defeq
  \sys{\res{\vect{u},\green{n_c},\green{retry}}\net'}{
      \proc{ J }{n_i}{}
      \paral
      \proc{ R }{n_r}{} \paral
      \proc{!B}{n_b}{} \paral
      \red{\proc{\skill}{n_r}{}} \paral
		  \green{\proc{!C}{n_c}{}}
  }
\]
where:
  \begin{align*}
	& J = req(y,z).((\spawn{n_r}{\out{r_1}{y,z}}) \parop retry.\spawn{n_r}{\out{r_1}{y,z}}) \\
  & C = \create{n_r}{(R \parop \spawn{n_i}{\eout{retry}})}
  \end{align*}

Here, $\net'$ could be graphically represented like $\net$ only with an extra link between
$n_r$ and another location $\green{n_c}$.
System \textbf{servDF} represents a distributed server where location $n_r$ may be subject to one failure, 
modeled by the primitive \tms{kill}. The system has a
recovery mechanism in place to deal with that potential failure: 
the \emph{controller} $\green{\proc{!C}{n_c}{}}$
is a
location that keeps trying, through the apposite
primitive \tms{create}, to recreate $n_r$ 
(the ! operator is akin to the $\pi$-calculus operator for replication, which makes available an unbounded number of copies of the trailing process),
together with a message to restart the handling of the request.
The interface is more sophisticated as it can now send a second request when asked to retry
if something goes wrong with the first one. If $n_r$ fails, the
controller can create another router, sending also a message $retry$ to communicate to
the interface $n_i$ to
start the second attempt.

At first glance, \textbf{servDF} seems to correctly handle the failure scenario,
but that is not the case. Indeed, consider an execution where the
failure happens after the request has already been handled. In such a case, the
request would be processed again and another response sent back by the interface, a
behavior that cannot be exhibited by \textbf{servD}.
If one considers \textbf{servD} as the specification to meet,
\textbf{servDF} does not satisfy it.

The following system correctly handles recovery, in that it meets the \textbf{servD} specification:
\[
  \textbf{servDFR} \defeq
  \sys{\res{\vect{u},\green{n_c},\green{retry},\green{c}} \net'}{
        \proc{ K }{n_i}{} \paral
      \proc{ R }{n_r}{} \paral
      \proc{ !B }{n_b}{} \paral
      \red{\proc{ \skill }{n_r}{}}\paral
        \green{\proc{ !C }{n_c}{}}
  }
\]
where:
\begin{align*}
K = req(y,z).((\spawn{n_r}{\out{r_1}{y,c}}) \parop
	      c(w).\out{z}{w} \parop
          retry.\spawn{n_r}{\out{r_1}{y,c}})
\end{align*}

In system \textbf{servDFR} the response from the backend is not directly sent to
the client, but goes through a private channel $c$ on which the interface listens
only once. This mechanism prevents emitting the answer to the request twice.
System \textbf{servDFR} can be understood as a masking $1$-fault tolerant system (according to the terminology in~\cite{Gartner99}),
equivalent to the ideal system \textbf{servD}. This claim could also be
  formally verified by leveraging in our setting the work done in~\cite{FrancalanzaH07}, in which a
  behavioral theory capable of discriminating systems up to $n$ faults is given.
We develop in Section~\ref{sec:behavioral-theory} 
a behavioral theory able to tell apart \textbf{servD} from
\textbf{servDF}, and able to prove equivalent \textbf{servD} and
\textbf{servDFR}.

%%% Local Variables:
%%% mode: latex
%%% TeX-master: "main-distributed"
%%% End:

\section{The Calculus}\label{sec:calculus}

\subsection{Names and Notations}

We assume given
mutually disjoint infinite denumerable sets $\ms C$, $\ms N$ and $\ms{I}$. $\ms
C$ is the set of \emph{channel name}s, $\ms N$ is the set of \emph{location
names}, and $\ms{I}$ is the set of \emph{incarnation variables}. 
We use the set of integers $\mathbb{Z}$ as the set of \emph{incarnation
  numbers}.
Generally, we span over
$\ms{C}$ with $x,y,z$ and their decorated versions, over $\ms{N}$ with $l,m,n$ and
their decorated versions, over $\ms{I}$ with $\iota$ and its decorated versions,
and over $\mathbb{Z}$ with $\lambda,\kappa$ and their decorated versions.
However, in
some examples we abuse the notation and use evocative words or other identifiers to span over
$\ms{C}$ and $\ms{N}$ to help the readability of the systems that
we describe.
An incarnation number is paired with a location name for recovery purposes, 
to distinguish the current instance of a location 
from its past failed instances.
We denote by $\ms{N}^{\godloc}$ the set $\ms{N} \cup \{\godloc \}$, where $\godloc \not\in \ms{N}$ and the symbol
$\godloc$ identifies a special location that cannot be killed.
As in the $\pi$-calculus, channel names can be free or bound in terms.
The same holds for location names. Incarnation variables can be bound, but
incarnation numbers cannot.
We denote by $\vect{u}$ a finite (possibly empty) tuple of elements.
We write $T\sub{\vect{v}}{\vect{u}}$ for the usual capture-avoiding substitution of elements of $\vect{u}$
by elements of $\vect{v}$ in term $T$, assuming tuples $\vect{u}$ and $\vect{v}$ have the same number of elements.
We write $u,\vect{v}$ or $\vect{v},u$ for the tuple $\vect{v}$ extended with element $u$ as first or last element.
Abusing notation, we sometimes identify a tuple $\vect{u}$ with the set of its elements.
We denote by $\mathbb{N}^{+}$ the set of strictly positive integers (by definition $0 \not\in \mathbb{N}^{+}$),
and by $\mathbb{N}$ the set of positive integers ($0 \in \mathbb{N}$).
We denote by $\mathbf{\hat{0}}$ the function $\mathbf{\hat{0}}: \ms{N}^{\godloc} \rightarrow \mathbb{Z}$ 
that maps any $n \in \ms{N}$ to $0$ and $\godloc$ to $1$.
If $f: \ms{A} \rightarrow \ms{B}$ is a function from $\ms{A}$ to $\ms{B}$, we write $f[a \mapsto b]$
for the function from $\ms{A}$ to $\ms{B}$ that agrees with $f$ on all elements of $\ms{A}$ different from $a$,
and maps $a$ to $b$, i.e.\ $\forall x \in \ms{A}\setminus\{a\}, f[a \mapsto b](x) = f(x)$, and
$f[a \mapsto b](a) = b$.

\subsection{Syntax}

Systems in our calculus are defined through three levels of
syntax, one for \emph{processes}, one for \emph{configurations}, one for \emph{systems}.

The syntax of \emph{processes} is defined as follows:

\begin{align*}
  \begin{array}{lll}
    P,Q & ::=
    &
      \nil 
	  ~\big{|}~ \out{x}{\vect{u}}.P
      ~\big{|}~ x(\vect{v}).P 
      ~\big{|}~ !x(\vect{v}).P 	  
	  ~\big{|}~ \res{w} P 
	  ~\big{|}~ \ms{if}~ r = s~\ms{then}~P~\ms{else}~Q 
      ~\big{|}~ P\parop Q \\
        &&
           \node{n,\iota}.P ~\big{|}~ \remove{n}.P ~\big{|}~ \spawn{n}{P}
		   ~\big{|}~  \skill  ~\big{|}~ \\
        &&  
           \create{n}{P} ~\big{|}~\link{n}.P ~\big{|}~ \sbreak{n}.P
  \end{array}
\end{align*}
where
\begin{align*}
  &\vect{u} \subset \ms{C} \cup \ms{N}^{\godloc} \cup \ms{I} \cup \mathbb{Z} \quad
  r,s \in \ms{C} \cup \ms{N}^{\godloc} \cup \ms{I} \cup \mathbb{Z} \quad
  \vect{v} \subset \ms{C} \cup \ms{N} \cup \ms{I}\\ %\quad
  &w \in \ms{C} \cup \ms{N} \quad
  x \in \ms{C} \quad
  n \in \ms{N} \quad
  \iota \in \ms{I}
\end{align*}
Terms of the form $x(\vect{u}).P$, $\res{w}P$, and $\node{n,\iota}.P$
are binding constructs for their
arguments $\vect{u}$, $w$ and $n, \iota$, respectively.

The syntax of processes is that of the $\pi$-calculus with matching~\cite{SangiorgiW01} and replicated receivers
(first line of productions), 
extended with primitives for distributed computing inspired from the Erlang programming language 
(second line of productions),
and primitives to activate locations, establish and remove links (third line of productions).

$\nil$ is the null process which can take no action. 
Two processes $\out{x}{\vect{v}}.P$ and $x(\vect{u}).Q$ \emph{residing on the same location} can communicate synchronously.
More precisely, process $\out{x}{\vect{v}}.P$ is an output on channel $x$ of a tuple of values $\vect{v}$ with continuation $P$. Dually, $x(\vect{u}).Q$ receives on channel $x$ a tuple of values and replaces them for elements of $\vect{u}$. $Q$ is the continuation.  
We write $\out{x}{\vect{u}}$
for $\out{x}{\vect{u}}.\nil$, and just $\eout{x}$ when $\vect{u}$ is empty.
We assume the calculus is well-sorted, so that the arity of receivers always matches that of
received messages. 
The construct $!x(\vect{u}).P$ is the replicated input construct, 
which, as in~\cite{FrancalanzaH08}, replicates itself once and then awaits for a local message on channel $x$ before replicating
again. Note that we use in our examples 
(as we did in the previous section) the short-hand
$!P$ for $\res{c} (\eout{c} ~\parop~ !c.(P \parop \eout{c}))$.
The construct $\res{w}P$ is the standard restriction construct, which creates a fresh
location or channel name.
If $\vect{u}$ is a (possibly empty) tuple of names, we write $\res{\vect{u}}P$
for $\res{u_1} \ldots \res{u_n}P$ if $\vect{u} = (u_1, \ldots, u_n)$.
If $\vect{u}$ is empty, $\res{\vect{u}}P$ is just $P$.
The construct $\ifte{r = s}{P}{Q}$ tests the equality of names $r$
and $s$ and continues as $P$ if the names match and as $Q$ otherwise.
The construct $P\parop Q$ is the standard parallel composition for processes.
Primitive $\node{n,\iota}.P$ gives access to the name of the current location and
its incarnation number, substituting them for $n$ and $\iota$ inside the
continuation $P$.
Primitive $\remove{n}.P$ removes the reference to the location with name $n$
from the local view of the current location and continues as $P$. This primitive in Erlang is used to remove a connection to a remote location.
Primitive $\spawn{n}{P}$ launches
process $P$ at the location named $n$, if the latter is accessible.
Primitive $\skill$ stops the current location in its current incarnation: 
no process can execute on a killed location; $\skill$ models both a programmed
stop and the crash of a location.
Primitive $\create{n}{P}$ creates a new location $n$, or reactivates a killed location
with a new incarnation number, and launches process $P$ on it. 
Primitive $\link{n}.P$ creates a connection between the current location and location $n$
and continues as $P$, while, $\sbreak{n}.P$ breaks the
link between the current location and location $n$ and continues as $P$; \tms{unlink} also models the failure of a link.

The syntax of \emph{configurations} is defined as follows:
\begin{align*}
	& L,M,N ::=  \nil~\big{|}~ \proc{P}{n}{\lambda} ~\big{|}~ \spawnc{(m,\kappa):P}{n}{\lambda} ~\big{|}~  N\paral M \quad
    \text{ where } n,m \in \ms{N}^{\godloc}\quad
    \lambda,\,\kappa \in \mathbb{N}^{+}
\end{align*}

A configuration can be the empty configuration $\nil$, 
a located process $\proc{P}{n}{\lambda}$,
a spawning message $\spawnc{(m,\kappa):P}{n}{\lambda}$,
or a parallel composition of configurations $N \paral M$.
A located process $\proc{P}{n}{\lambda}$ is a process $P$
running on location $n_\lambda$, where $n$ is the name of the location and $\lambda$ is an incarnation number.
A spawning message $\spawnc{(m,\kappa):P}{n}{\lambda}$ denotes a message sent
by location $n$ in its incarnation $\lambda$ to spawn process $P$ on the target location $m$
in its incarnation $\kappa$.
In examples, we may drop incarnation numbers of located processes if they are not relevant. 
Note that the special name $\godloc$ identifies 
a well-known location which we will assume to be  un-killable.
Location $\godloc$ is used merely for technical purposes to ensure we can simply
build appropriate contexts for running systems. Apart from being unkillable,
location $\godloc$ behaves just as any other location.
We denote by $\cset{L}$ the set of configurations.

The syntax of \emph{systems} is defined as follows:
\begin{align*}
	S,R ::= \sys{\net}{N}~|~\res{w}S \quad \text{ where } w \in \ms{N} \cup \ms{C}
\end{align*}

A system is the composition of a network $\net$ with a configuration $N$, 
or a system under a name (channel or location) restriction. 
We denote by $\cset{S}$ the set of systems.
A \emph{network}  $\net$ is a triple $\langle \mc{A},\mc{L}, \mc{V} \rangle$ 
where
\begin{itemize}
\item $\mc{A} : \ms{N}^{\godloc} \rightarrow \mathbb{Z}$ is a function
such that $\mc{A}(\godloc) = 1$ and such that 
the set $\ms{supp}(\mc{A}) \defeq \{ n \in \ms{N} \mid \mc{A}(n) \neq 0 \}$ is finite.
Function $\mc{A}$ may record three types of information on locations. 
If $\mc{A}(n) = \lambda \in \mathbb{N}^{+}$, then 
location $n$ is alive and its current incarnation number is $\lambda$. 
If $\mc{A}(n) = -\lambda, \lambda \in \mathbb{N}^{+}$ then location $n$ 
has been killed and its last incarnation number while alive was $\lambda$. 
If $\mc{A}(n) = 0$, then there is no location $n$ in the network,
alive or not. Because of the finiteness condition above, a network 
can only host a finite number of locations, alive or dead.
In examples, the $\mc{A}$ component of a network is represented as a finite set
of pairs of the form $n \mapsto \lambda$, where $n \in \ms{supp}(\mc{A})$ and
$\lambda = \mc{A}(m)$ (we usually omit the pair $\godloc \mapsto 1$ from network descriptions,
as it is systematically present, by definition).

\item $\mc{L} \subseteq \ms{N}^\godloc \times \ms{N}^\godloc$ is the set of links between locations. 
  $\mc{L}$ is a finite symmetric binary relation over location names. Since it is symmetric we model it as a set of unordered pairs of the form $n \connection m$, namely $n \connection m$ and $m \connection n$ are the same element. We require
  $\ms{dom}(\mc{L}) \defeq \{ n \in \ms{N}^\godloc \mid \exists m. n \connection m \in \mc{L} \}$ to be finite.
%  Since the relation is symmetric we assume that $(n,m)$ and $(m,n)$
%  
%In examples, the $\mc{L}$ component of a network is represented as a finite set of pairs
%of the form $n \connection m$
%, where $(n,m)\in \mc{L}$.
  
\item $\mc{V}:\ms N^{\godloc} \rightarrow (\ms{N}^{\godloc} \rightarrow \mathbb{N})$ is a function 
that maps location names to their \emph{local view}, which is such that the set
$\ms{supp}(\mc{V}) \defeq \{ n \in \ms{N} \mid \mc{V}(n) \neq \hat{\mathbf{0}} \}$ is finite and
$\ms{supp}(\mc{V}) \subseteq \ms{supp}(\mc{A})$. 
The local view of a location $n$ is a function $\mc{V}(n): \ms{N}^{\godloc} \rightarrow \mathbb{N}$
such that the set $\{ m \in \ms{N} \mid \mc{V}(n)(m) \neq 0 \}$ is finite.
If $\mc{V}(n)(m) = \kappa \in \mathbb{N}^{+}$, then location $n$ believes
location $m$ to be alive at incarnation $\kappa$. 
If $\mc{V}(n)(m) = 0$, then location $n$ holds no belief on the status of location $m$.
In examples, the $\mc{V}$ component of a network is represented as a finite set of pairs of the form
$n \mapsto \mc{V}(n)$, where $\mc{V}(n)$ is represented as a finite set of pairs of the form $m \mapsto \kappa$.
\end{itemize}

For convenience we use $\net_\mc{A}$,
$\net_\mc{L}$, and $\net_\mc{V}$ to denote the individual
components of a network $\net$, and we use the following
notations for extracting information from $\net$:
\begin{itemize}
\item $\net\vdash n_\lambda : \mathsf{alive} $ if $\net_\mc{A}(n) = \lambda$ and $\lambda \in \mathbb{N}^{+}$.
\item $\net\vdash n : \mathsf{dead} $ if $\net_\mc{A}(n) \not\in \mathbb{N}^{+}$ 
\item $\net\vdash n \connection m$ if $n\connection m \in \net_{\mc{L}}$
%  or $m\connection n \in \net_{\mc{L}}$}
% \item $\net\vdash n_\lambda \connected m_\kappa$ if  $(n, m) \in \net_\mc{L}$, 
%       $\net \vdash n_\lambda:\ms{alive}$ and $\net \vdash m_\kappa:\ms{alive}$
\end{itemize}

\begin{exa}[Network Representation]
	In the example in Section~\ref{sec:running-example}, the network $\net$, on which the system \textbf{servD} runs, 
	can be defined as follows:
  \begin{align*}
      & \net_{\mc{A}} = \{  n_i \mapsto 1, n_r \mapsto 1, n_b \mapsto 1\} \\
      & \net_{\mc{L}} = \{ n_i \connection n_r, n_r \connection n_b \} \\
      & \net_{\mc{V}} = \{ n_i \mapsto \fnull, n_r \mapsto \fnull, n_b \mapsto \fnull \}
    \end{align*}
  assuming that locations have no belief on other locations.
\end{exa}

We now define update operations over a network $\net$: 

\begin{defi}[Network Updates]

  Network update operations are defined as follows:
  \begin{itemize}
  \item $\net \oplus n \connection m = \langle
     \net_\mc{A}, \net_\mc{L}\cup\{n\connection m\}, \net_\mc{V} \rangle$
	 
  \item $\net \ominus n \connection m = \langle \net_\mc{A},
    \net_\mc{L}\setminus\{n\connection m\}, %m\connection n\},
    \net_\mc{V}\rangle$
	
  \item $\net \oplus (n,\lambda) = \langle \net_\mc{A}[n \mapsto \lambda],
    \net_\mc{L}, \net_\mc{V}[n \mapsto \mathbf{\hat{0}}] \rangle$
	
  \item $\net \ominus (n,\lambda) = \langle \net_\mc{A}[n \mapsto
    -\lambda], \net_\mc{L}, \net_{{\mc{V}}} \rangle$
	
  \item $\net \oplus n \views (m,\lambda) = \langle\net_\mc{A}, \net_\mc{L},
    \net_{\mc{V}}[n \mapsto \net_{\mc{V}}(n)[m \mapsto \lambda] ] \rangle  $, if $ n \neq m$
	
  \item $\net \ominus n \views m = \langle \net_\mc{A}, \net_\mc{L},
    \net_{\mc{V}}[n \mapsto \net_{\mc{V}}(n)[m \mapsto 0]] \rangle$, if $n \neq m$
  
  \item $\net \ominus n \views n = \net \oplus n \views (n,\lambda) = \net$
  \end{itemize}
\end{defi}
$\net \oplus n \connection
m$ and $\net \ominus n \connection m$
add and remove a link, respectively, between $n$ and $m$. 
$\net \oplus
(n,\lambda)$ activates location $n$ with incarnation number $\lambda$, 
and resets its view to the empty one.
$\net \ominus (n,\lambda)$ kills a location in its incarnation $\lambda$. 
$\net \oplus n \views (m,\lambda)$ adds $(m,\lambda)$ to the view of $n$, and 
$\net \ominus n \views m$ removes any belief on location $m$ from the view of $n$.

We use below the notions of \emph{closed} and \emph{well-formed} system, which we now define:

\begin{defi}
	A system $S = \sys{\res{\vvect{w}} \net}{M} \in \cset{S}$ is \emph{closed} iff it does not have any free incarnation variable.
	It is \emph{well-formed} iff the belief that each location $n$ has on any
    remote location $m$ is less or equal than the current incarnation number of $m$, that
    is $\net_{\mc{V}}(n)(m)\leq |\net_{\mc{A}}(m)|$, and for any occurrence of the spawning message $\spawnc{(m,\kappa):P}{n}{\lambda}$ in $M$,
	we have $\lambda \leq |\net_{\mc{A}}(n)|$ and $\kappa \leq |\net_{\mc{A}}(m)|$, where $|\cdot|$ computes the absolute value.
\end{defi}

The notion of well-formed system ensures that spawning messages in a system are consistent with the state of its network,
and in particular that they do not come from, nor target, locations at future
incarnations. From now on, we will only consider well-formed systems.

The definition of free incarnation variables in systems and that of free names in
  configurations is completely standard. The notion of free names in a system is
  slightly unconventional because of the presence of functions inside the
  network. However, free names can be easily defined by leveraging two auxiliary
  functions that return, respectively, the support set and the domain set of a
  given function. Details are given in Appendix~\ref{app:free-names}. Throughout the entire paper we use the Barendregt convention,
  that is if some terms occur in some context, then in these terms all bound variables are chosen to be different from the free ones.

\subsection{Reduction Semantics}\label{sec:red-sem}

The operational semantics of our calculus is defined via a reduction semantics
given by a binary relation $\step \subseteq \cset{S} \times \cset S$ between closed well-formed systems,
and a structural congruence relation $\equiv \; \subseteq \; \cset S^2\cup \cset L^2$, 
that is a binary equivalence relation between systems and between configurations. 
Evaluation contexts are ``systems with a hole $\cdot$'' defined by
the following grammar:
\begin{align*}
	\ctx{C}  ~::=~  \sys{\res{\tilde w}\net}{\ctx E}
  && \ctx{E} ~::=~  \cdot ~\big{|}~ (N \paral \ctx{E})
  && \text{where:}
	&& \vect{w} \subset \ms{N} \cup \ms{C}
\end{align*}

\begin{figure}[t]
  \scriptsize
  \begin{mathpar}
    \inferrule*[left={[S.Par.C]}]{}{
      N \paral M \equiv M \paral N
    }
    \and
    \inferrule*[left={[S.Par.A]}]{}{
      (L \paral M) \paral N \equiv L \paral ( M \paral N)
    }
    \and
    \inferrule*[left={[S.Par.N]}]{}{
      (N \paral \nil) \equiv N
    }
    \and
    \inferrule*[left={[S.Res.C]}]{}{
      \res u \res v S \equiv \res v \res u S
    }
    \and
    \inferrule*[left={[S.Res.Nil]}]{
	  u \notin \fn{S}
    }{
      \res u S \equiv S
    }
    \and
    \inferrule*[left={[S.$\alpha$]}]{
      S  =_\alpha R
    }{
      S \equiv R
    }
    \and
    \inferrule*[left={[S.Ctx]}]{
      N \equiv M
    }{
      \ctx{C}[N] \equiv \ctx{C}[M]
    }
  \end{mathpar}
  \caption{Structural Congruence Rules}
  \label{fig:cong-rls}
\end{figure}

Relation $\equiv$ is the smallest equivalence relation 
defined by the rules in Fig.~\ref{fig:cong-rls}, 
where $=_{\alpha}$ stands for equality up to alpha-conversion,
$M,N, L \in \cset{L}$, and $S,R \in \cset{S}$.
Most rules are mundane. 

Rule \textsc{S.Ctx} turns $\equiv$ into a congruence for the parallel and restriction operators.
Alpha-conversion on systems, which appears in Rule \textsc{S.$\alpha$}, is slightly
unusual since it acts also on $\net$, which contains functions.
However it can be defined straightforwardly (see Appendix~\ref{app:free-names} for
details).

\begin{figure}[t]
  \begin{flushleft}
    Assuming $\Delta \vdash n_\lambda:\ms{alive}$
  \end{flushleft}
  \scriptsize
  \[
    \begin{array}{ll}

      \inferrule*[lab=spawn-l]{
      }{
      \sys{\net}{\proc{\spawn{n}{P}}{n}{\lambda}}
      \step
      \sys{\net}{\proc{P}{n}{\lambda}}
      }
      &
	  \inferrule*[lab=bang]{
	   	  }{
	  \sys{\net}{\proc{!x(\vect{u}).P}{n}{\lambda}}
	   	  \step
	  \sys{\net}{\proc{x(\vect{u}).(P \parop~ !x(\vect{u}).P)}{n}{\lambda}}
	   	  }
      \\[3ex]

        \inferrule*[lab=new]{ u\notin \fn{\net}\cup\{ n  \} }{
        \sys{\net}{\proc{\res u P}{n}{\lambda}}
        \step
        \sys{\res u \net}{\proc{P}{n}{\lambda}}
        }

      &
        \inferrule*[lab=fork]{  }{
        \sys{\net}{\proc{P\parop Q}{n}{\lambda}}
        \step
        \sys{\net}{\proc{P}{n}{\lambda} \paral \proc{Q}{n}{\lambda}}
        }

      \\[3ex]

      \inferrule*[lab=if-eq]{ }{
      \sys{\net}{\proc{\ms{if}~r=r~\ms{then}~P~\ms{else}~ Q}{n}{\lambda}}
      \step
      \sys{\net}{\proc{P}{n}{\lambda}}
      }
      &
        \inferrule*[lab=if-neq]{ r\neq s }{
        \sys{\net}{\proc{\ms{if}~r=s~\ms{then}~P~\ms{else}~Q}{n}{\lambda}}
        \step
        \sys{\net}{\proc{Q}{n}{\lambda}}
        }\\[3ex]

      \inferrule*[lab=node]{ }{
      \sys{\net}{\proc{\node{m,\iota}.P}{n}{\lambda}}
      \step
      \sys{\net}{\proc{P\{n,\lambda/m,\iota\}}{n}{\lambda}}
      }

      &
        \inferrule*[lab=forget]{
        }{
        \sys{\net}{\proc{\remove{m}.P}{n}{\lambda}}
        \step
        \sys{\net \ominus n \views m}{\proc{P}{n}{\lambda}}
        }

      \\[3ex]

	  \inferrule*[lab=msg]{
      }{
      \sys{\net}{\proc{\out{x}{\vect{v}}.Q}{n}{\lambda} \paral \proc{x(\vect{u}).P}{n}{\lambda}}
      \step
      \sys{\net}{\proc{Q}{n}{\lambda} \paral \proc{P\sub{\vect{v}}{\vect{u}}}{n}{\lambda}}
      }

    \end{array}
  \]
  \caption{Local Rules}
  \label{fig:red-loc-rls}
\end{figure}

The reduction relation $\step$ is defined by the rules in Figs.~\ref{fig:red-loc-rls}, 
\ref{fig:red-dist-rule} and \ref{fig:ctx-rls}.
Fig.~\ref{fig:red-loc-rls} depicts the \emph{local} reduction rules, i.e., those rules
that involve only a single location and that essentially do not modify the
network.
Rule \rl{spawn-l} defines the local launch of process $P$.
Rule \rl{bang} defines the expansion of a replicated input process.
Rule \rl{new} performs the scope extrusion of a name from a process to the system level.
Rule \rl{fork} turns a parallel composition into parallel processes in the same location.
Rules \rl{if-eq}, \rl{if-neq} define the semantics of the branching construct.
Rule \rl{node} gets hold of the current location name and its incarnation number for
further processing.
Rule \rl{forget} deletes from the local view of the current location the belief it may hold about 
a given location $m$.
Finally, rule \rl{msg} defines the receipt of a local message by an input
process.
We introduced rules \rl{new}, \rl{bang} and \rl{fork} as computational steps instead of as structural congruence rules
for it simplifies our proofs.

\begin{figure}[t]
  \scriptsize
  % Assuming $\Delta \vdash n_\lambda:\ms{alive}$
  % \inferrule*[lab=spawn-s,rightstyle=\scriptsize,right={$
  % \begin{array}{l}
  %   \net_n(m) = \kappa\\
  %   \net \vdash n_\lambda \connected m_\kappa
  % \end{array}
  % $}]{
  % }{
  %   \sys{\net}{\proc{\spawn{m}{P}}{n}{\lambda}}
  %   \step
  %   \sys{\net \oplus m\views (n,\lambda)}{\proc{P}{m}{\kappa}}
  % }\\[3ex]
  %   
  %   \inferrule*[lab=spawn-f,rightstyle=\scriptsize,right={$
  %   \begin{array}{l}
  %     \net_n(m) = \kappa\\
  %     \net \not\vdash n_\lambda \connected m_\kappa
  %   \end{array}
  %   $}]{
  % }{
  %   \sys{\net}{\proc{\spawn{m}{P}}{n}{\lambda}}
  %   \step
  %   \sys{\net  \ominus n \views m}{\nil}
  % }\\[3ex]
  \[
  \begin{array}{ll}
      \inferrule*[lab=link]{ \net \vdash n_\lambda:\ms{alive} \\
	   \net \not\vdash n \connection m
	   }{
      \sys{\net}{\proc{\link{m}.P}{n}{\lambda}}
      \step
      \sys{\net \oplus n \connection m}{\proc{P}{n}{\lambda}}
      }
    &
      \inferrule*[lab=unlink]{ \net \vdash n_\lambda:\ms{alive} \\ 
	  \net \vdash n \connection m}{
      \sys{\net}{\proc{\sbreak{m}.P}{n}{\lambda}}
      \step
      \sys{\net \ominus n \connection m}{\proc{P}{n}{\lambda}}
      }\\[3ex]
      \inferrule*[lab=create-s]{ \Delta \vdash n_\lambda:\ms{alive} \\
        \net \vdash m :\ms{dead}  \\\\
        \Delta_{\mc{A}}(m)= -\kappa \\
		\kappa \in \mathbb{N}
      }{
      \sys{\net}{\proc{\create{m}{P}}{n}{\lambda}}
      \step
      \sys{\net \oplus (m,\kappa + 1)}{\proc{P}{m}{\kappa+1}}
      }
    &
      \inferrule*[lab=create-f]{ \Delta \vdash n_\lambda:\ms{alive} \\
	   \net \not\vdash m:\ms{dead} }{
      \sys{\net}{\proc{\create{m}{P}}{n}{\lambda}}
      \step
      \sys{\net}{\nil}
      }\\[3ex]
      \inferrule*[lab=kill]{ \Delta \vdash n_\lambda:\ms{alive} \\
	   n \neq \godloc }{
      \sys{\net}{\proc{\skill}{n}{\lambda}}
      \step
      \sys{\net \ominus (n,\lambda)}{\nil}
    }
    & \\[3ex]
	
      \inferrule*[lab=spawn-c-s]{ \Delta \vdash n_\lambda:\ms{alive} \\ 
	  \net \vdash n \connection m \\\\
	  n \neq m \\
    \net_{\mc{V}}(n)(m) = \kappa \in \mathbb{N}
    }{
    \sys{\net}{\proc{\spawn{m}{P}}{n}{\lambda}}
    \step
    \sys{\net}{\spawnc{(m,\kappa):P}{n}{\lambda}}
    }
	&
	
      \inferrule*[lab=spawn-c-f]{ \Delta \vdash n_\lambda:\ms{alive} \\ 
	  \net \not\vdash n \connection m \\
	  n \neq m
    }{
    \sys{\net}{\proc{\spawn{m}{P}}{n}{\lambda}}
    \step
    \sys{\net \ominus n \views m}{\nil}
    }
	\\[3ex]
	
    \inferrule*[lab=spawn-s]{ \net_{\mc{A}}(m) = \kappa \\
    \kappa > 0 \\
	  (\kappa^{\ast} = \kappa \vee \kappa^{\ast} = 0) \\\\
	  \net \vdash n \connection m \\
    \net_{\mc{V}}(m)(n) \leq \lambda
    }{
    \sys{\net}{\spawnc{(m,\kappa^{\ast}):P}{n}{\lambda}}
    \step
    \sys{\net \oplus m \views (n,\lambda)}{\proc{P}{m}{\kappa}}
    }
	  &
    \inferrule*[lab=spawn-f]
    {
    (\net_{\mc{A}}(m) \neq \kappa \wedge \kappa \neq 0) \\\\
	\vee \; \net \not\vdash n \connection m \;
    \vee \; \lambda < \net_{\mc{V}}(m)(n)
    }{
    \sys{\net}{\spawnc{(m,\kappa):P}{n}{\lambda}}
    \step
    \sys{\net \ominus n \views m}{\nil}
    }\\[3ex]
  \end{array}
  \]
  \caption{Distributed Rules}
  \label{fig:red-dist-rule}
\end{figure}

Fig.~\ref{fig:red-dist-rule} depicts the \emph{distributed} rules, i.e.,
rules that involve locations and spawning messages, or modify the network.

Rules \rl{link} and
\rl{unlink} define respectively the establishment and the removal of a link.
Rule \rl{create-s} defines the successful creation of a location if it never existed or its reactivation if it was crashed. 
The newly activated location has an incarnation number that is the successor of the previous one 
($0$ by convention if the location was not present in the network). 
Note that we do not require the existence of an alive
link between the current location and the one to be activated since we want 
this operation to also model the possibility of interventions external to the system, 
such as those performed by human administrators.
% Note also that a location cannot be
% created anew each time, otherwise it would be impossible to resume the
% execution of a service under a well-known name.
% The use of incarnation numbers provides support for recovery schemes, as
% anticipated in the Introduction.
Rule \rl{create-f} defines the silent
failure of a create operation, which can fail because the location to activate
may already be alive.
Rule \rl{kill} defines the killing of a location.
% Rule \rl{spawn-c}, for \emph{spawn-commit}, generates a spawn message, 
% where the target locality comes with the incarnation number found in the local view of the spawning locality. 
% The spawn message will later be (resp. fail to be) consumed by the target locality 
% (rule \rl{spawn-s}, resp. rule \rl{spawn-f}). 

Rule \textsc{spawn-c-s}, for successful  \emph{spawn-commit}, generates a spawning message, 
where the target location comes with the incarnation number found in the local view of the spawning location. 
The spawning message will later be (resp. fail to be) received by the target location 
(rule \textsc{spawn-s}, resp. rule \textsc{spawn-f}). 
Rule \textsc{spawn-c-f} accounts for an unsuccessful spawn commit, due to the lack of a link
between the spawning location and the target one.

Note that, because of the non atomicity of the transfer
of the spawning message to the target location, the rules have to account for a possible mismatch 
between the incarnation number $\lambda$ of the spawning location $n$, 
and the incarnation number recorded at the target location $m$ for location $n$. 
In other words, target location $m$ may receive a spawning message from
an old incarnation of $n$, i.e.\ one which has crashed after the release of the spawning message. 
Thus, the spawn operation ultimately succeeds (rule \textsc{spawn-s}) 
only if the incarnation number $\lambda$ of the spawning location $n$ 
in the spawning message is greater or equal than the incarnation number $\Delta_{\mc{V}}(m)(n)$ currently attributed by $m$ to $n$, 
preventing the receipt of spawning messages from a known crashed location. 
Arguably, this is what happens in practice in Erlang 
and the side conditions involving incarnation numbers in rules \textsc{spawn-s} and \textsc{spawn-f} 
are checks which are performed by the Erlang environment 
to prevent duplication of spawning messages in case of crash and subsequent recovery of a node.
% Rule \rl{spawn-s} defines a successful spawn, conditional upon the fact that  a
% link exists between the spawning and target locations,
% and that the spawning location rightly believes
% the target location, with the incarnation recorded in its view, to be alive,
% or has no belief on the target location in its local view.
% This last constraint is captured by the side
% condition $\net_n(m) = \kappa$, which we formally define as follow:
% \[
%     \net_n(m) =
%       \begin{cases}
%       \kappa
%       \hspace*{1cm}
%         \ms{if~} n = m \ms{~~and~~} \net_{\mc{A}}(n) = \kappa
%       \\
%       \kappa
%       \hspace*{1cm}
%         \ms{if~} n \neq m \ms{~~and~~} \net_{\mc{V}}(n)(m) = \kappa
%       \\
%       \kappa
%       \hspace*{1cm}
%         \ms{if~} n \neq m \ms{~~and~~} \net_{\mc{V}}(n)(m) = 0 \ms{~~and~~} \net_{\mc{A}}(m) = \kappa
%       \\
%       0
%       \hspace*{1cm}
%         \ms{otherwise}
%     \end{cases}
% \]
% Note that a successful spawn updates the target local view with the belief that the spawning location
% is alive with the incarnation number it had when it initiated the spawn.
% Rule \rl{spawn-f} defines a failed spawn, which may fail due to a wrong view, to a missing
% link between the two locations, or because the remote location is not alive.
% The view of the sender is updated by removing the belief on the
% target. This behavior is inspired by Erlang whose runtime, in a manner transparent
% to the user, updates the view of the spawning location if it receives no acknowledgement
% from the target as part of its distributed protocol.

\begin{figure}
  \scriptsize
  \begin{mathpar}
    \inferrule*[lab=par]{
      \sys{\net}{N
        \step
        \sys{\res{\vect u}\net'}{N'}}
      \quad
      \vect{u} \cap \fn{M} =\emptyset
    }{
      \sys{\net}{N\paral M}
      \step
      \sys{\res{\vect u}\net'}{N' \paral M}
    }
    \qquad
    \inferrule*[lab=res]{
      S \step S'
    }{
      \res u S
      \step
      \res u S'
    }
    \qquad
    \inferrule*[lab=str]{
      S \equiv S' \quad S' \step R'  \quad R'\equiv R
    }{
      S
      \step
      R
    }
      \end{mathpar}
  \caption{Contextual Rules}
  \label{fig:ctx-rls}
\end{figure}

Fig.~\ref{fig:ctx-rls} shows the contextual rules of our calculus. Rules
\rl{par} and \rl{res} define, respectively, parallel
execution and execution under restriction. Rule \rl{str} allows the use of
structural congruence before and after a reduction step.
Rule \rl{par} is slightly unconventional in its use of restriction.
When we consider the case where $\vect{u}$ is empty in rule \rl{par}, 
we obtain a more standard-looking contextual rule for the parallel operator.
However, we also have to consider cases when the active component
promotes a restriction at system level (via an application of rule \rl{new}).
In this case we have to avoid name capture by the idle component.

\begin{exa}
  Here we show some reductions in the interaction between \textbf{servD}, from Section~\ref{sec:running-example}, and its
  client $\proc{\spawn{n_i}{(\out{req}{h,z} \parop z(w).Q})}{n_k}{}$ on network $\net$,
  defined as follows:
  {
    \footnotesize
    \begin{align*}
      & \net_{\mc{A}} = \{  n_i \mapsto 1, n_r \mapsto 1, n_b \mapsto 1, n_k \mapsto 1 \}
      & \net_{\mc{L}} = \{ n_i \connection n_r, n_r \connection n_b, n_k \connection n_i \} \\
      & \net_{\mc{V}} = \{ n_i \mapsto \fnull, n_r \mapsto \fnull, n_b \mapsto \fnull, n_k \mapsto \fnull \}
    \end{align*}
  }
  To ease the parsing of the reduction we underline the terms that get reduced and indicate the key reduction rule applied.
  {\scriptsize
  \begin{align*}
    & \sys{\res{\vect{u}}\net}{
          \underline{\proc{\spawn{n_i}{(\out{req}{h,z} \parop z(w).Q})}{n_k}{}} \paral
          \proc{ req(y,z).\spawn{n_r}{\out{r_1}{y,z}}  }{n_i}{}
          \paral \proc{ R }{n_r}{}
          \paral \proc{ B }{n_b}{}
          } &
	& \step && \textsc{spawn-c-s} \\
    & \sys{\res{\vect{u}}\net}{
               \underline{\spawnc{(n_i,0):\out{req}{h,z} \parop z(w).Q}{n_k}{}} \paral
               \proc{ req(y,z).\spawn{n_r}{\out{r_1}{y,z}}  }{n_i}{}
          \paral \proc{ R }{n_r}{}
          \paral \proc{ B }{n_b}{}
          } &
	& \step && \textsc{spawn-s} \\
    & \sys{\res{\vect{u}}\net'}{
               \underline{\proc{\out{req}{h,z} \parop z(w).Q}{n_i}{}} \paral
               \proc{ req(y,z).\spawn{n_r}{\out{r_1}{y,z}}  }{n_i}{}
               \paral \proc{ R }{n_r}{}
               \paral \proc{ B }{n_b}{}
               } &
	& \step && \textsc{fork} \\
	    & \sys{\res{\vect{u}}\net'}{
	               \underline{\proc{\out{req}{h,z}}{n_i}{}}
                 \paral \proc{z(w).Q}{n_i}{}
	               \paral \underline{\proc{ req(y,z).\spawn{n_r}{\out{r_1}{y,z}}  }{n_i}{}}
	               \paral \proc{ R }{n_r}{}
	               \paral \proc{ B }{n_b}{}
	               } &
	& \step &&  \textsc{msg}\\
    & \sys{\res{\vect{u}}\net'}{
      \proc{z(w).Q}{n_i}{}
      \paral 
	               \proc{\spawn{n_r}{\out{r_1}{h,z}}}{n_i}{}    
	               \paral \proc{ R }{n_r}{}
	               \paral \proc{ B }{n_b}{}
	               } 
  \end{align*}}
  where network $\net'$ is similar to $\net$ but with the following view:
  {
    \footnotesize
    \begin{align*}
      & \net'_{\mc{V}} = \{ n_i \mapsto \{ n_k \mapsto 1 \}, n_r \mapsto \fnull, n_b \mapsto \fnull, n_k \mapsto \fnull \}
    \end{align*}
  }

  On the first reduction the client successfully spawns a message,
  bearing the request $\out{req}{h,z}$, towards the interface of
  \textbf{servD}. After the spawn, the spawn message is correctly
  delivered to its receiver. The freshly spawned process first action
  is to fork. Finally, the process descending from the fork, bearing
  the request, and the process awaiting for a request on the
  interface, synchronize, so that the request is now ready to be
  forwarded to the router.
\end{exa}

A check on the consistency of our reduction semantics is given by the following proposition.

\begin{prop}
	If $S$ is a closed well-formed system, and $S \step T$, then $T$ is a closed well-formed system.
\end{prop}
\begin{proof}
	A simple induction on the derivation of $S \step T$, 
	noting for well-form\-ed\-ness that the only rules updating the view of a location, \textsc{forget} and
	\textsc{spawn-s}, preserve well-formedness and ensure that incarnation numbers attributed to locations
	in views are actual, and that the only rule
	creating a new spawning message, \textsc{spawn-c}, henceforth preserves well-formedness.
\end{proof}

Our semantics for the parallel composition operator, coupled with the systematic presence of 
the distinguished location $\godloc$ in any network, spares us the need 
to introduce a parallel composition operator between systems.
The intuition is that
we can always extend a system with a process located on $\godloc$ (to ensure it
is alive) performing the desired changes on the public part of the network.
Indeed, given a network $\net = \langle \mc{A}, \mc{L}, \mc{V} \rangle $, we can change it to
network $\net' = \langle  \mc{A'}, \mc{L'}, \mc{V'} \rangle$, 
provided the following consistency constraints are respected:
\begin{itemize}
\item The new function $\mc{A}'$, for each location $n$, must agree with $\mc{A}$ on
  private names and must map each free location $n$ to either
  the same incarnation number or to a future one. Notice that, e.g., $-3$ is in the future of $2$ because from $2$ we can get to $-3$ if the location
  fails, is re-activated and fails again. Similarly, $-3$ is also in the future of
  $3$ because from $3$ we can get to $-3$ if the location fails.
\item The new link set $\mc{L'}$ must agree with $\mc{L}$ on links which have at least a private end.
\item The new view function $\mc{V'}$ must i) agree with $\mc{V}$ on private names; ii)
  for each location $n$, map each belief that $n$ has about $m$ to a non-negative incarnation number smaller or equal than
  the absolute value of the incarnation number of $m$ in $\mc{A'}$.
\end{itemize}
This is formalized in Proposition~\ref{prop:extension}, whose
  proof is available in Appendix~\ref{app:proof-network} as well.
The constraints above are required to be sure that the target system is
reachable by our reduction semantics.
Note that a
parallel composition operator acting on systems would have to enforce
analogous consistency constraints between the composed systems.

%%% Local Variables:
%%% mode: latex
%%% TeX-master: "main-distributed"
%%% End:

\section{Behavioral Theory}\label{sec:behavioral-theory}

In this section we present the behavioral theory for our
calculus. We first define a contextual equivalence
in the classical form of a weak barbed congruence, a notion originally proposed 
by Milner and Sangiorgi in \cite{MilnerS92}. We then present a labelled transition system semantics (LTS)
for our calculus and we show that weak bisimilarity between (closed well-formed) 
D$\pi$FR systems coincides with weak-barbed congruence (our full-abstraction result).
This result is valuable for it provides a compositional proof technique for proving or disproving 
contextual equivalence between D$\pi$FR systems. E.g., later in Section~\ref{sec:examples} we will
use weak bisimilarity to prove equivalent the ideal server \textbf{servD} with
the more concrete server \textbf{servDFR}, hence proving the latter correct with
respect to the former.
The developments and the proof of full abstraction are relatively standard but the subtlety
lies in correctly accounting for the contribution of the network in the LTS semantics,
even though the network is not a running process in our calculus, 
and in dealing with the non-standard handling of restriction.

% introduce the elements necessary to define the notion of
% (weak) barbed congruence. We then introduce a labelled transition system semantics
% for our calculus. Finally we show that for D$\pi$FR systems
% weak bisimilarity  coincides with weak barbed congruence.
%
% This result, in the literature referred as \emph{full abstraction}, is valuable because two
% systems are distinguishable only if there is a computational context able to
% observe a behavioral difference and not because they differ in theirs internal
% structure or in theirs internal communications. Being able to reason on system
% behaviors, rather than their internal structure, is a powerful tool to reason
% about their correctness. For e.g., later in Section~\ref{sec:examples} we will
% use weak bisimilarity to prove equivalent the ideal server \textbf{servD} with
% the more concrete server \textbf{servDFR}, hence proving the latter correct with
% respect to the former.
%
% \red{Mitigate the impact of standard by highlighting the difficulties}
% The framework of the proof which we use is standard; we refer the reader
% to~\cite{FrancalanzaH08,AmadioCS98,SangiorgiW01} for additional discussion.

\subsection{Weak Barbed Congruence}

We define a standard notion of contextual equivalence
called  \emph{weak barbed congruence}, originally proposed in \cite{MilnerS92}.
We denote by $\wstep$ the reflexive and transitive closure of the reduction relation $\step$.
We rely on a notion of observables on systems, called \emph{barbs},
formally defined as follows:

\begin{defi}[Barb]\label{def:barb}
  A system $S$ exhibits a
  barb on channel $x$ at location $n$ in its incarnation $\lambda$, in symbols $S\barb{x@n_{\lambda}}$, iff
  $S \equiv \sys{\res{\vect{u}}\net}{\proc{\out{x}{\vect{v}}.P}{n}{\lambda} \paral N} $, 
  for some $\vect{u},\vect{v}, P, N$, where
  $x,n \notin \vect u$, and $\Delta \vdash n_\lambda:\ms{alive}$. 
  Also, $S\wbarb{x@n_{\lambda}}$ iff $S \Rightarrow S'$ and $S'\barb{x@n_{\lambda}}$.
\end{defi}

We now define standard properties expected for a contextual equivalence.

\begin{defi}[System Congruence]
	A binary relation $\relation{R}$ over closed systems is a \emph{system congruence} iff
	it is an equivalence relation and
	whenever ${\sys{\res{\vect{u}}\net_1}{N} ~~\relation{R}~~ \sys{\res{\vect{v}}\net_2}{M}}$, 
	for any names $\vect{w}$ and for any configuration $L$ such that $fn(L) \cap \vect{u} = fn(L) \cap \vect{v} = \emptyset$,
	we have: 
	$$\sys{\res{\vect{w}}\res{\vect{u}}\net_1}{N \parallel L} ~~~\relation{R}~~~ \sys{\res{\vect{w}}\res{\vect{v}}\net_2}{M \parallel L}$$
\end{defi}

\begin{defi}[Weak Barb-Preserving Relation]
	A binary relation $\relation{R}$ over closed systems is
	 \emph{weak barb-preserving} iff whenever  $S~\relation{R}~R$ and $S\barb{x@n}$ then
	  $R\wbarb{x@n}$.
\end{defi}

\begin{defi}[Weak Reduction-Closed Relation]
	A binary relation $\relation{R}$ over closed systems is
	  \emph{weak reduction-closed} iff whenever $S~\relation{R}~R$ and $S\step S'$ then
	  $R \wstep R'$ for some $R'$ such that $S'~\relation{R}~R'$.
\end{defi}

We can now define weak barbed congruence.

\begin{defi}[Weak Barbed Congruence]\label{def:barbed-congruence}
  \emph{Weak barbed congruence}, noted $\bcong$, is the
  largest weak barb-preserving, reduction-closed, system congruence.
\end{defi}

As a first result, we can check that structural congruence is included in weak barbed congruence:

\begin{prop}
	Structural congruence $\equiv$ is a weak barb-preserving re\-duction-closed system congruence.
\end{prop}
\begin{proof}
	Structural congruence is a barb-preserving system congruence by definition.
	Structural congruence is weak reduction-closed thanks to 
	Propositions~\ref{prop:reductionsaretausteps} and
  \ref{prop:taustepsarereductions} in
	Appendix~\ref{app:full-abstraction}.
\end{proof}

\newcommand{\lcong}{\bcong_l}

Much as in~\cite{NicolaGP07}, a simpler kind of barbs gives rise to the same
barbed congruence. The alternate definition of barb we consider in this section
is one where a barb just displays the location of messages.

\begin{defi}[Location Barb]\label{def:barblocation}
  We say that a system $S$ exhibits a
  barb at location $n$, in symbols $S\barb{n}$, iff
  $S \equiv \sys{\res{\vect{u}}\net}{\proc{\out{x}{\vect{v}}.P}{n}{\lambda} \paral N} $, 
  for some $x,\lambda, \vect{u},\vect{v}, P, N$, where
  $x,n \notin \vect u$, and $\Delta \vdash n_\lambda:\ms{alive}$. 
  Also, $S\wbarb{n}$ iff $S \Rightarrow S'$ and $S'\barb{n}$.
  We denote by $\lcong$ the weak barbed congruence obtained by considering barbs at locations instead of barbs at both channels and locations.
\end{defi}

The two weak barbed congruences $\bcong$ and $\lcong$ coincide

\begin{restatable}[Equivalence of barbed congruences]{prop}{barbsequiv}
	$\lcong = \bcong$
\end{restatable}
  \begin{proof}
	In what follows we use the following notation : if $U \equiv \sys{\res{\vect{u}}\net}{N}$, and
	$L$ is such that $\fn{L} \cap \vect{u} = \emptyset$, then $U \paral L$ denotes the system
	$\sys{\res{\vect{u}}\net}{N \paral L}$.
	
	That $\bcong \subseteq \lcong$ is clear. We show the converse.
	Let closed systems $S$ and $T$ be such that $S \lcong T$ and $S \barb{x@n_\lambda}$.
	By definition of barbs at channels and locations, we have $S \equiv \sys{\res{\vect{u}}\net}{\proc{\out{x}{\vect{v}}.P}{n}{\lambda} \paral N}$
	for some $\vect{u},\vect{v},P,N$. By definition of location barbs, we have $S\barb{n}$.
	Since $S \lcong T$, we must have $T \wstep T_1 \barb{n}$ for some $T_1$. Assume for the sake of contradiction that 
	$\neg (T \wbarb{x@n_\lambda})$, and consider the systems $S \paral L$ and $T \paral L$, where:
	\begin{align*}
		& L = \proc{\create{m}{\link{n}.{ \spawn{n}{(\node{\texttt{\_},\iota}.\ift{\iota = \lambda}{x(\vect{y}).\spawn{m}{\eout{t}})}}}}}{\godloc}{1} \\
		& \text{with }m,t \notin \fn{S} \cup \fn{T}
	\end{align*}
	Since $\lcong$ is a system congruence, we must have $S \paral L ~\lcong~ T \paral L$. 
	Now, by construction, we have 
	\begin{align*}
		& S \paral L \wstep \sys{\res{\vect{u}}\Upsilon}{N \paral \proc{P}{n}{\lambda} \paral \proc{\eout{t}}{m}{1}} = S'
	\end{align*}
	where $\Upsilon$ is $\net$ extended with live location $m$ at incarnation number $1$ and link $m \connection n$.
	We thus have $S \paral L \wstep S'$ and $S'\barb{m}$.
	But since $\neg (T \wbarb{x@n_\lambda})$, and $t,m$ are fresh for $T$, 
	there can be no $T'$ such that $T \paral L \wstep T'$ and $T'\wbarb{m}$, contradicting the fact that $S ~\lcong~ T$.
\end{proof}

  One may notice that in the definition of the less precise location barbs,
  in the barb itself, there is no mention of the incarnation number of the
  observed locality and wonder if an alternative definition with mention of it
  may have different observational power. The quick answer to this is no.
  Indeed, with the current definition of location barbs, even without mentioning
  the incarnation number in the observable, it is possible to distinguish two
  systems that differ because of the presence of a same locality with different
  incarnation numbers. The intuition is that it is always possible to build a
  context that grabs the incarnation number, tests it and then branches
  according to its value. A context of this kind is also visible in the above proof,
  when we need to make sure that the incarnation number of locality $n$ is the
  same on both systems, and we achieve so by first grabbing the incarnation
  number through the \tms{node} primitive and then we test it with the \tms{if}
  construct.

From now on, we will use the more detailed observables $\barb{x@n_{\lambda}}$ for convenience,
to simplify certain arguments in our proofs.

\subsection{A Labeled Transition Semantics}

In this subsection we present a labeled transition semantics for our calculus in
order to have a co-inductive characterization of weak barbed congruence. Working
with the labeled transition semantics is more convenient as it avoids the need
to work with a universal quantification over contexts.
Labels $\alpha$ in our LTS semantics take the following forms:
\[
  \begin{array}{ll}
    \alpha ::= & \tau ~\big{|}~
                 \res{\vect{w}}\out{x}{\vect{u}}@n_{\lambda} ~\big{|}~ 
				 x(\vect{u})@n_{\lambda} ~\big{|}~
                \ms{kill}(n,\lambda) ~\big{|}~ 
				\ms{create}(n,\lambda) ~\big{|}~\\
               &
                  \oplus n_\lambda \mapsto m ~\big{|}~ 
                 \ominus n_\lambda \mapsto m ~\big{|}~ 
				n_\lambda \views m
  \end{array}
\]
The first three labels are classical: silent action, output action (possibly with restricted names), and input action.
Output and input actions mention the name and incarnation number of the location performing the action.
Labels $\ms{kill}(n,\lambda)$ and $\ms{create}(n,\lambda)$ indicate respectively the
killing and activation of location $n$ at its incarnation $\lambda$. 
Labels $\oplus n_\lambda \mapsto m$ and $\ominus n_\lambda \mapsto m$ signal respectively the creation and destruction of a link
between $n$ and $m$, initiated by $n$ at incarnation $\lambda$. 
Finally, $n_\lambda \views m$ signals that location $n$ at incarnation $\lambda$ holds the correct belief about location $m$.

Free names in labels are defined as follows:
  \begin{align*}
    &\fn{\tau} = \emptyset  && \fn{\res{\vvect{w}}\out{x}{\vect{u}}@n_{\lambda}} = (\{ x , n \} \cup \vect{u})\setminus \vvect{w}
    && \fn{x(\vect{u})@n_{\lambda}} = \vect{u} \cup \{x,n\} \\
    & \fn{\ms{kill}(n,\lambda)} = \{n \} && \fn{\ms{create}(n,\lambda)} = \{ n \} \\
    & \fn{\oplus n_\lambda \mapsto m} = \{n,m \} && \fn{\ominus n_\lambda \mapsto m} = \{n,m \}
    && \fn{n_\lambda \views m} = \{n,m \}
  \end{align*}

A transition labeled by $\alpha$ is denoted $\lstep{\alpha}$.
We denote by $\lwstep{\tau}$ the reflexive and transitive closure of $\lstep{\tau}$.
For $\alpha \neq \tau$, we denote by $\lwstep{\alpha}$ the relation $\lwstep{\tau} \lstep{\alpha}\lwstep{\tau}$.

\begin{figure}[t]
  \scriptsize
  \begin{flushleft}
    Assuming $\Delta \vdash n_\lambda:\ms{alive}$
  \end{flushleft}
  \[
    \begin{array}{ll}

	  \inferrule*[lab=l-bang]{
	   	  }{
	  \sys{\net}{\proc{!x(\vect{u}).P}{n}{\lambda}}
	   	  \lstep{\tau}
	  \sys{\net}{\proc{x(\vect{u}).(P \parop~ !x(\vect{u}).P)}{n}{\lambda}}
	   	  }
	  
      &
	  
        \inferrule*[lab=l-fork]{  }{
        \sys{\net}{\proc{P\parop Q}{n}{\lambda}}
        \lstep{\tau}
        \sys{\net}{\proc{P}{n}{\lambda} \paral \proc{Q}{n}{\lambda}}
        }
		  
      \\[3ex]

      \inferrule*[lab=l-new]{
      u\notin \fn{\net}\cup\{n,\lambda\}
      }{
        \sys{\net}{\proc{\res u P}{n}{\lambda}}
        \lstep{\tau}
        \sys{\res u \net}{\proc{P}{n}{\lambda}}
        }

      &

      \inferrule*[lab=l-node]{ }{
      \sys{\net}{\proc{\node{m,\iota}.P}{n}{\lambda}}
      \lstep{\tau}
      \sys{\net}{\proc{P\sub{n,\lambda}{m,\iota}}{n}{\lambda}}
      }

      \\[3ex]

      \inferrule*[lab=l-if-neq]{
      r\neq s
      }{
	  \sys{\net}{\proc{\ms{if}~r=s~\ms{then}~P~\ms{else}~Q}{n}{\lambda}}
	  \lstep{\tau}
	  \sys{\net}{\proc{Q}{n}{\lambda}}
	  }

      &

      \inferrule*[lab=l-forget]{
      }{
      \sys{\net}{\proc{\remove{m}.P}{n}{\lambda}}
      \lstep{\tau}
      \sys{\net \ominus n \views m}{\proc{P}{n}{\lambda}}
      }

      \\[3ex]

	  \inferrule*[lab=l-if-eq]{ }{
      \sys{\net}{\proc{\ms{if}~r=r~\ms{then}~P~\ms{else}~ Q}{n}{\lambda}}
      \lstep{\tau}
      \sys{\net}{\proc{P}{n}{\lambda}}
      }
	  
	  &
	  
      \inferrule*[lab=l-kill]{
      n \neq \godloc
      }{
      \sys{\net}{\proc{\skill}{n}{\lambda}}
      \lstep{\tau}
      \sys{\net \ominus (n,\lambda)}{\nil}
      }
	  
    \end{array}
  \]
  \caption{LTS: Local Rules}
  \label{fig:lts-local-rules}
\end{figure}

\begin{figure}[t]
  \scriptsize
  \begin{flushleft}
    %Assuming $\Delta \vdash n_\lambda:\ms{alive}$
  \end{flushleft}
  \[
    \begin{array}{ll}

      \inferrule*[lab=l-in]{ \Delta \vdash n_\lambda:\ms{alive} }{
      \sys{\net}{\proc{x(\vect{v}).P}{n}{\lambda}}
      \lstep{x(\vect{u})@n_\lambda}
      \sys{\net}{\proc{P\sub{\vect{u}}{\vect{v}}}{n}{\lambda}}
      }

      &

      \inferrule*[lab=l-out]{ \Delta \vdash n_\lambda:\ms{alive} }{
        \sys{\net}{\proc{\out{x}{\vect{v}}.P}{n}{\lambda}}
        \lstep{\out{x}{\vect{v}}@n_\lambda}
        \sys{\net}{\proc{P}{n}{\lambda}}
      }
	  
	  \\[3ex]

          \inferrule*[lab=l-link]{
            \Delta \vdash n_\lambda:\ms{alive} \\ 
      \net \not\vdash n \connection m
      }{
      \sys{\net}{\proc{\link{m}.P}{n}{\lambda}}
      \lstep{\tau}
      \sys{\net \oplus n \connection m}{\proc{P}{n}{\lambda}}
      }

	  &
	  
      \inferrule*[lab=l-unlink]{
        \Delta \vdash n_\lambda:\ms{alive} \\ 
      \net \vdash n \connection m
      }{
      \sys{\net}{\proc{\sbreak{m}.P}{n}{\lambda}}
      \lstep{\tau}
      \sys{\net \ominus n \connection m}{\proc{P}{n}{\lambda}}
      }\\[3ex]

      \inferrule*[lab=l-create-s]{
        \Delta \vdash n_\lambda:\ms{alive} \\ 
      \net \vdash m :\ms{dead} \\
      \Delta_{\mc{A}}(m)= -\kappa \\
      }{
      \sys{\net}{\proc{\create{m}{P}}{n}{\lambda}}
      \lstep{\tau}
      \sys{\net \oplus (m,\kappa + 1)}{\proc{P}{m}{\kappa+1}}
      }

      &

      \inferrule*[lab=l-create-f]{
        \Delta \vdash n_\lambda:\ms{alive} \\ 
        \net \not\vdash m:\ms{dead}
        }{
      \sys{\net}{\proc{\create{m}{P}}{n}{\lambda}}
      \lstep{\tau}
      \sys{\net}{\nil}
        }
        \\[3ex]

        \inferrule*[lab=l-spawn-l]{
          \Delta \vdash n_\lambda:\ms{alive}
        }{
        \sys{\net}{\proc{\spawn{n}{P}}{n}{\lambda}}
        \lstep{\tau}
        \sys{\net}{\proc{P}{n}{\lambda}}
        }\\[3ex]

      \inferrule*[lab=l-spawn-c-s]{ \Delta \vdash n_\lambda:\ms{alive} \\ 
	  \net \vdash n \connection m \\\\
	  n \neq m \\
    \net_{\mc{V}}(n)(m) = \kappa \in \mathbb{N}
    }{
    \sys{\net}{\proc{\spawn{m}{P}}{n}{\lambda}}
    \lstep{\tau}
    \sys{\net}{\spawnc{(m,\kappa):P}{n}{\lambda}}
      }
      &
	
      \inferrule*[lab=l-spawn-c-f]{ \Delta \vdash n_\lambda:\ms{alive} \\ 
	  \net \not\vdash n \connection m \\
	  n \neq m
    }{
    \sys{\net}{\proc{\spawn{m}{P}}{n}{\lambda}}
    \lstep{\tau}
    \sys{\net \ominus n \views m}{\nil}
      }
      \\[3ex]

      \inferrule*[lab=l-spawn-s,rightstyle=\scriptsize]{ 
	  \net_{\mc{A}}(m) = \kappa \\
	  (\kappa^{\ast} = \kappa \vee \kappa^{\ast} = 0) \\\\
	  \net \vdash n \connection m \\
      \net_{\mc{V}}(m)(n) \leq \lambda
      }{
      \sys{\net}{\spawnc{(m,\kappa^{\ast}):P}{n}{\lambda}}
      \lstep{\tau}
      \sys{\net \oplus m \views (n,\lambda)}{\proc{P}{m}{\kappa}}
      }

      &

      \inferrule*[lab=l-spawn-f,rightstyle=\scriptsize]{ 
	  (\net_{\mc{A}}(m) \neq \kappa  \neq 0) \; \vee \; \net \not\vdash n \connection m \; \vee \; \lambda < \net_{\mc{V}}(m)(n)
      }{
      \sys{\net}{\spawnc{(m,\kappa):P}{n}{\lambda}}
      \lstep{\tau}
      \sys{\net \ominus n \views m}{\nil}
      }
    \end{array}
  \]
  \caption{LTS: Concurrent and Distributed Rules}
  \label{fig:lts-rls-conc-dist}
\end{figure}

Transition relation $\lstep{\alpha}$ in our LTS semantics is defined inductively by several sets of inference rules.
The first set of rules, shown in Fig.~\ref{fig:lts-local-rules}, contains the equivalent of all the local rules of the reduction relation
in Fig.~\ref{fig:red-loc-rls} obtained by replacing $\step$ by $\lstep{\tau}$, except for rule \rl{msg}.
Fig.~\ref{fig:lts-rls-conc-dist} depicts classical rules for local message
exchange, which are as in the standard early instantation-style LTS for
$\pi$-calculus~\cite{SangiorgiW01} as well as
the equivalents of distributed rules in~Fig.~\ref{fig:red-dist-rule}, obtained by replacing $\step$ by $\lstep{\tau}$.

\begin{figure}
  \scriptsize
  \[
    \hspace*{-3em}
    \begin{array}{ll}
      \inferrule*[lab=l-create-ext]{ \Delta \vdash n:\ms{dead} \\
        \Delta_{\mc{A}}(n) = - \kappa
    }{
      \sys{\net}{N}
      \lstep{\ms{create}(n,\kappa+1)}
      \sys{\net \oplus (n,\kappa+1)}{N}
    }\hspace*{2cm}
    &

    \inferrule*[lab=l-kill-ext]{ 
	\net \vdash n_\lambda:\ms{alive}
    }{
      \sys{\net}{N}
      \lstep{\skill(n,\lambda)}
      \sys{\net \ominus (n,\lambda)}{N}
    }
      \\[3.5ex]

      \inferrule*[lab=l-unlink-ext]{ 
	  \net \vdash n_\lambda :\ms{alive} \\
        \Delta \vdash n \connection m
    }{
      \sys{\net}{N}
      \lstep{\ominus n_{\scaleto{\lambda}{3pt}} \mapsto m}
      \sys{\net \ominus n \connection m}{N}
      }\hspace*{2cm}
    &

      \inferrule*[lab=l-link-ext]{
      \net \vdash n_\lambda :\ms{alive} \\
      \Delta \not\vdash n \connection m
    }{
      \sys{\net}{N}
      \lstep{\oplus n_{\scaleto{\lambda}{3pt}} \mapsto m}
      \sys{\net \oplus n \connection m}{N}
      }    \\[3.5ex]

      \inferrule*[lab=l-view]{ 
      \net \vdash n_\lambda : \ms{alive} \\
      (\net_{\mc{V}}(n)(m) = \net_\mc{A}(m) \neq 0 \; \vee \; \net_{\mc{V}}(n)(m)=0)
	   }{
      \sys{\net}{N}
      \lstep{n_{\scaleto{\lambda}{3pt}} \views m}
      \sys{\net}{N}
      }
    \end{array}
    \]
  \caption{LTS: Net Rules}
  \label{fig:lts-net-rules}
\end{figure}

Fig.~\ref{fig:lts-net-rules} depicts the rules modeling the effects of actions
from a context on the public part of the network, in particular the creation of
a new location (\lrl{create-ext}{}), the killing of a location
(\lrl{kill-ext}{}), the linking of two locations (\lrl{link-ext}{}) or the
unlinking of two locations (\lrl{unlink-ext}{}). Finally, rule \lrl{view}{}
is used to impose equality of views of locations for two equivalent systems.
Net rules do not have a counterpart in the reduction semantics as the effect
of each rule can be achieved by an appropriate context.

\begin{figure}[t]
  \scriptsize
  \[
    \begin{array}{ll}

      \inferrule*[lab=l-par\textsubscript L]{
      \sys{\net}{N
      \lstep{\alpha}
      \sys{\res{\vect u}\net'}{N'}}
      \quad
      \vect{u} \cap \fn{M} =\emptyset
      }{
      \sys{\net}{N\paral M}
      \lstep{\alpha}
      \sys{\res{\vect u}\net'}{N' \paral M}
      }

      &

        \inferrule*[lab=l-par\textsubscript R]{
        \sys{\net}{N
        \lstep{\alpha}
        \sys{\res{\vect u}\net'}{N'}}
        \quad
        \vect{u} \cap \fn{M} =\emptyset
        }{
        \sys{\net}{M\paral N}
        \lstep{\alpha}
        \sys{\res{\vect u}\net'}{M \paral N'}
        }
        
      \\[3ex]
      
      \inferrule*[lab=l-sync\textsubscript L]
      {
      \sys{\net}{N}
      \lstep{\out{x}{\vect{u}}@n_\lambda}
      \sys{\net}{N'}
      \\\\
      \sys{\net}{M}
      \lstep{x(\vect{u})@n_\lambda}
      \sys{\net}{M'}
      }{
      \sys{\net}{N \paral M}
      \lstep{\tau}
      \sys{\net}{ N' \paral M'}
      } 

      &
        \inferrule*[lab=l-sync\textsubscript R]
        {
        \sys{\net}{N}
        \lstep{\out{x}{\vect{u}}@n_\lambda}
        \sys{\net}{N'}
        \\\\
        \sys{\net}{M}
        \lstep{x(\vect{u})@n_\lambda}
        \sys{\net}{M'}
        }{
        \sys{\net}{M \paral N}
        \lstep{\tau}
        \sys{\net}{ M' \paral N'}
        }
      \\[3ex]
      
	    \inferrule*[lab=l-res]{
      S
      \lstep{\alpha}
      S'
      \quad
      u\notin \fn{\alpha}
	    }{
      \res{u}S
      \lstep{\alpha}
      \res{u}S'
      }
      
      &

        \inferrule*[lab=l-res\textsubscript{O}]{
        S
        \lstep{\res{\vect{v}}\out{x}{\vect{u}}@n_\lambda}
        S'
        \quad
        w \in \vect{u}\setminus \vect{v},x,n
        }{
        \res w S
        \lstep{\res{w}\res{\vect{v}}\out{x}{\vect{v}}@n_\lambda}
        S'
        }

      \\[3ex]

      \inferrule*[lab=l-$\alpha$]
      { S =_\alpha T \\ T \lstep{\alpha} T' \\ T' =_\alpha S'}
      {S \lstep{\alpha} S'}
    \end{array}
  \]
  \caption{LTS: Composition Rules}
  \label{fig:lts-ctx}
\end{figure}

Fig.~\ref{fig:lts-ctx} depicts composition rules for the labeled transition
semantics, we only discuss some relevant examples, the others are straightforward.
Rules \lrl{par}{l} is the rule for parallel composition allowing independent evolution
of the left component.
The side condition on the idle (right) component is required to avoid name capture
when the other component introduces a restriction.
Rule \lrl{res}{\textsubscript{o}} is analogous to the classical \rl{open} rule in $\pi$-calculus.
What is unusual is that there is no corresponding \rl{close} rule in our LTS semantics, because
rule \lrl{res}{\textsubscript{o}} operates at the system level, and we have no operation for composing systems.
Rule \lrl{res}{\textsubscript{o}} just signals that a system can send a message at a given address,
bearing private names in its payload.

\subsection{Full Abstraction}

This subsection presents the full abstraction result. We begin by recalling the definitions of
\emph{strong} and
\emph{weak  bisimilarity}.

\begin{defi}[Strong Bisimilarity] 
  A binary relation over closed systems $\rel S \subseteq \cset{S}^{2}$ is a \emph{strong simulation} iff
  whenever $(S,R)\in \rel S$, and $S \lstep{\alpha} S'$ then $R\lstep{\alpha} R'$ for some $R'$ with
    $(S',R')\in \rel S$.
A binary relation $\rel S$ over closed systems is a \emph{strong bisimulation} if both $\rel S$ and $\rel S^{-1}$ are strong simulations.
\emph{Strong bisimilarity}, denoted by $\sim$, is the largest strong bisimulation.
\end{defi}

\begin{defi}[Weak Bisimilarity] A binary relation over closed systems
  $\rel S \subseteq \cset{S}^{2}$ is a \emph{weak simulation} iff whenever 
  $(S,R)\in \rel S$ and $S \lstep{\alpha} S'$,
  then $R\lwstep{\alpha} R'$ for some $R'$ with $(S',R')\in \rel S$.
  A binary relation $\rel S$ over closed systems
    is a \emph{weak bisimulation} if both $\rel S$ and $\rel S^{-1}$
    are weak simulations.
  \emph{Weak bisimilarity}, denoted by $\wbsim$, is the largest weak  bisimulation.
\end{defi}

The fact that $\sim$ and $\wbsim$ are equivalence relations is a standard result for any labelled transition system.
Our main result states that weak bisimilarity fully characterizes weak barbed congruence:

\begin{thm}[Full Abstraction]
  $S \bcong R$ iff $S \wbsim R$.
\end{thm}
\begin{proof}
    The nature of the proof consists in showing that (weak) bisimilarity is
  included in (weak) barbed congruence, i.e., $\wbsim \subseteq \bcong$, and also that
  (weak) barbed congruence is included in (weak) barbed bisimilarity, i.e., $\bcong
  \subseteq \wbsim$. The first inclusion, in the literature referred to as
  soundness of bisimilarity, is proved by showing that if two systems are
  (weak) bisimilar then they are also (weak) barbed congruent.
  The second inclusion, in the literature referred to as completeness of
  bisimilarity, is proved by showing that if two systems are (weak) barbed
  congruent then they are also bisimilar. The spirit of the second part of the proof
  is to be able to build contexts that force a system to perform the expected
  labeled action. Soundness (Proposition~\ref{prop:bisimsound}) and
  Completeness (Proposition \ref{prop:completeness}) are proved in Appendix~\ref{app:full-abstraction}.
\end{proof}

For those readers familiar with the $\pi$-calculus behavioral theory, our full abstraction result may seem surprising, but note that
we are only considering congruence with respect to operators acting at the level of systems or configurations, namely the parallel operator $\paral$, and the
restriction operator $\nu$. In particular we are not considering congruence with respect to the input operator (it is well known that in $\pi$-calculus weak bisimilarity
is not a congruence with respect to the input operator). 

We conclude the section with two results, showing respectively that
  processes on dead localities cannot ever reduce and that two
  strong equivalent systems must have the same public network.
  \begin{prop}
    Given a well-formed system $\sys{\res{\vect u}\net}{N \paral \proc{P}{n}{\lambda}}$ such
    that $\net \vdash n_\lambda :\ms{dead}$ we have
    \[
      \sys{\res{\vect u}\net}{N \paral \proc{P}{n}{\lambda}} \bsim \sys{\res{\vect u}\net}{N}
    \]
  \end{prop}
  \begin{proof}  The result follows by observing that
    \[
      \rel{R}=\set{\langle \sys{\res{\vect u}\net}{N \paral \proc{P}{n}{\lambda},
          \sys{\res{\vect u}\net}{N}} \rangle \mid \sys{\res{\vect
            u}\net}{N}\in\cset S, \net \vdash n_\lambda :\ms{dead} }
    \]
    is a strong bisimulation since all the labeled transitions involving a
    process require for it to be alive and, hence, no transition can be derived
    from process $\proc{P}{n}{\lambda}$. We remark that a $\create{n}{P}$ operation builds a new location $n$ with a higher incarnation number without affecting the state of $n_{\lambda}$. 
  \end{proof}
  \begin{prop}[Strong Bisimilar Systems Have The Same Public Network]\label{prop:net-eq-bsim}
    If we have $\sys{\res{\vect u}\net}{N}\bsim \sys{\res{\vect v}\net'}{M}$
    then the following conditions hold:
    \begin{itemize}
    \item $\forall n\not\in \vect{u}. \net_\mc{A}(n) = \net'_\mc{A}(n) $
    \item $\forall n\not\in \vect{u}, m\not\in \vect{u}. (n,m)\in\net_\mc{L}
      \Leftrightarrow (n,m)\in \net'_\mc{L} $
    \item $\forall n\not\in \vect{u}, m\not\in \vect{u}. \net_\mc{V}(n)(m) =
      \net'_\mc{V}(n)(m) $
    \end{itemize}
  \end{prop}
  \begin{proof}
    We just make the case for the first condition, the others are similar.
    Assume to have two systems $S=\sys{\res{\vect u}\net}{N}$,
    $R=\sys{\res{\vect v}\net'}{M}$ such that $S\bsim R$. Now suppose, towards a
    contradiction, that there exists $n\not\in \vect{u}$ such that $\net_\mc{A}(n)\neq \net'_\mc{A}(n)$. Let us first consider the case where $\net_\mc{A}(n) \leq 0$. Hence, $S$ can perform a labeled transition derived using rules \rl{l-create-ext} and \rl{l-res}, with label $\ms{create}(n,|\mc{A}(n)|+1)$.
    Such a transition cannot be matched by $R$, against the hypothesis that $S \bsim R$.
    The case where $\net_\mc{A}(n) > 0$ is similar, using rule \rl{l-kill-ext} instead of rule \rl{l-create-ext}.
%    whether $\net \vdash n: \ms{alive}$ or not, $S\lstep{\alpha} S'$ such
%    that $R \hspace*{0.8em}\not\hspace*{-0.8em}\lstep{\alpha}$, but that is a
    %    contradiction since $S\bsim R$.
    Hence, $\forall n\not\in \vect{u}. \net_\mc{A}(n) = \net'_\mc{A}(n) $.
  \end{proof}

\section{Behavioral Theory in Action} \label{sec:examples}

In this section we show our behavioral theory in action. First
we apply it to the running example of Section~\ref{sec:running-example}. Then,
we contrast our \textsf{spawn} primitive against a \textsf{go} primitive analogous to 
the one from D$\pi$F. We then rephrase a key example from D$\pi$F in D$\pi$FR 
where our behavioral theory agrees with the one of D$\pi$F. 
Finally we show, also using an example from D$\pi$F,
that our behavioral theory is more discriminating than that of D$\pi$F, 
because of the possibility in D$\pi$FR to reestablish links and
to restart locations.

\begin{exa}[\textbf{servD} and \textbf{servDFR} are bisimilar]\label{ex:running-ex}
  To prove $\textbf{servD} \wbsim \textbf{servDFR}$ it suffices to show a candidate
  bisimulation relation and then play the bisimulation game on its elements.
  Consider relation $$\rel{R} = \{(\textbf{servD},\textbf{servDFR})\} \cup \rel{S}_0 \cup
  \rel{S}_1 \cup \rel{S}_2$$
  where
  {
    \begin{align*}
    & \rel{S}_0 = \{(\textbf{servD}, R_0) \mid \textbf{servDFR} \lwstep{\tau}R_0\}\\
    & \rel{S}_1 = \{(S_1,R_1) \mid (S_0,R_0)\in \rel{S}_0, S_0\lwstep{req(x,y)@n_i} S_1, R_0\lwstep{req(x,y)@n_i} R_1\}\\
    & \rel{S}_2 = \{(S_2,R_2) \mid (S_1,R_1)\in \rel{S}_1, S_1\lwstep{\out{z}{w_\lambda}@n_i} S_1, R_1 \lwstep{\out{z}{w_\lambda}@n_i} R_2\}
  \end{align*}
  }
  Intuitively, from an external perspective, \textbf{servD} only inputs the
  request and exhibits the answer; hence, we need to prove that
  \textbf{servDFR} is able to match those two actions and has no other
  observable behavior.
  The proof is available in Appendix~\ref{app:bisimulation}.
  \hfill $\Diamond$
\end{exa}

\begin{exa}[Non-atomicity of the \textsf{spawn} primitive]
	We now contrast our \tms{spawn}
	primitive against the \tms{go} primitive in D$\pi$F~\cite{FrancalanzaH08}.
	The semantics of the \textsf{spawn} primitive given by the four rules \textsc{spawn-c-s}, \textsc{spawn-c-f}, \textsc{spawn-s} and \textsc{spawn-f} 
	is different from the semantics of the \textsf{go} primitive in D$\pi$F  \cite{FrancalanzaH08}. 
	In our setting, a \textsf{go} primitive analogous to the one
        %\textsf{go} primitive
        in D$\pi$F can be defined by the following reduction rules:

	{\small
	\[
	    \begin{array}{l}
	      \inferrule*[lab=go-s]{
	      \Delta \vdash n_\lambda,m_\kappa:\ms{alive}\qquad
	      \net \vdash n_\lambda \connection m_\kappa\qquad
	      \net_n(m) = \kappa
	      }{
	        \sys{\net}{\proc{\spwn{m}{P}}{n}{\lambda}}
	        \step
	        \sys{\net \oplus m\views (n,\lambda)}{\proc{P}{m}{\kappa}}
	      }\\[3ex]

	      \inferrule*[lab=go-f]{
	      \Delta \vdash n_\lambda:\ms{alive}\qquad
	      (\Delta \vdash m:\ms{dead}
	      \vee \net_n(m)\neq \net_{\mc{A}}(m)
	      \vee \net \not\vdash n_\lambda \connection m_\kappa)}{
	      \sys{\net}{\proc{\spwn{m}{P}}{n}{\lambda}}
	      \step
	      \sys{\net  \ominus n \views m}{\nil}
	      }
	    \end{array}
	\]
	}

	Rule \rl{go-s} defines a successful \tms{go}, conditional upon the fact that
	locations $n$ and $m$ are both alive and connected, 
	and that the spawning location rightly believes the target location to be alive
	with the incarnation number recorded in its view, or has no belief on the
	target location in its local view. This last constraint is captured by the side
	condition $\net_n(m) = \kappa$, which we formally define as follow:

	{\scriptsize
	\[
	    \net_n(m) =
	      \begin{cases}
	      \kappa
	      \hspace*{1cm}
	        \ms{if~} n = m \ms{~~and~~} \net_{\mc{A}}(n) = \kappa
	      \\
	      \kappa
	      \hspace*{1cm}
	        \ms{if~} n \neq m \ms{~~and~~} \net_{\mc{V}}(n)(m) = \kappa
	      \\
	      \kappa
	      \hspace*{1cm}
	        \ms{if~} n \neq m \ms{~~and~~} \net_{\mc{V}}(n)(m) = 0 \ms{~~and~~} \net_{\mc{A}}(m) = \kappa
	      \\
	      0
	      \hspace*{1cm}
	        \ms{otherwise}
	    \end{cases}
	  \]
	}

	Rule \rl{go-f} defines a failed \tms{go}. Failure may be due to the remote location not being alive,
        to a wrong view, or
        to a missing
	link between the two locations.
	The atomicity of the \tms{go} operation avoids 
	the possibility of having the spawning message delivered when the sending
  location is already dead.
	The \tms{go} primitive appears to be much simpler than our \tms{spawn} primitive, but
	unfortunately the \tms{spawn} operation and the \tms{go} operation lead to observationally different behaviors. 
	To see this consider the variant D$\pi$FR$_\ms{go}$ of D$\pi$FR, where the \tms{go} operation
	has been added. The LTS semantics for D$\pi$FR$_\ms{go}$ is identical to that of D$\pi$FR, except for the 
	addition of  rules for silent transitions corresponding to the \textsc{go-s} and \textsc{go-f} rules above.
	The notions of weak simulation, weak bisimulation and weak bisimilarity  are the same for D$\pi$FR and D$\pi$FR$_\ms{go}$.
	Define the encoding function $\ltrans{\cdot}$ from D$\pi$FR$_\ms{go}$ terms to D$\pi$FR terms that just replaces any occurrence of
	a \textsf{go} instruction in a  D$\pi$FR$_\ms{go}$ term by the corresponding \textsf{spawn} instruction.
	Now we have the following result:
	\begin{restatable}[\textsf{go} does not simulate \textsf{spawn}]{prop}{atomicspawn}\label{prop:atomic}
		Let $U$ be the following closed system in D$\pi$FR$_\ms{go}$:
		\begin{align*}
		& U =  \sys{\net}{ \proc{\spwn{m}{\eout{s}}}{n}{1} } & & &
		& \ltrans{U} =  \sys{\net}{ \proc{\spawn{m}{\eout{s}}}{n}{1} } & \\
		& \net_{\mc{A}} = \{ n \mapsto 1, m \mapsto 1 \} &
		& \net_{\mc{L}} = \{ n \connection m \} &
		& \net_{\mc{V}} = \{ n \mapsto \fnull,  m \mapsto \fnull \}
		\end{align*}
		$U$ cannot weakly simulate $\ltrans{U}$.
	\end{restatable}
        \begin{proof}
	Define $\net_1, \net_2, \net_3, \net_4, \net_5$ as the following networks:
	\begin{align*}
		& \net_1 = \net \ominus (n,1) &
		& \net_2 = \net \ominus (m,1) &
		& \net_3 = \net_2 \oplus (m,2)\\
    & \net_4 = \net_1 \ominus (m,1) &
		& \net_5 = \net_4 \oplus (m,2)
	\end{align*}
	Assume for the sake of contradiction that there exists a weak simulation $\scal$ such that $\langle \ltrans{U}, U \rangle \in \scal$.
	Now consider the following transition from $\ltrans{U}$, obtained by applying rule \textsc{l-spawn-c-s} :
	\begin{align*}
		\ltrans{U} \lstep{\tau} T_1 = \sys{\net}{ \spawnc{(m,0):\eout{s}}{n}{1} } 
	\end{align*}
	Because $\scal$ is a weak simulation, we must have $U \wstep U_1$ and $\langle T_1, U_1 \rangle \in \scal$ for some $U_1$.
	There are in fact only two possibilities for $U_1$ since $U$ has a single silent transition, obtained by applying rule \textsc{l-go-s}:
	\begin{enumerate}
		\item $U \lstep{\tau} U_1 = \sys{\net}{ \proc{\eout{s}}{m}{1} }$: in this case, consider 
		the transition, obtained by applying rule \textsc{l-kill-ext}:
		$$T_1 \lstep{ \ms{kill}(m,1)} T_2 = \sys{\net_2}{ \spawnc{(m,0):\eout{s}}{n}{1} }$$
		Since $U_1$ has no silent transition, the only possibility to match this transition from $T_1$ is
		that obtained from $U_1$ by applying \textsc{l-kill-ext}:
		$$ U_1 \lstep{ \ms{kill}(m,1)} U_2 = \sys{\net_2}{ \proc{\eout{s}}{m}{1} }$$
		and $\langle T_2, U_2 \rangle \in \scal$.
		Consider now the following transition from $T_2$ obtained by applying rule \textsc{l-create-ext}:
		\begin{align*}
			& T_2 \lstep{ \ms{create}(m,2)} T_3 = \sys{\net_3}{ \spawnc{(m,0):\eout{s}}{n}{1} }
		\end{align*}
		Since $U_2$ has no silent transition, the only possibility to match this transition from $T_2$
		is that obtained from $U_2$ by applying \textsc{l-create-ext}:
		\begin{align*}
			U_2 \lstep{\ms{create}(m,2)} U_3 = \sys{\net_3}{ \proc{\eout{s}}{m}{1} }
		\end{align*}
		and $\langle T_3, U_3 \rangle \in \scal$.
		Now $U_3$ has no silent transition and has no weak barb on $\eout{s}$ ($m_1$ is not alive in $\net_3$). But this is impossible for
		$T_3$ has a weak barb on $\eout{s}$, as can be seen from the following transitions, obtained from $T_3$ applying \textsc{l-spawn-s} and 
		\textsc{l-out}:
		\begin{align*}
			T_3 \lstep{\tau} \sys{\net_3}{ \proc{\eout{s}}{m}{2} } \lstep{\eout{s}@m_2} \sys{\net_3}{ \proc{\nil}{m}{2} }
		\end{align*}
    		\item $U_1 = U$: in this case, consider the following transition from $T_1$, obtained by applying rule \textsc{l-kill-ext}:
		\begin{align*}
			T_1 \lstep{\ms{kill}(n,1)} T_2 = \sys{\net_1}{ \spawnc{(m,0):\eout{s}}{n}{1}  }
		\end{align*}
		We must have $U \lwstep{\ms{kill}(n,1)} U_2$ and $\langle U_2, T_2 \rangle \in \scal$ for some $U_2$.
		Since there is only a single silent transition from $U$, we have two possibilities for $U_2$, obtained by applying rule
		\textsc{l-kill-ext}, or by applying rule \textsc{l-go-s} followed by rule \textsc{l-kill-ext}:
		\begin{enumerate}
			\item $U \lstep{ \ms{kill}(n,1)} U_2 = \sys{\net_1}{ \proc{ \spwn{m}{\eout{s}}}{n}{1}  }$: in this case, 
			consider the following transition from $T_2$ obtained by applying rule \textsc{l-spawn-s}:
			\begin{align*}
				T_2 \lstep{\tau} T_3 = \sys{\net_1}{ \proc{ \eout{s}}{m}{1}}
			\end{align*}
			Since there is no silent transition from $U_2$ ($n_1$ is not alive in $\net_1$), we must have $\langle U_2, T_3 \rangle \in \scal$.
			But this is impossible for $U_2$ has no barb on $\eout{s}$ and we have the following transition from $T_3$, obtained
			by applying rule \textsc{l-out}:
			\begin{align*}
				T_3 \lstep{\eout{s}@m_1} \sys{\net_1}{ \proc{\nil}{m}{1} }
			\end{align*}
			\item $U \lstep{\tau} \sys{\net}{ \proc{ \eout{s} }{m}{1}} \lstep{\ms{kill}(n,1)} U_2 = \sys{\net_1}{ \proc{ \eout{s} }{m}{1}}$:
			in this case, we reason as in the case above, with $\net_4$ and $\net_5$ in place of $\net_2$ and $\net_3$, to conclude 
			to an impossibility.
		\end{enumerate}
	\end{enumerate}
	All the cases above thus lead to an impossibility, which means no weak simulation $\scal$ can exist.
\end{proof}

	A consequence of this result is that the \textsf{spawn} primitive and the
	\textsf{go} primitive are not weakly bisimilar and are not even inter-similar
	(one being able to simulate the other and vice-versa). This suggests that a \tms{go} primitive
	as defined above hides too much atomicity for a distributed setting, 
	where location and link failures can disrupt interactions in various ways,
	which are more faithfully captured by our \tms{spawn} primitive.
  We conclude by remarking that the \textsf{spawn} primitive is always able
    to weakly simulate the \textsf{go} primitive, indeed a successful \textsf{go} can be
    matched by a \textsf{spawn} primitive by
    applying rules \rl{spawn-c-s} and \rl{spawn-s}, while a failed \textsf{go} primitive can be matched by
  applying rule \rl{spawn-c-f}.
	\hfill $\Diamond$
\end{exa}

\begin{exa}[Distributed Server]

  Here, we rephrase \cite[Example 11]{FrancalanzaH08}, where Francalanza and
  Hennessy show that the behavioral theory they present is able to
  distinguish \textbf{servFHD}, a distributed server only able to reach its backend $n$ by a direct
  connection, and \textbf{servFHD2Rt}, that, in addition to the direct connection, has also an
  indirect connection that goes through a third location $m$.

  System \textbf{servFHD} in D$\pi$FR is defined as
  \[
    \res{d,w_y} \net \vartriangleright
      \proc{req(x,y).\spawn{n}{\out{d}{x,y}}}{l}{}\paral
      \proc{d(x,y).\spawn{l}{\out{x}{w_y}}}{n}{}
  \]
  while system \textbf{servFHD2Rt} is defined as
  \[
    \res{d,w_y}\net \vartriangleright
    \left(
      \begin{array}{l}
        \left[ req(x,y).\res{s}
        \left(
        \begin{array}{l}
          \spawn{n}{\out{d}{s,y}} \parop \\
          \spawn{m}{\spawn{n}{\out{d}{s,y}}} \parop \\
          s(w).\out{x}{w}
        \end{array}
        \right)
        \right]^{l}
        \paral\\[8mm]
        \left[
        d(s,y).\left(
        \begin{array}{l}
          \spawn{l}{\out{s}{w_y}} \parop\\
          \spawn{m}{\spawn{l}{\out{s}{w_y}}}
        \end{array}
        \right)
        \right]^{n}
      \end{array}
    \right)
  \]
    where network $\net$ is defined as
  \begin{align*}
    \net = \tuple{\set{n \mapsto 1,m\mapsto 1,l \mapsto 1}, \set{n \connection m, n\connection l, m \connection l}, \set{n \mapsto \fnull, m \mapsto \fnull, l \mapsto \fnull}}
  \end{align*}

  In both the cases, we use $w_y$ to emphasize that the answer of the
  server depends on the query $y$. Notice in \textbf{servFHD2Rt} the
  use of forwarder $s(x).\out{y}{x}$ to ensure that if the requests
  via both the links succeed, only one answer is made available.
  
  Now, \textbf{servFHD} and \textbf{servFHD2Rt}, as in~\cite{FrancalanzaH08}, can
  be distinguished in D$\pi$FR by the following context
  \[
  \proc{\sbreak{n}.\out{req}{z,h}}{l}{} %\paral \proc{}{n}{}
  %\proc{\sbreak{n}}{l}{} \paral \proc{\out{req}{z,h}}{n}{}
  \]
  as \textbf{servFHD} would stop working after the \tms{unlink} reduces, while
  \textbf{servFHD2Rt} would keep working correctly since it could route the
  request through $m$.
  \hfill$\Diamond$
\end{exa}

\begin{exa}[Network Observations]\label{ex:network}
  Here we rephrase~\cite[Example 12]{FrancalanzaH08} and show a crucial
    difference between the observational theory of D$\pi$F and that of D$\pi$FR which involves three different features, related respectively
    to: i) the presence of our \tms{create} primitive; ii) the presence of our
    \tms{link} primitive; and iii) their combination.

  In \cite[Example 12]{FrancalanzaH08}, Francalanza and Hennessy show that for any configuration $N$ the
  three systems, $\sys{\res{k}\net_1}{N}$, $\sys{\res{k}\net_2}{N}$ and $\sys{\res{k}\net_3}{N}$, 
  where $\net_1$, $\net_2$, and $\net_3$ are as given below, 
  are equivalent according to their behavioral theory.
  
  { \scriptsize
  \[
    \begin{array}{lclcl}
     \net_1 &=
      &
        \langle \set{l\mapsto 1, k\mapsto -1}, \set{l \connection k}, \set{l \mapsto
        \mathbf{\hat{0}}} \rangle &=
      &
	  \raisebox{-2mm}{%
      \begin{tikzpicture}
        \node[label={[label distance=-1mm] $l_1$}] (l1) at (-3,-1) {\large $\circ$};
        \node[label={[label distance=-1mm] $k_{-1}$}] (l2) at (0,-1) {\large $\bullet$};

      \draw[latex-latex, thick] (l1) -- (l2);
    \end{tikzpicture}
	  }\\
      \net_2 &=
      &
        \langle \set{l \mapsto 1, k \mapsto -1}, \emptyset, \set{l \mapsto
        \mathbf{\hat{0}}} \rangle &=
      &	 
	  \raisebox{-2mm}{%
        \begin{tikzpicture}
      \node[label={[label distance=-1mm] $l_1$}] (l1) at (-3,-1) {\large $\circ$};
      \node[label={[label distance=-1mm] $k_{-1}$}] (l2) at (0,-1) {\large $\bullet$};
    \end{tikzpicture}
	}\\
      \net_3 &=
      &
        \langle \set{ l \mapsto 1, k \mapsto 1}, \emptyset, \set{l \mapsto
        \mathbf{\hat{0}}, k \mapsto
        \mathbf{\hat{0}}} \rangle &=
      &
	\raisebox{-2mm}{%
      \begin{tikzpicture}
        \node[label={[label distance=-1mm] $l_1$}] (l1) at (-3,-1) {\large $\circ$};
        \node[label={[label distance=-1mm] $k_1$}] (l2) at (0,-1) {\large $\circ$};

      \end{tikzpicture}
	}\\
    \end{array}
\]
}

The impossibility of distinguishing the three systems in~\cite{FrancalanzaH08} is due to the absence of
recovery of both links and locations. Even if $N$ is a
configuration extruding the private name (e.g., pick $N$ to be $\proc{\out{a}{k}}{l}{1}$) 
and put in a context where another
process can receive the name $k$, then this process cannot establish a connection
or reactivate $k$ and hence cannot observe any difference.

In more detail, in the first system, even if an observer can establish a link to
$k$ the location is dead and cannot be revived. In the second system $k$ is dead
and unreachable by any hypotethical observer. Indeed, the only
way to establish new links in D$\pi$F is to create a fresh location, but such
fresh location will only be connected to the set of locations reachable by the
parent location. Since here $l$ is not connected to $k$, $l$ cannot interact with $k$, and in particular it is impossible to establish a link with it.
Finally, in the third system, even if $k$ is alive, it is still not reachable due
to the lack of a connection as above and hence there is no way to discover that
$k$ is alive.

In D$\pi$FR, since we allow for creation of links without requiring prior
connections and we allow for recovery of locations, if $N$ extrudes $k$ then we can easily distinguish the
three networks. We do it by pairing them with a configuration that receives the extruded name $k$ and then exploits it to assess the status of the link, trying to link or to unlink the location, and of the location, trying to restart or to kill it.

%   Distinguishing them pairwise is enough to show that they are not equivalent. A
%   context that gets the location name $k$, destroys the connection with it and
%   then makes disappear a fresh observable $t$ would behave differently in the first
%   and second system, as in the second one there is no connection to destroy and
%   so it cannot make disappear $t$.
% % In the second system a context that after grabbing $k$ establishes a
% % connection with it and then reactivates it; again, would behave differently in
% % the other two networks. Finally,
%   A context that grabs $k$, establishes a connection with it, spawns on it a
%   fresh observable $t$ and then kills it would distinguish the second and third
%   system, as in the second location $k$ is already dead. }

Thanks to the full-abstraction result, this can be seen directly on our LTS with the following labeled derivations:

{
  \scriptsize
  \begin{align*}
  & \sys{\res{k}\net_1}{\proc{\out{a}{k}}{l}{1}} \lstep{\res{k}{\out{a}{k}@l_1}} \sys{\net_1}{\nil} \lstep{\ominus l \connection k}\\
  & \sys{\res{k}\net_2}{\proc{\out{a}{k}}{l}{1}}  \lstep{\res{k}{\out{a}{k}@l_1}}
    \sys{\net_2}{\nil} \lstep{\oplus l \connection k}  \sys{\net_2'}{\nil} \lstep{\ms{create}(n,2)}\\
  & \sys{\res{k}\net_3}{\proc{\out{a}{k}}{l}{1}} \lstep{\res{k}{\out{a}{k}@l_1}}  \sys{\net_3}{\nil} \lstep{\skill~k}
  \end{align*}
}
Notice that each derivation cannot be matched by the other systems.
\hfill $\Diamond$
\end{exa}

%%% Local Variables:
%%% mode: latex
%%% TeX-master: "main-distributed"
%%% End:

\section{Capturing Distributed Systems Characteristics}\label{sec:discussion}

Here we discuss how our calculus captures various features which we
believe typical of distributed systems.

  \textbf{Communication.} First of all, communication between remote parties must
be asynchronous. 
That is, interaction between remote processes proceeds by a non-atomic 
exchange of messages between the nodes that support them; messages
can possibly be lost (because of link failure) or reordered while transiting to their destination.
Moreover, each communication must target a single location, either local or
remote. This ensures that the complexity of basic message exchange is
commensurate with that of simple asynchronous communication used in the
Internet, and that no hidden cost, due for instance to leader election or
routing protocols, is implied for the implementation of a simple message
exchange (see e.g. \cite{FournetG96} for a discussion).

In our calculus, interactions are local. Remote communication is obtained using
the \textsf{spawn} operation to send to a target node a process performing an output action. 
As a result remote communication is indeed asynchronous, and there is a single location where the receiver can
reside (since such location is specified in the \textsf{spawn}). This avoids the
need of a type system to ensure that possible receivers are located on a
same node, in contrast to calculi based on channel-based remote communication
such as~\cite{Amadio97}.

\textbf{Dynamic nodes and links}. During execution,
new nodes and links can be established and existing nodes and links can be removed,
either because of failures or by design. This feature is necessary to account
for actual distributed systems whose configurations may vary at run-time,
notably because of failure and performance management (e.g.\
scaling decisions in cloud systems).
The explicit presence of links is
important because partial connections often affect large
distributed systems.  Dealing with link failures in addition of node failures
can lead to a subtly different
behavioral theory than dealing with node failures only, as shown in
\cite{FrancalanzaH08}.

\begin{exa}
  The following configuration, which can be added as a context to any system where $m$ is a free name, 
  is an example of how a system can be extended with an
  unbounded number of fresh locations
  \[
    \proc{\res{c}(\eout c \parop !c.(\res{n}\create{n}{\nil} \parop \eout c))}{m}{}
    \]
  Indeed, the synchronization between the output $\eout c$ and the replicated input will create a new location with a fresh name $n$, and recreate the output for a further iteration, resulting in an infinite computation.  
\end{exa}

\textbf{Imperfect Knowledge}.
In distributed systems, the only
way for locations to know something about the context that surrounds them is to
communicate. If a location $n$ receives a spawning message from a remote location $l$
then $n$ learns something on the context, namely that at some point in time $l$ was alive and working since it sent a spawning message to $n$. Nonetheless,
$n$ cannot infer anything on the current status of $l$ or the
status of the connection. Indeed, location $l$ could have stopped right after sending
the spawning message or the link could have broken right after the spawning message was received, or both.
Erlang systems, like many others, have an optimistic approach: after
a first two-way interaction two locations establish a mutual knowledge of their respective incarnations,
typically by means of a shared socket connection.
From that point on, they keep using that shared connection until their view changes, rather
than setting up a new connection for each spawning message exchange.
Reflecting this in our calculus plays a role in the semantics of our  spawn
primitive whenever the view
of the location is not in sync with the real state of the system.
This in turn plays a role in our behavioral theory, as the following example illustrates.

\begin{exa}
	Consider a variant \textbf{servDFV} of \textbf{servDFR} where $n_c$ is
  linked to $n_i$ instead of $n_r$ and the network is such that the router $n_r$
  is in its incarnation $\kappa$ and the local view of the interface location
  $n_i$ contains $n_r \mapsto \kappa$. The controller process running on $n_c$
  is defined as follows:
	\[
	  C' = \res{x} \create{n_r}{(R \parop \spawn{n_c}{\eout{x}})} \parop x.\spawn{n_i}{\eout{retry}}
	\]
        instead of
        \[
        C = \create{n_r}{(R \parop \spawn{n_i}{\eout{retry}})}
        \]
        as in \textbf{servDFR}.
        
	Now, \textbf{servDFV}, differently from \textbf{servDFR}, is not equivalent to \textbf{servD} because, in case of
  failure of $n_r$, the controller triggers the interface to restart the request
	instead of the router $n_r$, thus failing to update $n_i$'s local view with the knowledge of $n_r$'s new incarnation.
	A spawning message from the interface at this point would fail as its local view contains the previous incarnation of $n_r$.
	If not for the wrong belief of $n_i$, \textbf{servDFV} would have
	been equivalent to \textbf{servD}. In the companion repository~\cite{RepositoryErlang} we discuss an implementation of this system and we show that this behavior arises
	in reality too.
\end{exa}

\textbf{Networks in Equivalent Systems}.
We remarked in the Introduction that one ought to be able to prove from the behavioral theory that two weakly barbed congruent systems have networks with the same public part 
(i.e.\ those nodes and links whose names are not restricted).
Let us first consider the case of strongly bisimilar systems:
%We can see in our case that strongly bimilar systems 
they do indeed have identical public network parts. This follows from the rules
in Fig.\ref{fig:lts-net-rules} and the rule \textsc{l-res} as shown in Proposition~\ref{prop:net-eq-bsim}.
In particular, incarnation numbers of public nodes in barbed congruent systems must coincide.
This may seem to be overly discriminating but in fact it is warranted: 
for instance, recovery protocols or failure handling protocols such as SWIM~\cite{DasGM02},
which rely on incarnation numbers, would operate differently in dissimilar systems.
Another consequence, visible from rule \textsc{l-view}, is that two weak bisimilar systems must weakly agree on their public correct local views,
meaning that, in the local view of a given public location $n$, they either have the same correct beliefs on the status of a location, 
or they hold no belief on the status of a location, or one has a correct view on the status of a location and the other holds no belief on this location.
In other terms, two equivalent systems cannot disagree in their local
views, with one having a correct belief and the other one having an incorrect
one.

Let us now consider
weakly barbed congruent systems. A consequence of our full abstraction result is that
%only holds in a weak form, in the sense that
the public parts of networks of weakly barbed congruent systems agree only up to reductions, namely there exist sequences of reductions leading to systems with the same public parts of networks. This is weaker than the requirement imposed by
%This in turn shows that
the decision in~\cite{FrancalanzaH08} to consider public networks as types, and requiring equivalent systems to have the same type.
%is in fact too restrictive,
Indeed the latter approach, because of types, differentiates between systems which would otherwise be weakly barbed congruent.

\textbf{Persistence}.
Various forms of recovery, such as checkpoint-rollback schemes, require some
persistent storage to store information that will be preserved even if
failures occur. D$\pi$FR has by construction a location $\godloc$ which,
since it cannot be killed, models a persistent location. However the only purpose of $\godloc$ is to offer a
way to extend a system with a running process, without having to worry about
finding an alive location. Nonetheless, persistent memories can be encoded in D$\pi$FR
by leveraging the restriction operator. 

\begin{exa}
	As an example, consider the following piece of code, 
	which implements a simple read and write persistent server

  \[
    \footnotesize
	  \sys{\res{l,m}\net}{
	    \proc{!I}{l}{1} \paral
	    \proc{M}{m}{1}
	  }
	\]
	where
  {
    \footnotesize
    \begin{align*}
      &
        \net = \langle \set{l \mapsto 1, m \mapsto 1}, \set{l \connection m, n \connection m}, \set{l \mapsto \mathbf{\hat{0}}, m \mapsto \mathbf{\hat{0}}} \rangle \quad I \defeq \create{n}{addr(x).(!W \parop !R)}\\
      &
        M \defeq (!\spawn{n}{\out{addr}{m}}) \parop \out{data}{\vvect{u_0}}
        \qquad
        W \defeq write(\vect{u}).\spawn{x}{data(\vect{v}).\out{data}{\vect{u}}}\\
      & R \defeq read(l,x).\spawn{x}{data(\vect{u}).(\out{data}{\vect{u}} \parop
        \spawn{n}{\spawn{l}{\out{x}{\vect{u}}}})}
    \end{align*}
  } % The server $n$ may crash but it has a persistent memory $M$ where it can
  % store information that survives crashes of $n$ and remains accessible to $n$
  % in its different incarnations.
  
  Process $I$, for \emph{init},
  creates a server running on location $n$. The server first awaits on channel
  $addr$ to know the address of its memory. Once acquired, it offers to its
  clients the possibility to write (process $!W$) or to read (process
  $!R$) data. The data stored by the server are persistent, indeed, even if the
  server $n$ fails it will be re-instantiated by $\proc{!I}{l}{1}$, and if a
  read request is forwarded the data will be retrieved from the memory, i.e.,
  location $m$. Location $m$, being a private location, cannot be killed or
  unlinked ever by any context, hence we can leverage it to encode a persistent
  memory. The behavior of $M$ itself is simple: it repeatedly sends the name of
  the private memory $m$ to its server $n$, and holds information at the channel
  $data$, which can be retrieved or updated by $n$ via the $read$ or $write$
  operations.
\end{exa}

\section{Erlang Experiments}\label{sec:experiment}

\lstdefinestyle{md}{
  columns=fullflexible,
  literate={~}{{\raisebox{0.5ex}{\texttildelow}}}{1},
  framexleftmargin=2mm, % Adjust left margin
  framexrightmargin=2mm, % Adjust right margin
  framextopmargin=2mm,  % Adjust top margin
  framexbottommargin=2mm,  % Adjust bottom margin
  frame=single,
  framerule=0pt,
  backgroundcolor=\color{lightgray!20},
  basicstyle=\ttfamily,
  rulecolor=\color{gray},
  tabsize=4,
  breaklines=false
}

In the Introduction we claimed that D$\pi$FR faithfully reflects the behavior of
Erlang systems in case of failure. Erlang comes with a
documentation~\cite{ErlangDocs} and several other resources that do their best
to explain the principles behind the language and the semantics of each
primitive. Nonetheless, there are plenty of corner cases that are not discussed
and whose behavior is not explicitly documented. For instance the documentation
for the spawn targeting a remote node does not mention that in case of failure
the view of the local node is updated by removing any knowledge of the remote
node. Nor is it mentioned that if the local node holds a wrong belief on the
remote one then the spawn is doomed to fail, even if both nodes are alive and
connected.

To shed light on these obscure corner cases and clarify the behavior of Erlang
systems in presence of failures, we carried out some experiments on a
(simulated) distributed Erlang environment. To simulate different machines
connected by a network we leveraged Docker~\cite{Docker}. Docker is a tool that
allows one to test and deploy applications relying on the concept of
containerization. A container is a bundle packing together everything that is
needed to run an application (code, libraries, system tools, etc.) which is then
run isolated from other containers, sharing the hosting system's kernel.
Containers can communicate with each other through a network, and
Docker provides facilities to manage it. For the sake of our investigation, we
used containers to run Erlang nodes and then observe their behavior in corner
cases.

A description of all the experiments we did, including used code and
scripts, is available in our companion
repository~\cite{RepositoryErlang}.  We describe below one of these
examples, to give to the reader an intuition about how such experiments
can be done, while referring to~\cite{RepositoryErlang} for a
description of the other experiments, including Erlang implementations
of our examples and case studies.

We describe an experiment showing how views can impact a system behavior.
The scenario we consider is the following: two locations $n$ and $m$, running on
two separate containers, connected by a network link $n \connected m$.
Graphically this system could be represented as:
{
  \[
  \begin{tikzpicture}

    \node[label={[label distance=-1mm]\footnotesize $n$}] (l1) at (-3,0)
    {\Large $\circ$};
    \node[label={[label distance=-1mm]\footnotesize $m$}]
    (l2) at (0,0) {\Large $\circ$};

    \draw[latex-latex, thick] (l1) -- (l2);

  \end{tikzpicture}
  \]
}

We then consider the following sequence of events:
\begin{enumerate}
\item location $m$ spawns a process on location $n$
\item location $m$ is unlinked from location $n$
\item location $m$ dies
\item location $m$ is recreated
\item location $m$ is linked again to location $n$
\item location $n$ spawns a process on location $m$
\end{enumerate}

A system in D$\pi$FR with such a configuration, whose behavior includes a sequence of events as above can be defined as follows:
\[
  \textbf{WrongSp} \defeq \sys{\net}{
    \proc{\spawn{n}{P}}{m}{1} \paral \proc{\spawn{m}{Q}}{n}{1} \paral
    \proc{\sbreak{n}}{m}{1} \paral \proc{\skill}{m}{1} \paral
    \proc{\create{m}{\link{n}}}{\godloc}{}
  }
\]
with
\[
  \net \equiv \langle \set{n \mapsto 1, m \mapsto 1}, \set{ n \connection m}, \hat{\textbf{0}} \rangle
\]

At first glance, the outcome of each of these operations, of the two spawns in particular, seems to be pretty straightforward to
determine.
Both spawns are carried out with both locations alive and connected, hence one would
expect both of them to succeed. However, there are circumstances under which the
last spawn may fail. In particular, this occurs when the sequence of events is
carried out rapidly and location $n$ does not have enough time to detect that
the previous incarnation of $m$ -- the one it interacted with during the first
spawn -- is dead. As a result, it will attempt to spawn the process on the old
incarnation instead of the new one.

We now show how our experiment can be performed with Erlang. The first step consists of
setting up the system. To do so, we use a ``docker-compose'' file, which is a YAML
file describing the system configuration. The configuration file we used can be found
at~\cite{RepositoryErlang}, under the \verb_wrong-spawn_ folder. The following command sets up the system.

% We've reduced the prompt to $ which is the usual prompt indicator in e.g. wiki-articles and added the \ for linebreaks
\begin{lstlisting}[style=md]
$ docker-compose up -d
\end{lstlisting}
The following commands attach a remote console to each of location $n$ and location $m$, so as to be able to perform operations from them.
\pagebreak[5]
\begin{lstlisting}[style=md]
$ docker exec -it loc_n.com erl -name test@loc_n.com \
  -setcookie cookie -remsh app@loc_n.com -hidden
\end{lstlisting}
\begin{lstlisting}[style=md]
$ docker exec -it loc_m.com erl -name test@loc_m.com \
  -setcookie cookie -remsh app@loc_m.com -hidden
\end{lstlisting}
Event (1) is executed from the remote console of location $m$ as follows
\begin{lstlisting}[style=md]
(app@loc_m.com)2> erlang:spawn('app@loc_n.com', erlang, self, []).
\end{lstlisting}
Events (2,3,4,5) are executed by the following commands, where \verb_$1_ is
replaced with the network name and \verb_$2_ is replaced with the container name.
\begin{lstlisting}[style=md]
docker network disconnect $1 $2
docker container stop $2
docker container start $2
docker network connect $1 $2
\end{lstlisting}
However, rather than executing them as single commands we conveniently bundle
them in the script \verb_restart.sh_, available in the repository. Executing
them with a script ensures little delay between each command's execution,
leaving us enough time to attempt another spawn from location $n$ towards
location $m$ before location $n$'s reaction to the absence of location $m$'s
heartbeat. The script is then executed as follows.
\begin{lstlisting}[style=md]
$ ./restart.sh wrong-spawn_net1 loc_m.com
\end{lstlisting}
Finally, event (6) is executed as follows.
\begin{lstlisting}[style=md]
(app@loc_n.com)2> erlang:spawn('app@loc_m.com', erlang, self, []).
<0.108.0>
(app@loc_n.com)3> =WARNING REPORT==== 12-Sep-2024::13:22:40.233966 ===
** Can not start erlang:self,[] on 'app@loc_m.com' **
\end{lstlisting}
As we can see, the spawn from location $n$ towards location $m$ fails because the spawn
message sent by location $n$ is targeting an old instance of location $m$.

The above experiment mixes the use of Erlang primitives to perform spawns with
the use of Docker primitives to alter the network state, as one can do in D$\pi$FR. We now show how this
behavior is faithfully captured in D$\pi$FR, by showing a possible reduction sequence for system \textbf{WrongSp}:

{
  \small
\[
  \begin{array}{cl}
    &
      \sys{\net}{
      \underline{\proc{\spawn{n}{P}}{m}{1}} \paral \proc{\spawn{m}{Q}}{n}{1} \paral
      \proc{\sbreak{n}}{m}{1} \paral  \proc{\skill}{m}{1} \paral
      \proc{\create{m}{\link{n}}}{\godloc}{}
      }\\
    \step
    &
      \sys{\net}{
      \underline{\spawnc{(n,1):P}{m}{1}} \paral \proc{\spawn{m}{Q}}{n}{1} \paral
      \proc{\sbreak{n}}{m}{1} \paral  \proc{\skill}{m}{1} \paral
      \proc{\create{m}{\link{n}}}{\godloc}{}
      }\\
    \step
    &
      \sys{\net_1}{
      \proc{P}{n}{1} \paral \proc{\spawn{m}{Q}}{n}{1} \paral
      \underline{\proc{\sbreak{n}}{m}{1}} \paral  \proc{\skill}{m}{1} \paral
      \proc{\create{m}{\link{n}}}{\godloc}{}
      }\\
    \step
    &
      \sys{\net_2}{
      \proc{P}{n}{1} \paral \proc{\spawn{m}{Q}}{n}{1} \paral  \underline{\proc{\skill}{m}{1}} \paral
      \proc{\create{m}{\link{n}}}{\godloc}{}
      }\\
    \step
    &
      \sys{\net_3}{
      \proc{P}{n}{1} \paral \proc{\spawn{m}{Q}}{n}{1} \paral
      \underline{\proc{\create{m}{\link{n}}}{\godloc}{}}
      }\\
    \step
    &
      \sys{\net_4}{
      \proc{P}{n}{1} \paral \proc{\spawn{m}{Q}}{n}{1} \paral
      \underline{\proc{\link{n}}{m}{2}}
      }\\
    \step
    &
      \sys{\net_5}{
      \proc{P}{n}{1} \paral \underline{\proc{\spawn{m}{Q}}{n}{1}}
      }\\
    \step
    &
      \sys{\net_5}{
      \proc{P}{n}{1} \paral \underline{\spawnc{(m,1):Q}{n}{1}}
      }\\
    \step
    &
      \sys{\net_6}{
      \proc{P}{n}{1}
      }\\
  \end{array}
\]
}

with

{
  \small
  \[
  \begin{array}{lll}
    \net & \equiv & \langle \set{n \mapsto 1, m \mapsto 1}, \set{ n \connection
                    m}, \hat{\textbf{0}} \rangle\\
    \net_1 & \equiv & \langle \set{n \mapsto 1, m \mapsto 1}, \set{ n
                      \connection m}, \set{n\mapsto \set{m\mapsto 1}} \rangle\\
    \net_2 & \equiv & \langle \set{n \mapsto 1, m \mapsto 1}, \emptyset,
                      \set{n\mapsto \set{m\mapsto 1}} \rangle\\
    \net_3 & \equiv & \langle \set{n \mapsto 1, m \mapsto -1}, \emptyset,
                      \set{n\mapsto \set{m\mapsto 1}} \rangle\\
    \net_4 & \equiv & \langle \set{n \mapsto 1, m \mapsto 2}, \emptyset, \set{n\mapsto \set{m\mapsto 1}} \rangle\\
    \net_5 & \equiv & \langle \set{n \mapsto 1, m \mapsto 2}, \set{n \connection m},
                      \set{n\mapsto \set{m\mapsto 1}} \rangle\\
    \net_6 & \equiv & \langle \set{n \mapsto 1, m \mapsto 2}, \set{n \connection m},
                      \hat{\textbf{0}} \rangle\\
  \end{array}
  \]
}

In the above trace, the error takes place in the last reduction when rule
\rl{spawn-f} is triggered, due to the spawn message targeting the wrong
incarnation number of location $m$, which in turn is due to the wrong view
location $n$ yields about location $m$ in $\net_5$.

\section{Related Work and Conclusion} \label{sec:related}

We have presented in this paper D$\pi$FR, a distributed $\pi$-calculus with dynamic locations and links, 
crash failures and a weak recovery model, supporting location recovery by means of incarnation numbers.
To the best of our knowledge, this is the first work that combines these different features, and the first to 
adopt a weak recovery model. We have conducted experiments which show our model can indeed represent faithfully 
the behavior of Erlang systems in presence of node failures and recoveries. We have developed a behavioral theory 
for D$\pi$FR, including a full abstraction result that characterizes weak barbed congruence for the calculus
by a weak bimilarity. We have shown  
that our behavioral theory extends the one from the D$\pi$F paper \cite{FrancalanzaH08}, but that
weak barbed congruence in D$\pi$FR is in general more discriminative than in D$\pi$F because of the recovery-specific features.

We highlighted in the Introduction how the previous process calculus analyses of distributed systems with crash failures and recovery 
broadly compared with respect to these key features. We provide below some additional discussion and consider other related works.

The work by Fournet et al.~on the join calculus \cite{FournetGLMR96} shows how to extend the join calculus 
with primitives for crash failures and recoveries
but does not take into account links and link failures and does not present a behavioral theory
for these extensions of the join calculus. The work by Amadio \cite{Amadio97} on the $\pi_{1l}$-calculus
presents an asynchronous $\pi$-calculus with unique receivers, located processes, 
and location failures. It develops a behavioral theory for this calculus by translation of the $\pi_{1l}$-calculus
into the $\pi_1$ calculus (an asynchronous $\pi$-calculus with unique
receivers), but it relies on perfect failure detectors (i.e., a failure detector
that only detects faulty processes, with no false positives~\cite{GuptaCG01}),
and does not support links and link failures.
In its discussion of weaker failure detectors, 
it does indicate how an extension could support a local view (of failed locations) but it does not 
elaborate the corresponding calculus and behavioral theory. In \cite{FournetG96,Amadio97} failure/recovery is akin to an off/on switch: 
recovering a failed process is just a matter of restarting a process whose execution has been suspended by a failure. 
This is arguably extremely simplistic, and the underlying ability to recover a failed process in the exact state 
prior to failure is at best non-trivial or costly to implement in practice.

The recovery model in \cite{BergerH00} and \cite{BocchiLTV23} improves on the on/off 
switch model from \cite{FournetG96,Amadio97}.
To model recovery they rely on timed systems and a
checkpointing primitive (their \textsf{save} primitives) that captures the
execution state of a process and saves it in permanent storage for restart
after failure. This is more realistic than the on/off model, however the checkpointing primitive is still an expensive construct and not necessarily found in practice. 
For instance, the Erlang environment does not include a checkpointing capability
in its basic features.
The approach in~\cite{BergerH00} considers only a fixed finite number of nodes. They develop 
a behavioral theory in the form of a sound higher-order bisimilarity.
In~\cite{BocchiLTV23}
 systems comprise a fixed finite number of nodes with links connecting each pair of nodes,
together with a function giving the failure status  of nodes and links at any given time.
Their failure model is more sophisticated than just node and link crash failures, 
as it also allows to model link slowdowns, delaying the transmission of messages
beyond the expected link latency. 
Their behavioral theory is limited to a notion of weak barbed bisimilarity 
(symmetric weak-barb preserving and reduction closed relation) 
and they do not present 
a labelled transition semantics for their model.
In contrast, we are only concerned with crash failures and eschew the need for a timed model. 
Dealing with dynamic node and link creation is a prime objective of our work,
which cannot be simply modelled by activating a location at a given time, 
and which has a strong bearing on our behavioral theory and our full-abstraction result.

A clear inspiration for this paper was the work by Francalanza and Hennessy 
\cite{FrancalanzaH08} for their handling of node and link failures, and their behavioral theory. 
Apart from dealing with recovery, which they do not consider, our development is markedly different from theirs, 
as we described in the Introduction.
Compared to theirs, our approach is untyped, with a simpler handling of scope extrusion, 
simpler labels in our LTS semantics, 
and no need to make explicit the partial view of a network available to an observer to obtain our full abstraction result. 
This has several effects on our behavioral theory. Because our networks are not types,
more systems can be seen to be barbed congruent for they only need to weakly agree on the public part of their networks,
and do not need to have them identical. At the same time, our behavioral theory is more discriminative due to
our ability to establish links without the connectedness constraints in \cite{FrancalanzaH08}.
Also, we introduce explicit local views, which correspond to the belief that a location has of its neighbours 
and their current incarnation, which also have a bearing on our behavioral theory. 
Our local views are handled similarly as in Erlang 
-- in particular they may not reflect the current state of the network --,
and they have no equivalent in \cite{FrancalanzaH08}. 
The notion of partial views for observers in \cite{FrancalanzaH08}
is different: it is a way to filter out information contained 
in their complex labels so as to obtain full abstraction.
One should note also that \cite{FrancalanzaH08} is concerned with network partitions, 
with network extensions in their D$\pi$F calculus designed to be partition-preserving.
This is not a concern in D$\pi$FR. Our primitives for location and link creation and recovery
can heal network partitions.

Another work dealing with explicit links between computational nodes is \cite{NicolaGP07}, 
which presents a tuple-space based process calculus with dynamic locations and links.
The paper does not consider location failure and recovery but provides a behavioral theory with a full abstraction result.
Links in this work appear as terms of the process calculus, handling new link creation and name restriction
in a manner similar to ours. In contrast, we do not consider links as explicit process terms but gather all 
the link information, together with location incarnation and status, in our networks. We found this lead to
simpler reduction and LTS rules, and avoids the need to have a partial parallel composition operating only
on compatible systems (if network information is distributed in terms in parallel with locations as in \cite{NicolaGP07},
the definition of parallel composition must take into account the fact that the systems being composed may hold conflicting network information such 
as differing node and link status).

Formal models for distributed systems with failures can also be found in recent verification tools 
for distributed algorithms and distributed systems
such as  Disel \cite{SergeyWT18}, Gobra \cite{WolfACOPM21}, Perennial \cite{ChajedTKZ19}, 
Psync \cite{DragoiHZ16}, TLC \cite{GriffinLSY20},
Verdi \cite{WilcoxWPTWEA15}. They can either rely on a specific language (such as Gobra, for Go programs),
or domain specific languages for formally specifying algorithms (such as PSync or Disel), or be more general purpose, relying 
on a mixture of logic and more operational models (such as Perennial, Verdi or TLC).
Verdi in particular is interesting for it supports a variety of failure models, 
including a model of crash failures and recoveries which is quite close to ours, and makes use of simulation relations in its proof techniques.
However, to the best of our knowledge, they (Verdi included) do not provide as we do 
a compositional theory of system equivalence in presence of crash failures and recoveries.

A formal analysis of fault-tolerant behaviors by means of simulation relations is provided in \cite{DemasiCMA17}. 
Their analysis aims to find algorithms on finite Kripke structures for verifying 
simulation relations capturing different notions of fault-tolerance (masking, non-masking, failsafe) 
taken from \cite{Gartner99}. They characterize faults by an explicit colouring (good or bad) of states in a Kripke structure. 
In contrast, D$\pi$FR systems can have infinite states, and a fault, error or failure
would correspond to a deviation from a specified behavior,
with location and link crashes as primitive faults in a system.  
Nevertheless, it would be interesting to develop in our setting similar characterizations of fault-tolerance, 
as well as to study notions of recoverability inspired by those in \cite{BocchiLTV23}.

The calculus presented in this paper provides us with a basis for further studies. 
One of the consequences of the closeness to Erlang we have adopted in this work is the fact that the way we deal with local views
corresponds to the Erlang policy. It would be interesting to study a variant of D$\pi$FR where programs can specify
their own policies for managing local views. This can be achieved by adding further primitives to complement the forget primitive
to explicitly manipulate local view information. 
This extension would allow one to encode different failure detector schemes, as well as distributed protocols relying on an explicit management
of local views, such as epidemic protocols.
It would certainly be interesting to further 
expand D$\pi$FR to cater for other failure models, including the kinds of grey failures tackled by Bocchi et al. \cite{BocchiLTV23}, 
or Byzantine failures,
probably making it parametric in failure models along the lines of Verdi  \cite{WilcoxWPTWEA15}. 
It would also be interesting to revisit ideas from \cite{FrancalanzaH07} 
about the semantical characterization of, and proof techniques for, fault tolerance.
Of particular interest would be the ability in our setting to drastically reduce the size of bisimulations
in presence of unbounded occurrences of failures and recoveries as in failure and recovery loops.

% In its current form, we think
% it is close enough to the behavior of Erlang systems, and its primitives sufficient to account for a significant subset of Erlang
% failure handling constructs, to allow us to study reversible debugging for distributed Erlang systems with failures
% and recoveries, building on the substantial amount of work developed in the past ten years around reversibility and reversible debugging in Erlang \cite{LaneseNPV18,LanesePV21}.

%%% Local Variables:
%%% mode: latex
%%% TeX-master: "main-distributed"
%%% End:

\clearpage

\bibliographystyle{alphaurl}
\bibliography{bibliography}

\clearpage

\appendix

\section{Free Names and Alpha-Conversion on Systems}\label{app:free-names}

The definition of free incarnation variables in configurations and systems is completely standard.

Free incarnation variables in processes are defined inductively as follows:
\begin{align*}
	& \fv{\nil} = \emptyset &
	& \fv{\out{x}{\vect{u}}.P} = \fv{P} \cup (\vect{u} \cap \ms{I}) \\
	& \fv{x(\vect{v}).P} = \fv{P}\setminus \vect{v} &
	& \fv{!x(\vect{v}).P} = \fv{P}\setminus \vect{v} \\
	& \fv{\res{w} P} = \fv{P} \setminus \{ w \} &
	& \fv{\ms{if}~ r = s~\ms{then}~P~\ms{else}~Q} = \fv{P} \cup \fv{Q} \cup (\{r,s\} \cap \ms{I}) \\
	& \fv{P \parop Q} = \fv{P} \cup \fv{Q} &
	& \fv{\node{n,\iota}.P} = \fv{P}\setminus \{ \iota \} \\
	& \fv{\remove{n}.P} =  \fv{P} &
	& \fv{\spawn{n}{P}} =  \fv{P} \\
	& \fv{\skill} = \emptyset &
	& \fv{\create{n}{P}} = \fv{P} \\
	& \fv{\link{n}} =  \emptyset &
	& \fv{\sbreak{n}} = \emptyset
\end{align*}

Free incarnation variables in configurations are defined inductively as follows:
\begin{align*}
	& \fv{\proc{P}{n}{\lambda}} = \fv{P}&
	& \fv{\spawnc{(m,\kappa):P}{n}{\lambda}} = \fv{P}\\
        & \fv{N \paral M} = \fv{N} \cup \fv{M} &
        & \fv{\nil} = \emptyset
\end{align*}

Free incarnation variables in networks and systems are defined inductively as follows:
\begin{align*}
	& \fv{\sys{\net}{N}} = \fv{N} \\
	& \fv{\res{\vect{u}} S} = \fv{S} \setminus \vect{u}
\end{align*}

The notion of free names in configurations is completely standard too, while in systems is
slightly unconventional because of the presence of a network, but it can be
defined straightforwardly as follows.

Free names in processes are defined inductively as follows:
\begin{align*}
	& \fn{\nil} = \emptyset &
	& \fn{\out{x}{\vect{u}}.P} = (\vect{u} \cap (\ms{N} \cup \ms{C})) \cup \{ x \} \cup \fn{P} \\
	& \fn{x(\vect{v}).P} = \{ x \} \cup (\fn{P}\setminus \vect{v}) &
	& \fn{!x(\vect{v}).P} = \{ x \} \cup (\fn{P}\setminus \vect{v}) \\
	& \fn{\res{w} P} = \fn{P} \setminus \{ w \} &
	& \fn{\ms{if}\, r = s\,\ms{then}\,P\,\ms{else}\,Q} = \begin{array}{l}\fn{P} \cup \fn{Q} \\ \quad\mathrel{\cup} (\{r,s\} \cap (\ms{N} \cup \ms{C}))\end{array}\\
	& \fn{P \parop Q} = \fn{P} \cup \fn{Q} &
	& \fn{\node{n,\iota}.P} = \fn{P}\setminus \{ n \} \\
	& \fn{\remove{n}.P} =  \fn{P} \cup \{ n \} &
	& \fn{\spawn{n}{P}} =  \fn{P} \cup \{ n \} \\
	& \fn{\skill} = \emptyset &
	& \fn{\create{n}{P}} = \fn{P} \cup \{ n \} \\
	& \fn{\link{n}} =  \{ n \} &
	& \fn{\sbreak{n}} = \{ n \}
\end{align*}

Free names in configurations are defined inductively as follows:
\begin{align*}
	& \fn{\proc{P}{n}{\lambda}} = \fn{P} \cup \{ n \} &
	& \fn{\spawnc{(m,\kappa):P}{n}{\lambda}} = \fn{P} \cup \{m,n \} \\
        & \fn{N \paral M} = \fn{N} \cup \fn{M} &
        & \fn{\nil} = \emptyset
\end{align*}

Free names of networks and systems are defined inductively as follows:
\begin{align*}
	& \fn{\net} = \ms{supp}(\net_{\mc{A}}) \cup \ms{dom}(\net_{\mc{L}})\\
	& \fn{\sys{\net}{N}} = \fn{\net} \cup \fn{N} \\
	& \fn{\res{\vect{u}} S} = \fn{S} \setminus \vect{u}
\end{align*}
To define alpha-conversion on systems, we define capture-avoiding substitution on networks.
A capture avoiding substitution $\sub{v}{u}$ on network $\net$
is defined as follows: if $\net = (\mc{A}, \mc{L}, \mc{V})$, then
$\net\sub{v}{u} = (\mc{A}\sub{v}{u}, \mc{L}\sub{v}{u}, \mc{V}\sub{v}{u})$ where:
\begin{align*}
	& \mc{A}\sub{v}{u} =
    \begin{cases}
      \mc{A}[v \mapsto \mc{A}(u)][u \mapsto 0] \quad \quad \ms{if} \quad u,v \in \ms{N} ~\ms{and}~ v \not\in\ms{supp}(\mc{A}) \\
      \mc{A} \quad \ms{otherwise}
    \end{cases} \\
	& \mc{L}\sub{v}{u} =
    \begin{cases}
      (\mc{L} \setminus \mc{L}(u)) \cup \mc{L}(u)\sub{v}{u} \quad \ms{if} \quad u,v \in \ms{N} ~\ms{and}~ v \not\in\ms{supp}(\mc{A}) \\
      \mc{L} \quad \ms{otherwise}
    \end{cases} \\
        & \mc{V}\sub{v}{u} = \set{\mc{V}(n) \sub{v}{u} | n \in \ms{dom}(\mc{V}) \cup \set{v}} \quad \ms{if} \quad u,v \in \ms{N} ~\ms{and}~ v \not\in\ms{supp}(\mc{A})
%	& \mc{V}\sub{v}{u} =
%    \begin{cases}
%      \mc{V}[v \mapsto \mc{V}(u)][u \mapsto \mathbf{\hat{0}}] \quad \quad \ms{if} \quad u,v \in \ms{N} ~\ms{and}~ v \not\in\ms{supp}(\mc{A}) \\
%      \mc{V} \quad \ms{otherwise}
%    \end{cases}
\end{align*}
with:
\begin{align*}
	& \mc{L}(n) = \{ (a,b) \in \mc{L} \mid a = n ~\ms{or}~b = n \} \\
	& \mc{L}(n)\sub{m}{n} = \{ (a\sub{m}{n}, b\sub{m}{n}) \mid (a,b) \in \mc{L}(n) \} \\
	& a\sub{m}{n} = \begin{cases}
                    m \quad \ms{if} \quad a = n \\
                    a \quad \ms{otherwise}
  \end{cases}\\
	& \mc{V}(n)\sub{v}{u} =
    \begin{cases}
      \mc{V}(u) \quad \quad \ms{if} \quad n=v \\
      {\mathbf{\hat{0}}} \quad \quad \quad \ms{if} \quad n=u \\ 
      \set{n\sub{v}{u} \mapsto \lambda | n \mapsto \lambda \in \mc{V}(n)} \ \ \ms{otherwise}
    \end{cases}
\end{align*}
The effect of a capture avoiding substitution $\sub{v}{u}$ on a system
$\sys{\net}{N}$ is defined inductively as follows.

\begin{align*}
	& (\sys{\net}{N})\sub{v}{u} = \sys{\net\sub{v}{u}}{N\sub{v}{u}} \\
	& (\res{w} S) \sub{v}{u} = 
		\begin{cases}
		  \res{w} S\sub{v}{u} & \ms{if}~u,v \neq w \\
      \res{w} S & \ms{if}~u= w \\
      \res{w'} S\sub{w'}{w} \sub{v}{u} & \ms{if}~v= w~\ms{with}~w'~\ms{fresh}\\
			% \bot \quad \quad \quad \quad \quad \ms{otherwise}
		\end{cases}
\end{align*}
Equality modulo alpha-conversion $=_\alpha$ on systems is now defined as the smallest equivalence relation on systems defined by the following rules:
\begin{mathpar}
  \inferrule*[]{ } % S\sub{v}{u} \neq \bot}
  {
    \res{u}S =_{\alpha} \res{v} S\sub{v}{u}
  }
	
	\and
	
	\inferrule*[]{ S =_\alpha T}{
    \res{u}S =_\alpha \res{u} T}
	
\end{mathpar}

%%% Local Variables:
%%% mode: LaTeX
%%% TeX-master: "main-distributed"
%%% End:

\clearpage

\section{Modifying Network: Proof}\label{app:proof-network}

\begin{restatable}{prop}{modifyingnetwork}\label{prop:extension}
  Let $S = \sys{\res{\vect{u}}\net}{N}$ be a closed system,
  and $\net=\tuple{\mc{A}, \mc{L}, \mc{V}}$,
  $\net'=\tuple{\mc{A'},\mc{L'},\mc{V}'}$ be networks such that the following conditions hold:
  \begin{enumerate}
    \item $\forall n \in \vect{u}.~ \mc{A}(n) = \mc{A'}(n)$
    \item $\forall n \in \ms{supp}(\mc{A'}). \mc{A}(n) = \mc{A}'(n) \vee |\mc{A}(n)| < |\mc{A}'(n)| \vee ( \mc{A}'(n) = -\mc{A}(n) \wedge \mc{A}(n) \geq 0 )$ 
    \item $\forall n \connection m \in (\mc{L'} \setminus\mc{L}) \cup (\mc{L}\setminus \mc{L}'). ~ n,m\notin \vect{u}\wedge \exists l \in \set{n,m}.(\mc{A}(l) > 0 \vee |\mc{A}(l)| < |\mc{A'}(l)|)$
    \item $\forall n\in\mc{N}, m \in \vect{u}.~ \mc{V}(m) = \mc{V'}(m), \mc{V}(n)(m) = \mc{V'}(n)(m)$
    \item
        $\forall n,m \in \mc{N}, \mc{V}(n)(m) \neq \mc{V'}(n)(m) = \lambda \wedge \lambda \neq 0. (\forall l\in\set{n,m}. \mc{A}(l) > 0 \vee |\mc{A}(l)| < |\mc{A'}(l)|) \vee \lambda \leq |\mc{A'}(m)|$
          \item
        $\forall n,m \in \mc{N}, \mc{V'}(n)(m) \neq \mc{V'}(n)(m) = 0. ~\mc{A}(n) > 0 \vee |\mc{A}(n)| < |\mc{A'}(n)|$
  \end{enumerate}
  Then, there exists a configuration $L$ with $fn(L)\cap \vect{u} =\emptyset$
  such that: $\; \sys{\res{\vect{u}}\net}{N \paral L} \wstep \sys{\res{\vect{u}}\net'}{N}$.
\end{restatable}

  Before presenting the proof we first comment on the
  required conditions and then provide some informal intuition on the nature of the
  proof.

  Condition (1) requires $\mc{A}$ and $\mc{A'}$ to agree on private
  names and condition (2) requires for the new incarnation number of the target
  location to be equal or in the future of the previous one.  Condition (3) requires $\mc{L}$ and $\mc{L'}$ to agree on links with at least a private end 
  and for at least one of the two locations
  to be either already alive or to be alive during the transformation from
  $\mc{A}$ to $\mc{A'}$. Condition (4) requires $\mc{V}$ and $\mc{V'}$ to agree on private names, while conditions (5) and (6) require that in order for a view to change either both locations are alive at some point during the transformation (condition (5)) or that the location bearing the view is alive and it forgets about the other location (condition (6)).

  We now give an intuition of the proof.  

  To appropriately update the network we build a context performing the desired
  changes, which modifies its three elements: i) the alive set; ii) the link
  set; and iii) the view. The context to modify each element originates from the
  special location $\godloc$, since we know that this location is always alive.

  Let us begin with the alive set. To update the incarnation number of a given
  location we kill it, by spawning on it a \tms{kill} process, and recreate it the correct number of times from
  $\godloc$. For this operation to be carried out successfully, it is necessary
  for the special location $\godloc$ and the target one to be connected and in
  general this may or may not be the case. To obviate this problem, together
  with the appropriate number of killing and creation processes, we also spawn
  the following two processes
  \[
    \proc{\link{n}}{\godloc}{} \paral \proc{\sbreak{n}}{\godloc}{}
  \]
  where $n$ is the target location.
  These two processes can either set up a connection, for the time it
  is needed to update the desired incarnation number of $n$, and then
  destroy it, in case this was not present, or destroy the connection
  and then restore it immediately after, bearing no changes on the
  system. In other words, these two processes alter temporarily the
  state of the connection, making both the presence and the absence of a link available (at different times).

  Now, let us discuss the changes to the link set. The changes can be of two
  kinds, either a link between two locations is set up or it is destroyed. We
  discuss the first scenario, the second one is identical.
  Thanks to the condition on $\mc{L}'$ we know that the link in $\mc{L}$ does
  not exist and that at some point in the transformation from $\mc{A}$ to $\mc{A}'$
  at least one of the two ends will be alive. The strategy we follow is the same
  as in the previous case: we add in parallel a process on the special location
  $\godloc$ that spawns on one of the two ends a process setting up the
  connection. However, since we are only guaranteed that one of the two ends will be alive, but we do not know which one, we target both of them.

  We also
  put in parallel two sets of processes which, as in the above case, ``alter''
  the link between the special location $\godloc$ and the two ends. By doing so,
  when one of the two ends will be alive during the transition from $\mc{A}$ to  $\mc{A'}$, the special location $\godloc$ will be able to send on it the
  process setting up the connection. The processes on $\godloc$ targeting the
  other end can simply fail by triggering the \tms{spawn} when the link between
  $\godloc$ and the target location does not exist.

  Finally, let us discuss the changes to the view. To update $\mc{V}$ into
  $\mc{V'}$ it is necessary to either remove elements from a location's view or
  to add new ones. To remove them it suffices to spawn on the target location $n$ a
  process removing the belief about location $m$ from the special location
  $\godloc$. To update the knowledge of $n$ with the belief of $m$ at
  incarnation $\lambda$ it suffices to put in parallel a spawn message, in
  case the belief must be set up to some $\lambda$ that is obsolete, or
  to add a process on the special location $\godloc$ that will migrate to $m$
  and then \tms{spawn} a message on $n$, in case $\lambda$ belongs to
  the future of the system (we can not add directly a spawn message, since it would result in a system not well-formed). Notice that also in this case the
  existence of the necessary links, i.e., between location $\godloc$ and both locations $n$ and $m$, is guaranteed thanks to the presence of
  processes that alter the state of these links as needed.

\begin{proof}
  Assume to have $\sys{\res{\vect{u}}\net}{N}$, where
  $\net=\tuple{\mc{A},\mc{L},\mc{V}}$, and $\net'=\tuple{\mc{A'},\mc{L'},\mc{V'}}$.

  To transform $\mc{A}$ into
  $\mc{A'}$ we need to transform all the incarnation numbers of $\mc{A}$ into
  the ones of $\mc{A}'$. Here, we leverage function $\ms{transform}$ that for
  a location $n$ returns us an environment that kills and creates location $n$ to change its
  incarnation number from $\mc{A}(n)$ to $\mc{A}'(n)$.
  We formally define it as
  \begin{align*}
     \ms{transform}(n,\lambda,\kappa)~\ms{when}~\lambda < 0 \wedge \lambda \neq \kappa ::= &~\proc{\create{n}{\nil}}{\godloc}{} \paral \ms{transform}(n,|\lambda|+1,\kappa)\\
     \ms{transform}(n,\lambda,\kappa)~\ms{when}~\lambda > 0 \wedge \lambda \neq \kappa ::= &~\proc{\spawn{n}{\skill}}{\godloc}{} \paral \ms{transform}(n,-\lambda,\kappa)\\
     \ms{transform}(n,\lambda,\lambda) ::= &~\nil 
  \end{align*}
  Then, to effectively change all the incarnation numbers it suffices to build
  the following context
  \[
    L_\mc{A} = \prod_{n\in \ms{supp}(\mc{A}'). \mc{A}'(n)\neq 0} \ms{transform}(n, \mc{A}(n),\mc{A}'(n)) \paral \proc{\link{n}}{\godloc}{} \paral \proc{\sbreak{n}}{\godloc}{}
  \]

  To change $\mc{L}$ into $\mc{L'}$ we need to add and/or remove links.
  %to perform the same operations as
  %above.
  To remove links it suffices to build the following context
  \[
      L^-_\mc{L} =
      \prod_{(n,m)\in\mc{L}\setminus \mc{L'}}
      \begin{array}{l}
        \proc{\spawn{n}{\sbreak{m}.\nil}}{\godloc}{} \paral
        \proc{\link{n}.\nil}{\godloc}{} \paral \proc{\sbreak{n}.\nil}{\godloc}{}
        \paral \\
        \proc{\spawn{m}{\sbreak{n}.\nil}}{\godloc}{} \paral
        \proc{\link{m}.\nil}{\godloc}{} \paral \proc{\sbreak{m}.\nil}{\godloc}{}
      \end{array}
  \]

  Then, to add the links it suffices to build the following context
  \[
      L^+_\mc{L} = \prod_{(n,m)\in\mc{L'}\setminus \mc{L}}
      \begin{array}{l}
        \proc{\spawn{n}{\link{m}.\nil}}{\godloc}{} \paral
        \proc{\link{n}.\nil}{\godloc}{} \paral \proc{\sbreak{n}.\nil}{\godloc}{} \paral\\
        \proc{\spawn{m}{\link{n}.\nil}}{\godloc}{} \paral
          \proc{\link{m}.\nil}{\godloc}{} \paral \proc{\sbreak{m}.\nil}{\godloc}{}
      \end{array}
  \]

  To modify the views we need to consider three possibilities, and we build a context for each of them:
  \begin{itemize}
  \item the belief of location $n$ about $m$ in $\mc{V'}$ is 0
\[    L_{\mc{V}_1} = \prod_{\forall n,m \in \mc{N}^\godloc \setminus \vect{u}. \mc{V'}(n)(m) = 0}
      \proc{\spawn{n}{\remove{m}}}{\godloc}{}
\]

  \item the belief of location $n$ about $m$ in $\mc{V'}$ is an incarnation
    number $\lambda$ such that $\mc{V'}(n)(m) \leq |\mc{A}(m)|$

    \[L_{\mc{V}_2} = \prod_{\forall n,m \in \mc{N}^\godloc \setminus \vect{u}.
     0 < \mc{V'}(n)(m) \leq |\mc{A}(m)|} \proc{\spawn{n}{\remove{m}}}{\godloc}{} \paral \spawnc{(n,0):\nil}{m}{\mc{V}'(n)(m)}
    \]
    
  \item the belief of location $n$ about $m$ in $\mc{V'}$ is an incarnation
    number $\lambda$ such that $|\mc{A}(m)| < \mc{V'}(n)(m) \leq |\mc{A'}(m)|$

    \[L_{\mc{V}_3} = \prod_{\forall n,m \in \mc{N}^\godloc \setminus \vect{u}.
       |\mc{A}(m)| < \mc{V}'(n)(m) \leq |\mc{A}'(m)|} \proc{\spawn{m}{\spawn{n}{\nil}}}{\godloc}{}\]
    
  \end{itemize}

  In addition, we need to create links to ensure communications can be done, and remove them afterwards.
  \begin{align*}
    & L_{\mc{V}_{links}} = \prod_{n,m \in \ms{dom}(\mc{A}) \cup \ms{dom}(\mc{A}')}
    %{\forall n,m \in \mc{N}^\godloc \setminus (\vect{u}\cup \mc{L}'). \mc{V}'(n)(m) \neq 0}
      \begin{array}{l}
        \proc{\link{m}}{\godloc}{} \paral \proc{\sbreak{m}}{\godloc}{}\paral\\
        \proc{\link{n}}{\godloc}{} \paral \proc{\sbreak{n}}{\godloc}{}\paral\\
        \proc{\spawn{m}{\link{n}}}{\godloc}{} \paral \proc{\spawn{m}{\sbreak{n}}}{\godloc}{}
      \end{array}
  \end{align*}

  The first context, $L_{\mc{V}_1}$, handles the case in which the belief of location $n$
  about location $m$ is set to 0. Notice that we cannot simply put in parallel
  a process $\proc{\remove{m}}{n}{|\mc{A}'(n)|}$ since it may break the notion
  of well-founded system as $|\mc{A}'(n)|$ may possibly be greater than
  $|\mc{A}(n)|$. In other words, we cannot use processes from the future to set
  views.

  The second context, $L_{\mc{V}_2}$, handles the case in which the belief of
  location $n$ about location $m$ is set to an old incarnation number of $m$. In
  this case we can simply extend our system with a message with the desired
  incarnation number. The \tms{forget} process is required to erase any
  knowledge $n$ may have about $m$ which would impede the correct reception of
  the message.

  The third context, $L_{\mc{V}_3}$, handles the case in which the belief of
  location $n$ about location $m$ is set to an incarnation number for $m$ that
  still does not exist in the system. For this reason we cannot extend it with a
  message, otherwise we would invalidate the notion of well-founded systems, but
  rather with a process, which we create from $\godloc$ because it possibly is a
  ``future process'', that by reducing at the right time will set the view as we
  desire.

  The fourth context, $L_{\mc{V}_{links}}$, is required to activate a connection between
  $n$ and $m$, again to not use future processes we are obliged to use
  $\godloc$, and to deactivate it afterwards.

  Finally, we can set $L = L_\mc{A} \paral L^-_\mc{L}
  \paral L^+_\mc{L} \paral L_{\mc{V}_1} \paral L_{\mc{V}_2} \paral L_{\mc{V}_3} \paral L_{\mc{V}_{links}}$.
\end{proof}

  At first glance, the above result may seem weak since it states that there
  exists a reduction accomplishing the desired changes but does not say anything
  about possible other reductions which may not accomplish them.

%  This scenario is akin in nature to when in $\pi$ one extends a system $P$ with
%  a system $Q$, i.e., having $P \paral Q$. Putting the two systems in parallel
%  does not force the possible interactions between them nor the
%  order in which they will occur. However, when testing the equivalence of two
%  systems under the same context we are also considering all the possible
%  reductions that the composed system can do.

  However, when testing the equivalence of two systems, since the
  equivalence is universally quantified over all the possible reductions, the
  guaranteed existence of one computational path leading to the desired changes
  is enough to ensure that the two systems are also being tested under those
  changes.

%%% Local Variables:
%%% mode: latex
%%% TeX-master: "main-distributed"
%%% End:

\clearpage

\section{Full Abstraction}\label{app:full-abstraction}

\subsection{Soundness}\label{app:ssec:soundness}

% \begin{definition}[Strong Simulation] A binary relation $\rel S \subseteq \cset{S} \times \cset{S}$ over closed systems is a strong simulation iff
%   whenever $(P,Q)\in \rel S$,
%   \begin{itemize}
%   \item $P \lstep{\alpha} P'$ implies $Q\lstep{\alpha} Q'$ for some $Q'$ with
%     $(P',Q')\in \rel S$
%   \end{itemize}
% \end{definition}
%
% \begin{definition}[Strong Bisimulation] A relation $\rel S$ is a strong bisimulation if
%   both $\rel S$ and $\rel S^{-1}$ are strong simulations.
% \end{definition}
%
% \begin{definition}[Strong Bisimilarity]
%   Strong bisimilarity, denoted by $\sim$, is the largest srong bisimulation over systems.
% \end{definition}
%
% The fact $\sim$ is an equivalence relation is a standard result for any labelled transition system.

We start with a simple lemma on the preservation of free names by labelled transitions.
If $\alpha$ is a label from our LTS semantics, we define $\ms{po}(\alpha)$, $\ms{pi}(\alpha)$ and $\ms{pe}(\alpha)$ as follows:
\begin{align*}
	& \ms{po}(\alpha) = 
	     \begin{cases}
	     	\vect{u}\setminus\vect{w} \quad \quad \ms{if} \quad \alpha = \res{\vect{w}}\out{x}{\vect{u}}@n_\lambda \\
			\emptyset \quad \quad \ms{otherwise}
	     \end{cases}
	&& \ms{pi}(\alpha) = 
	     \begin{cases}
	     	\vect{u} \quad \quad \ms{if} \quad \alpha = x(\vect{u})@n_\lambda \\
			\emptyset \quad \quad \ms{otherwise}
	     \end{cases}\\
       	& \ms{pe}(\alpha) = 
	     \begin{cases}
	     	\vect{w} \quad \quad \ms{if} \quad \alpha = \res{\vect{w}}\out{x}{\vect{u}}@n_\lambda \\
			\emptyset \quad \quad \ms{otherwise}
	     \end{cases}      
\end{align*}

\begin{lem}\label{lemma:fnames}
	Let $S,S'$ be closed systems such that $S \lstep{\alpha} S'$.
	 Then $\fn{S'} \subseteq \fn{S} \cup \ms{pi}(\alpha) \cup \ms{pe}(\alpha)$.
\end{lem}
\begin{proof}
	By induction on the derivation of $S \lstep{\alpha} S'$. 
\end{proof}

We now prove two lemmas relating labelled transitions in the calculus LTS semantics and the structure of systems.
In the remainder, we extend $\equiv$ to output actions, and identify equivalent output actions, setting:
\begin{align*}
	& \res{v}\res{w}\out{x}{\vect{u}}@n_\lambda ~\equiv~ \res{w}\res{v}\out{x}{\vect{u}}@n_\lambda
	&& \res{w} \alpha ~\equiv~ \res{w} \omega \quad \ms{if} \quad \alpha ~\equiv~ \omega
\end{align*}

\begin{lem}[Input/output actions and systems]\label{lemma:act-sys}
	Let $S,S'$ be closed systems such that $S \lstep{\alpha} S'$. The following properties hold:
  \begin{enumerate}

  \item if $\alpha = \res{\vect{v}}\out{x}{\vect{u}}@n_\lambda$ then
  	$$ S \equiv \sys{\res{\vect{w}}\net}{\proc{\out{x}{\vect{u}}.P}{n}{\lambda} \paral N }
	\quad \text{for some}~ \net, \vect{w}, P, N  ~~\text{with}~~  x,n \notin \vect{w},~\vect{v} \subseteq \vect{w},
	~\net \vdash n_\lambda : \ms{alive}$$

  \item if $\alpha = x(\vect{u})@n_\lambda$ then
  	$$ S \equiv \sys{\res{\vect{w}}\net}{\proc{x(\vect{u}).P}{n}{\lambda} \paral N }
	\quad \text{for some}~ \net, \vect{w}, P, N  ~~\text{with}~~  \vect{u}, x,n \cap \vect{w} = \emptyset, 
	 ~\net \vdash n_\lambda : \ms{alive}$$
  \end{enumerate}
\end{lem}
\ifwithproof
  \begin{proof}  
  	We show the first assertion, the second one is handled similarly. 
	We reason by induction on the derivation of $S \lstep{\alpha} S'$, where $\alpha = \res{\vect{v}}\out{x}{\vect{u}}@n_\lambda$,
	considering the last rule used in the proof tree:
	\begin{itemize}
		\item Rule \lrl{out}{}: in this case, we have $S = \sys{\net}{\proc{\out{x}{\vect{u}}.P}{n}{\lambda}}$, 
		$\net\vdash n_\lambda : \ms{alive}$, as required.
		
		\item Rule \lrl{par}{L}: in this case, we have 
		$S = \sys{\net}{N \paral M}$, $\alpha = \out{x}{\vect{u}}@n_\lambda$, with 
		$\sys{\net}{N} \lstep{\alpha} \sys{\net'}{N'}$.
		By induction assumption, we have $N \equiv \proc{\out{x}{\vect{u}}.P}{n}{\lambda} \paral L$ for some $L$,
		with $\net \vdash n_\lambda:\ms{alive}$.
		Hence $S \equiv \sys{\net}{ \proc{\out{x}{\vect{u}}.P}{n}{\lambda} \paral L \paral M}$, 
		$\net \vdash n_\lambda:\ms{alive}$,
		as required.
		
		\item Rule \lrl{par}{R}: same as \lrl{par}{L}.
		
		\item Rule \lrl{res}{}: in this case we have 
		$S = \res{z} S'$ with $S' \lstep{\alpha} S''$, for some $S',S''$, and $z \notin \fn{\alpha}$.
		By induction assumption, we have 
		$S' \equiv \sys{\res{\vect{w}}\net}{\proc{\out{x}{\vect{s}}.P}{n}{\lambda} \paral N }$, with 
		$\net \vdash n_\lambda: \ms{alive}$,~$x,n \notin \vect{w}$, $\vect{v} \subseteq \vect{w}$.
		Thus $S \equiv \sys{\res{z,\vect{w}}\net}{ \proc{\out{x}{\vect{s}}.P}{n}{\lambda} \paral N }$, with
		$\net \vdash n_\lambda: \ms{alive}$, 
		and $x,n \notin z,\vect{w}$, and $\vect{v} \subseteq z,\vect{w}$ as required.		
		
  \item Rule \lrl{res}{o}: in this case, we have $S = \res{z} T$ with $T
    \lstep{\omega} T'$, $z \in \vect{u}\setminus \set{x,n}$, $\alpha = \res{z}
    \omega$. By induction assumption, we have $ T \equiv
    \sys{\res{\vect{w}}\net}{\proc{\out{x}{\vect{r}}.P}{n}{\lambda} \paral N }~$
     for some $\net,\vect{r}, \vect{w}, P, N ~~\text{with}~~ x,n \notin \vect{w},
    ~\vect{r}\setminus\{ z\} \subseteq \vect{w}, ~\net \vdash n_\lambda :
    \ms{alive} $. Hence, we have $ S \equiv
    \sys{\res{z,\vect{w}}\net}{\proc{\out{x}{\vect{r}}.P}{n}{\lambda} \paral N }
    $ for some $ \net,\vect{r}, \vect{w}, P, N ~~\text{with}~~ x,n \notin
    z,\vect{w}, ~\vect{v} \subseteq z,\vect{w}, ~\net \vdash n_\lambda :
    \ms{alive} $ as required. \qedhere
	\end{itemize}	
  \end{proof}
\else
\fi

\begin{prop}[Structural congruence is a strong bisimulation]\label{prop:str-congr-lts}
  We have $\equiv ~\subseteq~ \sim$.
\end{prop}
\ifwithproof
  \begin{proof}
	Since $\equiv$ is an equivalence relation, it suffices to prove that $\equiv$ is a strong simulation, namely that
	for any closed systems $S,R$, if $S\lstep{\alpha}S'$ and $S\equiv R$, then there exists $R'$ such that
	  $R\lstep{\alpha} R'$ and $S' \equiv R'$.
	We reason by induction on the derivation of $S \equiv R$, considering the last rule used in the proof tree:
	
	\begin{description}
		\item [Rule \rl{s.res.c}] In this case, $S = \res{u}\res{v}T$ and $R = \res{v}\res{u}T$. 
		Since $S \lstep{\alpha} S'$, this can only have been obtained by applying rule \lrl{res}{} twice, 
		rule \lrl{res}{o} twice, or a combination of rule \lrl{res}{} and rule \lrl{res}{o}. 
		We consider the four cases:
		\begin{description}
			\item [Rule \lrl{res}{} applied twice]  In this case,
			we have $T \lstep{\omega} T'$, for some $T'$, $u,v \notin \fn{\alpha}$, $\omega = \alpha$,
			and $S' = \res{u}\res{v}T'$. Now applying rule \lrl{res}{} twice we get
			$R \lstep{\alpha} \res{v}\res{u}T'$.
			Hence, by applying rule \rl{s.res.c} we have $R' \equiv S'$, as required.
			
			\item [Rule  \lrl{res}{o} applied twice] In this case, we have 
			$T \lstep{\omega} T'$ for some $T'$,
			$v \in \ms{po}(\omega)$, $u \in \ms{po}(\omega)\setminus\{v\}$, $\alpha = \res{u}\res{v}\omega$,
			and $S' = T'$. Applying rule \lrl{res}{o} twice we get:
			$R \lstep{\alpha} T'$, hence we have $R' \equiv S'$, as required.
			
    \item [Rule \lrl{res}{o} followed by rule \lrl{res}{} ] In this case we have
			$T \lstep{\omega} T'$ for some $T'$, 
			$v \in \ms{po}(\omega)$, $u \not\in \ms{po}(\omega)\setminus\{v\}$, $\alpha = \res{v}\omega$,
			and $S' = \res{u} T'$. Applying rule \lrl{res}{} we get:
			$\res{u}T \lstep{\omega} \res{u}T'$. Applying \lrl{res}{o} we get:
			$R = \res{v}\res{u}T \lstep {\alpha} \res{u}T' = R'$. 
			Hence we have $R' \equiv S'$, as required.
			
			\item [Rule \lrl{res}{} followed by rule \lrl{res}{o} ] Similar to the previous case.
		\end{description}

		\item [Rule \rl{s.res.nil}] In this case, $S = \res{u} R$, $u \notin \fn{R}$.
		Since $S \lstep{\alpha} S'$, this can only have been obtained by applying rule \textsc{l-res} as the last rule
		(rule \textsc{l-res\textsubscript{O}} is not a possibility for $u \notin \fn{R}$).
		Hence, we have $R \lstep{\omega} R'$, $S' = \res{u}R'$, 
		$u \not\in \fn{\omega}$.
		Now by Lemma~\ref{lemma:fnames}, we have $\fn{R'} \subseteq \fn{R} \cup \ms{pi}(\omega) \cup \ms{pe}(\omega)$.
%		Since $\ms{po}(\omega) \subseteq \fn{R}$ and $u \not\in \fn{R}$, then 
%		$u \notin\ms{po}(\omega)$, hence $\alpha = \omega$.
                Since $u \notin \fn{\omega}$, then
		$u \notin \ms{pi}(\omega)$. Also, since $u \notin \fn{R}$ then $u \notin \ms{pe}(\omega)$. Hence $u \notin \fn{R'}$. We can then apply rule \rl{s.res.nil}
		to get $S' \equiv R'$, as required.
		
		\item [Rule \rl{s.$\alpha$}] In this case $S =_\alpha R$. By rule $\rl{l-$\alpha$}$ we get directly
		$R \lstep{\alpha} S'$. Hence we have found $R' = S'$, as required.
		
		\item [Rule \rl{s.ctx}] In this case, we have 
		$S = \sys{\res{\vect{u}}\net}{L \paral N}$ and $R = \sys{\res{\vect{u}}\net}{L \paral M}$
		with $N \equiv M$. We reason by induction on the structure of $\vect{u}$: 
		\begin{itemize}
			\item $\vect{u} = \emptyset$: In this case, $S \lstep{\alpha} S'$ can only have been derived by an application
			of rule \lrl{par}{l}, rule \lrl{par}{r}, rule \lrl{sync}{r}, or rule \lrl{sync}{l}. We consider the
			four cases:
			\begin{description}
				\item [Rule \lrl{par}{l}] In this case, we have 
				$\sys{\net}{L} \lstep{\alpha} \sys{\res{\vect{v}}\net'}{L'}$ with $\vect{v} \cap \fn{N}=\emptyset$,
				and $S' = \sys{\res{\vect{v}}\net'}{L' \paral N}$. Applying rule \lrl{par}{l} we get
				$R \lstep{\alpha} \sys{\res{\vect{v}}\net'}{L' \paral N} = R'$, and by applying rule \rl{S.ctx}
				we have $R' \equiv S'$, as required.
				
				\item [Rule \lrl{par}{r}] In this case we have
				$\sys{\net}{N} \lstep{\alpha} \sys{\res{\vect{v}} \net'}{N'}$. We reason according to the last rule used
				to derive $N \equiv M$:
				\begin{description}
					\item [Rule \rl{s.par.n}] In this case, we have $N = (M \paral \nil)$. Now,
					$\sys{\net}{N} \lstep{\alpha} \sys{\res{\vect{v}} \net'}{N'}$ can only have been obtained 
					via an application of rule \lrl{par}{l}, with 
					$\sys{\net}{M} \lstep{\alpha} \sys{\res{\vect{v}} \net'}{M'}$, and
					$N' = M' \paral \nil$. But then applying \lrl{par}{r} we get:
					$R \lstep{\alpha} \sys{\res{\vect{v}}\net'}{L \paral M'} = R'$.
					Since $N' \equiv M'$ by rule \rl{s.par.n}, we have by rule \rl{s.ctx} $S' \equiv R'$,
					as required.
					
					\item [Rule \rl{s.par.c}] In this case, we have $N = U \paral V$ and $M = V \paral U$.
					Now the derivation $\sys{\net}{N} \lstep{\alpha} \sys{\res{\vect{v}} \net'}{N'}$ can only have been obtained
					by the application of one of the rules \lrl{par}{l}, \lrl{par}{r}, \lrl{sync}{l} or \lrl{sync}{r}.
					We consider the different cases:
					
					\begin{description}
						\item[Rule \lrl{par}{l}] In this case we have 
					$\sys{\net}{U} \lstep{\alpha} \sys{\res{\vect{v}}\net'}{U'}$, $\vect{v} \cap \fn{V} = \emptyset$ and
					$N' = U' \paral V$. Applying rule \lrl{par}{r} we obtain
					$\sys{\net}{M} \lstep{\alpha} \sys{\res{\vect{v}} \net'}{M'} $ with $M' = V \paral U'$.
					Applying rule \lrl{par}{r} we get
					$R = \sys{\net}{L \paral M} \lstep{\alpha} \sys{\res{\vect{v}} \net'}{L \paral (V \paral U')} = R'$.
					Since we have $S' = \sys{\res{\vect{v}} \net'}{L \paral (U' \paral V)}$,
					and by rule \rl{s.par.c} $V \paral U' \equiv U' \paral V$, and since $\equiv$ is a congruence, 
					we have by rule \rl{s.ctx} $S' \equiv R'$, as required.
        \item[Rule \lrl{par}{r}] As above.
        \item[Rule \lrl{sync}{l}] In  this case, we have $\alpha = \tau$, $\vec{v} = \emptyset$,
					$\sys{\net}{U} \lstep{\overline{\omega}} \sys{\net}{U'}$,
					$\sys{\net}{V} \lstep{\omega} \sys{\net}{V'}$,
					where $\overline{\omega}$ is an output action and $\omega$ is the matching input action.
					But then we can apply rule \lrl{sync}{r} to obtain
					$\sys{\net}{M} \lstep{\tau} \sys{\net}{M'}$ where $M' = V' \paral U'$, and then apply
					rule \lrl{par}{r} to get
					$R = \sys{\net}{L \paral (V \paral U)} \lstep{\tau} \sys{\net}{L \paral (V' \paral U')}$
					Since we have 
					$S' = \sys{\net}{L \paral (U' \paral V')}$, $U' \paral V' \equiv V' \paral U'$ by rule \rl{s.par.c},
					and since $\equiv$ is a congruence, we obtain by rule \rl{s.ctx} $S' \equiv R'$, as required.
					
						\item[Rule \lrl{sync}{r}] As above.
					\end{description}
					\item [Rule \rl{s.par.a}] this case is handled similarly to the case of rule \rl{s.par.c} above.
				\end{description}
				
				\item [Rule \lrl{sync}{l}] In this case we have $\alpha = \tau$,
				$\sys{\net}{L} \lstep{\overline{\omega}} \sys{\net}{L'}$,
				$\sys{\net}{N} \lstep{\omega} \sys{\net}{N'}$,
				and $S' = \sys{\net}{L' \paral N'}$.
				 We reason according to the last rule used to derive $N \equiv M$:
				 \begin{description}
				 	\item[Rule \rl{s.par.n}] In this case, we have $N = (M \parop \nil)$.
					The transition $\sys{\net}{N} \lstep{\overline{\omega}} \sys{\net}{N'}$ can only have been obtained
					by an application of rule \lrl{par}{l}, with
					$\sys{\net}{M} \lstep{\omega} \sys{\net}{M'}$ for some $M'$.
					Thus we have $S' = \sys{\net}{L' \paral (M' \paral \nil)}$.
					Now by applying rule \lrl{sync}{r} we get
					$R = \sys{\net}{L \paral M} \lstep{\tau} \sys{\net}{L' \paral M'} = R'$.
					By rule \rl{s.par.n}, we have $M' \paral \nil \equiv M'$. Since $\equiv$ is a congruence we have 
					$L' \paral (M' \paral \nil) \equiv L' \paral M'$, and by rule \rl{s.ctx} we get
					$S' \equiv R'$, as required.
					
					\item[Rule \rl{s.par.c}] we reason exactly as in the subcase \rl{s.par.c} in the case of rule \lrl{par}{l}
					above, except that the only rules to consider are \lrl{par}{l} and \lrl{par}{r}.
					
					\item[Rule \rl{s.par.a}] this case is handled similarly to the case of rule \rl{s.par.c} above.
				 \end{description}
				
				\item[Rule \lrl{sync}{r}] this case is handled similarly to the case of rule \lrl{sync}{l} above.
			\end{description}
			
			\item $\vect{u} = v,\vect{w}$:
			In this case $R = \res{v} U$ with $T \equiv U$ and $S = \res{v} T \lstep{\alpha} S'$. 
			The latter can only have been obtained by rule \lrl{res}{} or by rule \lrl{res}{o}.
			We consider the two cases:
			\begin{description}
				\item[Rule \lrl{res}{}] In this case, we have 
				$T \lstep{\alpha} T'$,
				%$v \not\in \ms{pi}(\alpha)$,
				$S' = \res{v}T'$. By the induction hypothesis, we have
				$U \lstep{\alpha} U'$ for some $U' \equiv T'$. By rule \lrl{res}{} we get
				$R \lstep{\alpha} \res{v} U' = R'$ and since $\equiv$ is a congruence, 
				we have $R' \equiv S'$, as required.
				
				\item[Rule \lrl{res}{o}] In this case, we have
				$T \lstep{\omega} T'$,
				$v \in \ms{po}(\omega)$,
				$\alpha = \res{v}\omega$,
				$S' = T'$. By the induction hypothesis, we have
				$U \lstep{\omega} U'$ for some $U' \equiv T'$. By rule \lrl{res}{o} we get
				$R \lstep{\alpha} U' = R'$. Hence we have $R' \equiv S'$, as required. \qedhere
			\end{description}
		\end{itemize}
	\end{description}
  \end{proof}
\else
\fi

\begin{lem}[Reductions are silent steps]\label{prop:reductionsaretausteps}
  If $S \step S'$ then $S \lstep{\tau} \equiv S'$.
\end{lem}
\ifwithproof
  \begin{proof}
    We proceed by induction on the inference of $S \step S'$.
    \begin{description}
    \item[Case inferred by {\rl{if-eq}}] Then $S$ is
      $\sys{\net}{\proc{\ms{if}~u=u.P~\ms{else}~Q}{n}{\lambda}}$, and the transition $S
      \lstep \tau S'$ follows by \lrl{if-eq}{}.
    \item[Case inferred by {\rl{if-neq}}] Then $S$ is
      $\sys{\net}{\proc{\ms{if}~u=v.P~\ms{else}~Q}{n}{}}$, where $u\neq v$, and the transition $S
      \lstep \tau S'$ follows by \lrl{if-neq}{}.
    \item[Case inferred by {\rl{bang}}] Then $S$ is
      $\sys{\net}{\proc{!x(\vect{u}.P)}{n}{\lambda}}$, and the transition $S
      \lstep \tau S'$ follows by \lrl{bang}{}.
    \item[Case inferred by {\rl{node}}] Then $S$ is
      $\sys{\net}{\proc{\node{m,\kappa}.P}{n}{}}$, and the transition $S
      \lstep \tau S'$ follows by \lrl{node}{}.
    \item[Case inferred by {\rl{msg}}] Then $S$ is
      $\sys{\net}{\proc{\out{x}{y}}{n}{\lambda} \paral \proc{x(z).P}}{n}{\lambda}$, and the transition 
	  $S \lstep \tau S'$ follows by \lrl{out}{}, \lrl{in}{}, and \lrl{sync}{l}.
    \item[Case inferred by {\rl{new}}] Then $S$ is
      $\sys{\net}{\proc{\res x P}{n}{\lambda}}$, and the transition $S
      \lstep \tau S'$ follows by \lrl{new}{}.
    \item[Case inferred by {\rl{spawn-l}}] Then $S$ is
      $\sys{\net}{\proc{\spawn{n}{P}}{n}{\lambda}}$, and the transition $S \lstep \tau S'$
      follows by \lrl{spawn-l}{}.
    \item[Case inferred by {\rl{spawn-c-s}}] Then $S$ is
      $\sys{\net}{\proc{\spawn{m}{P}}{n}{\lambda}}$, and the transition $S \lstep \tau
      S'$ follows by \lrl{spawn-c-s}{}.
    \item[Case inferred by {\rl{spawn-c-f}}] Then $S$ is
      $\sys{\net}{\proc{\spawn{m}{P}}{n}{\lambda}}$, and the transition $S \lstep \tau
      S'$ follows by \lrl{spawn-c-f}{}.
    \item[Case inferred by {\rl{spawn-s}}] Then $S$ is
      $\sys{\net}{\spawnc{(m,\kappa):P}{n}{\lambda} }$, and the transition $S \lstep \tau
      S'$ follows by \lrl{spawn-s}{}.
    \item[Case inferred by {\rl{spawn-f}}] Then $S$ is
      $\sys{\net}{\spawnc{(m,\kappa):P}{n}{\lambda} }$, and the transition $S \lstep \tau
      S'$ follows by \lrl{spawn-f}{}.
    \item[Case inferred by {\rl{unlink}}] Then $S$ is
      $\sys{\net}{\proc{\sbreak{m}{}.P}{n}{\lambda}}$, and the transition $S
      \lstep \tau S'$ follows by \lrl{unlink}{}.
    \item[Case inferred by {\rl{link}}] Then $S$ is
      $\sys{\net}{\proc{\link{m}{}.P}{n}{\lambda}}$, and the transition $S
      \lstep \tau S'$ follows by \lrl{link}{}.
    \item[Case inferred by {\rl{kill}}] Then $S$ is
      $\sys{\net}{\proc{\skill}{n}{\lambda}}$, and the transition $S
      \lstep \tau S'$ follows by \lrl{kill}{}.
    \item[Case inferred by {\rl{forget}}] Then $S$ is
      $\sys{\net}{\proc{\remove{m}.P}{n}{\lambda}}$, and the transition $S
      \lstep \tau S'$ follows by \lrl{forget}{}.
    \item[Case inferred by {\rl{create-s}}] Then $S$ is
      $\sys{\net}{\proc{\create{m}{P}}{n}{\lambda}}$, and the transition $S
      \lstep \tau S'$ follows by \lrl{create-s}{}.
    \item[Case inferred by {\rl{create-f}}] Then $S$ is
      $\sys{\net}{\proc{\create{m}{P}}{n}{\lambda}}$, and the transition $S
      \lstep \tau S'$ follows by \lrl{create-f}{}.
    \item[Case inferred by {\rl{par}}] Then $S$ is
      $\sys{\res{\vect{u}}\net}{N \paral M}$, and the transition $S
      \lstep \tau S'$ follows by \lrl{par}{l} and the inductive hypothesis.
    \item[Case inferred by {\rl{res}}]  Then $S$ is
      $\sys{\res{u}\net}{N}$, and the transition $S
      \lstep \tau S'$ follows by \lrl{res}{} and the inductive hypothesis.
    \item[Case inferred by {\rl{str}}] Then we have $S \equiv T$,
	$T \step T'$ and $T' \equiv S'$. By induction hypothesis, we have
	$T \lstep{\tau} T'$. By Proposition~\ref{prop:str-congr-lts} we have 
	$S \lstep{\tau} S''$ with $S'' \equiv T'$. Hence $S \lstep{\tau} S'' \equiv S'$,
	as required. \qedhere
    \end{description}
  \end{proof}
\else
\fi

\begin{lem}[Silent steps are reductions]\label{prop:taustepsarereductions}
  If $S \lstep \tau S'$ then $S \step S'$.
\end{lem}
\ifwithproof
  \begin{proof} We proceed by induction on the inference of $S \lstep \tau S'$.
    \begin{description}
    \item[Case {\lrl{if-eq}{}}] Then $S$
      is $\sys{\net}{\proc{\ifte{u=u}{P}{Q}}{n}{}}$, and the reduction
        $S\step S'$ follows by \rl{if-eq}.
      \item[Case {\lrl{if-neq}{}}] Then $S$ is
        $\sys{\net}{\proc{\ifte{u=v}{P}{Q}}{n}{}}$, where $u \neq v$, and the
          reduction $S\step S'$ follows by \rl{if-neq}.
        \item[Case {\lrl{bang}{}}] Then $S$ is $\sys{\net}{\proc{!x(\vect{u}).P}{n}{\lambda}}$, and the reduction $S\step S'$ follows by
          \rl{bang}.
        \item[Case {\lrl{node}{}}] Then $S$ is $\sys{\net}{\proc{\node{m,\kappa}.P}{n}{\lambda}}$, and the reduction $S\step S'$ follows by
          \rl{node}.
        \item[Case {\lrl{new}{}}] Then $S$ is $\sys{\net}{\proc{\res x
              P}{n}{\lambda}}$, and the reduction $S\step S'$ follows by \rl{new}.
		\item[Case \lrl{spawn-c-s}{}] Then $S = \sys{\net}{ \proc{\spawn{m}{P}}{n}{\lambda}}$ and
		$S' = \sys{\net}{ \spawnc{(m:\kappa):P}{n}{\lambda}}$ with $\kappa = \net_{\mc{V}}(n)(m)$. The reduction
		$S \step S'$ follows by \textsc{spawn-c-s} (the premises in rule \textsc{spawn-c-s} are the same as in rule \textsc{l-spawn-c-s}).
		\item[Case \lrl{spawn-c-f}{}] Then $S = \sys{\net}{ \proc{\spawn{m}{P}}{n}{\lambda}}$ and
		$S' = \sys{\net \ominus n \views m}{ \nil}$. The reduction
		$S \step S'$ follows by \textsc{spawn-c-f} (the premises in rule \textsc{spawn-c-f} are the same as in rule \textsc{l-spawn-c-f}).
        \item[Case {\lrl{spawn-s}{}}] Then $S$ is
          $\sys{\net}{\spawnc{(m,\kappa):P}{n}{\lambda}}$, 
		  $S' = \sys{\res{\vvect{w}}\net \oplus m \views (n,\lambda)}{\proc{P}{m}{\kappa} }$. 
		  The reduction $S\step S'$
          follows by \textsc{spawn-s} 
		  (the premises in rule \textsc{spawn-s} are the same as in rule \textsc{l-spawn-s}).
        \item[Case {\lrl{spawn-f}{}}]: Then $S$ is
          $\sys{\net}{\proc{\spawn{m}{P}}{n}{\lambda}}$, and 
		  $S' = \sys{\net \ominus n \views m}{\nil}$ the reduction $S\step S'$
          follows by \rl{spawn-f} 
		   (the premises in rule \textsc{spawn-f} are the same as in rule \textsc{l-spawn-f}).
        \item[Case {\lrl{unlink}{}}] Then $S$ is
          $\sys{\net}{\proc{\sbreak{m}{}.P}{n}{\lambda}}$, and the reduction
          $S\step S'$ follows by \rl{unlink}.
        \item[Case {\lrl{link}{}}] Then $S$ is
          $\sys{\net}{\proc{\link{m}{}.P}{n}{\lambda}} $, and the reduction
          $S\step S'$ follows by \rl{link}.
        \item[Case {\lrl{kill}{}}] Then $S$ is
          $\sys{\net}{\proc{\skill}{n}{\lambda}}$, and the reduction $S\step
          S'$ follows by \rl{kill}.
        \item[Case {\lrl{create-s}{}}] Then $S$ is
          $\sys{\net}{\proc{\create{n}{P}}{n}{\lambda}}$, and the reduction $S\step
          S'$ follows by \rl{create-s}.
        \item[Case {\lrl{create-f}{}}] Then $S$ is
          $\sys{\net}{\proc{\create{n}{P}}{n}{\lambda}}$, and the reduction $S\step
          S'$ follows by \rl{create-f}.
        \item[Case {\lrl{sync}{l}}] Then $S$ is
          $\sys{\net}{N_1 \paral N_2}$, and we have 
		  \begin{align*}
		  	& \sys{\net}{N_1} \lstep{\out{x}{\vect{v}}@n_\lambda} \sys{\net}{N_1'}
			&& \sys{\net}{N_2} \lstep{x(\vect{v})@n_\lambda} \sys{\net}{N_2'}
		  \end{align*}
		  Now, by
          Lemma~\ref{lemma:act-sys} we know the following:
		  \begin{align*}
		  	& \sys{\net}{N_1} \equiv \sys{\net}{\proc{\out{x}{v}}{n}{\lambda} \paral M_1}
			&& \sys{\net}{N_2} \equiv \sys{\net}{\proc{x(u).P}{n}{\lambda} \paral M_2}
			&& S' = \sys{\net}{N'_1 \paral N'_2}
		  \end{align*}
		  Hence $S \equiv \sys{\net}{\proc{\out{x}{v}}{n}{\lambda} \paral \proc{x(u).P}{n}{\lambda} \paral M_1 \paral M_2}$.
          Applying \rl{msg},  \rl{par} and \rl{str}, we get 
		  $S \step S'$.
        \item[Case {\lrl{sync}{r}}] Similar to case \lrl{sync}{l}.
        \item[Case {\lrl{par}{l}}] Then $S$ is $\sys{\net}{N_1\paral N_2}$, 
		  $S'$ is $\sys{\res{\vect{v}}\net'}{N_1' \paral N_2}$,
		  where
          $\sys{\net}{N_1} \lstep \tau \sys{\res{\vect{v}}\net'}{N_1'}$, and
		  $\vect{v} \cap \fn{N_2} = \emptyset$. Now, by using
		  the inductive hypothesis and \rl{par} we have
          $\sys{\net}{N_1\paral N_2} \step \sys{\res{\vect{v}}\net'}{N_1'\paral N_2}$.
        \item[Case {\lrl{par}{r}}] Similar to case \lrl{par}{l}.
        \item[Case {\lrl{res}{}}] Then $S$ is $\res{u}T$, $S'$ is $\res{u}T'$, where
          $T \lstep{\tau} T'$. Now by using
          the inductive hypothesis and \rl{res}
          we get $S = \res{u}T \step \res{u}T' = S'$. \qedhere
        \end{description}
      \end{proof}
    \else
    \fi

\begin{restatable}[]{lem}{structureforpars}\label{lemma:structforpars}
	Let $S = \sys{\res{\vect{s}}\net}{N \paral L}$ be a closed system.
	If $S \lstep{\alpha} S'$, where $\alpha$ is a silent action, an output action or an input action,
	then one of the following assertions holds:
	\begin{enumerate}
		\item \label{forparsone} $\alpha = \tau$,
		$ \sys{\res{\vect{s}}\net}{N} \lstep{\tau}  \sys{\res{\vect{w}}\res{\vect{s}}\net}{N'}$,
		and $ S' \equiv \sys{\res{\vect{w}}\res{\vect{s}}\net}{N' \paral L}$, 
		with $\fn{L} \cap \vect{w} = \emptyset$, and the derivation of $S \lstep{\alpha} S'$ terminates with an application of rule \lrl{par}{l}, possibly followed by applications of rule \lrl{res}{}.
		
		\item \label{forparstwo} $\alpha = \tau$,
		$ \sys{\res{\vect{s}}\net}{L} \lstep{\tau}  \sys{\res{\vect{w}}\res{\vect{s}}\net}{L'}$,
		and $ S' \equiv \sys{\res{\vect{w}}\res{\vect{s}}\net}{N \paral L'}$,
		with $\fn{N} \cap \vect{w} = \emptyset$, and the derivation of $S \lstep{\alpha} S'$ terminates with an application of rule \lrl{par}{r}, possibly followed by applications of rule \lrl{res}{}.
		
		\item \label{forparsthree} $\alpha = \tau$,
		$ \sys{\res{\vect{s}}\net}{N} \lstep{\res{\vvect{w}}\out{x}{\vect{u}}@n_\lambda}  \sys{\res{\vvect{r}}\net}{N'}$,
		$\vvect{r} = \vect{s}\setminus\vvect{w}$ and $\vect{u} \cap \vvect{r} = \emptyset$,
		$ \sys{\net}{L} \lstep{x(\vect{u})@n_\lambda}  \sys{\net}{L'}$,
		  and $ S' \equiv \sys{\res{\vect{s}}\net}{N' \paral L'}$,
                  and the derivation of $S \lstep{\alpha} S'$ terminates with an application of rule \lrl{sync}{l}, possibly followed by applications of rule \lrl{res}{}.
		\item \label{forparsfour} $\alpha = \tau$,
		$ \sys{\net}{N} \lstep{x(\vect{u})@n_\lambda}  \sys{\net}{N'}$,
		$ \sys{\res{\vect{s}}\net}{L} \lstep{\res{\vvect{w}}\out{x}{\vect{u}}@n_\lambda}  \sys{\res{\vect{r}}\net}{L'}$,
		$\vvect{r} = \vect{s}\setminus\vvect{w}$ and $\vect{u} \cap \vvect{r} = \emptyset$,
		  and $ S' \equiv \sys{\res{\vect{w}}\res{\vect{s}}\net}{N' \paral L'}$,
                  and the derivation of $S \lstep{\alpha} S'$ terminates with an application of rule \lrl{sync}{r}, possibly followed by applications of rule \lrl{res}{}.
		\item \label{forparsfive} $\alpha = \res{\vect{w}}\out{x}{\vect{u}}@n_\lambda$,
		$ \sys{\res{\vect{s}}\net}{N} \lstep{\res{\vect{w}}\out{x}{\vect{u}}@n_\lambda}  \sys{\res{\vect{r}}\net}{N'}$,
		$\vect{r} = \vect{s} \setminus \vect{w}$,
		$\vect{u} \cap \vect{r} = \emptyset$,
		  and $ S' \equiv \sys{\res{\vect{r}}\net}{N' \paral L}$,
                  and the derivation of $S \lstep{\alpha} S'$ terminates with an application of rule \lrl{par}{l}, possibly followed by applications of rule \lrl{res}{} or \lrl{res}{o}. 
		
		\item \label{forparssix} $\alpha = \res{\vect{w}}\out{x}{\vect{u}}@n_\lambda$,
		$ \sys{\res{\vect{s}}\net}{L} \lstep{\res{\vect{w}}\out{x}{\vect{u}}@n_\lambda}  \sys{\res{\vect{r}}\net}{L'}$,
		$\vect{r} = \vect{s} \setminus \vect{w}$,
		$\vect{u} \cap \vect{r} = \emptyset$,
		  and $ S' \equiv \sys{\res{\vect{r}}\net}{N \paral L'}$,
                   and the derivation of $S \lstep{\alpha} S'$ terminates with an application of rule \lrl{par}{r}, possibly followed by applications of rule \lrl{res}{} or \lrl{res}{o}. 
		
		\item \label{forparsseven} $\alpha = x(\vect{u})@n_\lambda$,
		$ \sys{\res{\vect{s}}\net}{N} \lstep{x(\vect{u})@n_\lambda}  \sys{\res{\vect{s}}\net}{N'}$,
		$ S' \equiv \sys{\res{\vect{s}}\net}{N' \paral L}$,
		  and $\vect{u} \cap \vect{s} = \emptyset$,
                   and the derivation of $S \lstep{\alpha} S'$ terminates with an application of rule \lrl{par}{l}, possibly followed by applications of rule \lrl{res}{}. 
		
		\item \label{forparseight} $\alpha = x(\vect{u})@n_\lambda$,
		$ \sys{\res{\vect{s}}\net}{L} \lstep{x(\vect{u})@n_\lambda}  \sys{\res{\vect{s}}\net}{L'}$,
		$ S' \equiv \sys{\res{\vect{s}}\net}{N \paral L'}$,
		  and $\vect{u} \cap \vect{s} = \emptyset$,
                  and the derivation of $S \lstep{\alpha} S'$ terminates with an application of rule \lrl{par}{r}, possibly followed by applications of rule \lrl{res}{}.  
	\end{enumerate}
	
\end{restatable}

\begin{proof}
	By case analysis on $\alpha$.
	\begin{description}
		\item [Case $\alpha = \tau$] in this case we reason by induction on the derivation of $S \lstep{\tau} S'$,
		considering the last rule used in the proof tree:
		\begin{description}
			\item [Case \lrl{par}{l}] In this case, we have 
			$S = \sys{\net}{N \paral L}$, 
			$\sys{\net}{N} \lstep{\tau} \sys{\res{\vect{w}}\net'}{N'}$,
			and $S' = \sys{\net}{N} \lstep{\tau} \sys{\res{\vect{w}}\net'}{N' \paral L}$, with $\fn{L} \cap \vect{w} = \emptyset$,
			corresponding to assertion~\ref{forparsone}.
			\item [Case \lrl{par}{r}] In this case, we have 
			$S \equiv \sys{\net}{N \paral L}$, 
			$\sys{\net}{L} \lstep{\tau} \sys{\res{\vect{w}}\net'}{L'}$,
			and $S' = \sys{\net}{N} \lstep{\tau} \sys{\res{\vect{w}}\net'}{N \paral L'}$, with $\fn{N} \cap \vect{w} = \emptyset$,
			corresponding to assertion~\ref{forparstwo}.
			
			\item [Case \lrl{sync}{l}] In this case, we have 
			$S = \sys{\net}{N \paral L}$,
			$\sys{\net}{N} \lstep{\out{x}{\vect{u}}@n_\lambda} \sys{\net}{N'}$,
			$\sys{\net}{L} \lstep{x(\vect{u})@n_\lambda} \sys{\net}{L'}$,
			and $S' = \sys{\net}{N' \paral L'}$, corresponding to assertion~\ref{forparsthree}.
			
			\item [Case \lrl{sync}{r}] In this case, we have 
			$S = \sys{\net}{N \paral L}$,
			$\sys{\net}{N} \lstep{x(\vect{u})@n_\lambda} \sys{\net}{N'}$,
			$\sys{\net}{L} \lstep{\out{x}{\vect{u}}@n_\lambda} \sys{\net}{L'}$,
			and $S' = \sys{\net}{N' \paral L'}$, corresponding to assertion~\ref{forparsfour}.
			
			\item [Case \lrl{res}{}] In this case, we have 
			$S = \res{a} T$,
			$S' = \res{a} T'$,
			$T \lstep{\tau} T'$.
			By induction hypothesis, we have for $T$ one of the four cases~\ref{forparsone} to~\ref{forparsfour}
			in the lemma.
			We consider only the cases \lrl{par}{l} and \lrl{sync}{l}, the other ones are handled similarly.
			\begin{description}
				\item [Rule \lrl{par}{l}] In this case we have 
				$T = \sys{\res{\vect{s}}\net}{N \paral L}$,
				$ T' \equiv \sys{\res{\vvect{w}}\res{\vect{s}}\net}{N' \paral L}$,
				$ \sys{\res{\vect{s}} \net}{N} \lstep{\tau} \sys{\res{\vvect{w}}\res{\vect{s}}\net}{N'}$,
				with $ \fn{L} \cap \vvect{w} = \emptyset$.
				Applying rule \lrl{res}{}, we get
				$S' \equiv \sys{\res{a} \res{\vect{w}}\res{\vect{s}}\net}{N' \paral L}$,
                                %\equiv 
				%      \sys{\res{a}\res{\vect{r}}\res{\vect{s}} \net}{N' \paral L}$, 
				$ \sys{\res{a}\res{\vect{s}} \net}{N} \lstep{\tau} \sys{\res{\vvect{w}}\res{a}\res{\vect{s}}\net}{N'}$,
				with $ \fn{L} \cap \vvect{w} = \emptyset$,
				corresponding to assertion~\ref{forparsone}.
					 
				\item [Rule \lrl{sync}{l}] In this case we have
				$ T  = \sys{\res{\vect{s}}\net}{N \paral L}$,
				$ T' \equiv \sys{ \res{\vect{s}}\net }{N' \paral L'}$,
				$ \sys{\res{\vect{s}}\net}{N} \lstep{\res{\vvect{w}} \out{x}{\vect{u}}@n_\lambda} \sys{\res{\vect{r}}\net}{N'}$,
				$\sys{\net}{L} \lstep{x(\vect{u})@n_\lambda} \sys{\net}{L'}$,
				with $\vect{r} = \vect{s}\setminus\vvect{w}$ and $\vect{u} \cap \vect{r} = \emptyset$.
				Applying rule \lrl{res}{}, we get
				$S = \sys{\res{w}\res{\vect{s}}\net}{N \paral L} \lstep{\tau} \sys{\res{a}\res{\vect{s}}\net}{N' \paral L'} = S'$.
				If $a \in \vect{u}$, then applying \lrl{res}{o}, we get
				$ \sys{\res{a}\res{\vect{s}}\net}{N} \lstep{\res{a,\vvect{w}} \out{x}{\vect{u}}@n_\lambda} \sys{\res{\vect{r}}\net}{N'}$,
				with $\vect{u} \cap \vect{r} = \emptyset$ and $\vect{r} = a,\vect{s} \setminus a,\vvect{w}$,
				as required.
				If $a \notin \vect{u}$, then  applying \lrl{res}{} we get
				$ \sys{\res{a}\res{\vect{s}}\net}{N} \lstep{\res{\vvect{w}} \out{x}{\vect{u}}@n_\lambda} \sys{\res{a}\res{\vect{r}}\net}{N'}$,
				with $\vect{u} \cap a,\vect{r} = \emptyset$ and $a,\vect{r} = a,\vect{s} \setminus \vvect{w}$,
				corresponding to assertion~\ref{forparsthree}.
			\end{description}			
		\end{description}
		
		\item [Case $\alpha = \res{\vvect{w}}\out{x}{\vect{u}}@n_\lambda$] handled by induction on the derivation of $S \lstep{\alpha} S'$
		with cases similar to the cases $\tau$.\lrl{par}{l}, $\tau$.\lrl{par}{r} and $\tau$.\lrl{res}{} above.
		
		\item [Case $\alpha = x(\vect{u})@n_\lambda$] handled by induction on the derivation of $S \lstep{\alpha} S'$
		with cases similar to the cases $\tau$.\lrl{par}{l}, $\tau$.\lrl{par}{r} and $\tau$.\lrl{res}{} above. \qedhere

	\end{description}
\end{proof}

\begin{restatable}[Bisimilarity is a System Congruence]{prop}{bisisacong}\label{prop:bisimacong}
      Weak bisimilarity is a weak system congruence; that is, 
	  if $\sys{\res{\vect u}\net}{N} \wbsim \sys{\res{\vect v}\net'}{M}$
      then for all $\vvect{w}$, $L$, with $\fn{L} \cap (\vect{u},\vect{v}) = \emptyset$,
	  we have $\sys{\res{\vvect{w}}\res{\vect u}\net}{N \paral L} \wbsim
        \sys{\res{\vvect{w}}\res{\vect v}\net'}{M \paral L} $
\end{restatable}
    \ifwithproof
      \begin{proof}
		We prove that the relation
		\begin{align*}
			\rel{S} \defeq \left\{
			\langle \sys{\res{\vvect{w}}\res{ \vect s}\net_S}{N_S \paral L},~\sys{\res{\vvect{w}}\res{\vect r}\net_R}{N_R \paral L} \rangle
			~\mid~
      \begin{array}{c}
        \sys{\res{\vect{s}}\net_S}{N_S}  
        \wbsim \sys{\res{\vect{r}}\net_R}{N_R} \\
        \fn{L} \cap (\vect{s} \cup \vect{r}) = \emptyset
      \end{array}
      \right\}
		\end{align*}
        is a weak bisimulation up to $\equiv$. Since $\rel{S}$ is symmetric, it suffices to prove that $\rel{S}$ is a weak simulation
		up to $\equiv$.
		Define
		\begin{align*}
			& S = \sys{\res{\vect{s}}\net_S}{N_S} & 
			& R = \sys{\res{\vect{r}}\net_R}{N_R} \\
			&  S_L = \sys{\res{\vect{w}}\res{ \vect s}\net_S}{N_S \paral L} & 
			& R_L = \sys{\res{\vect{w}}\res{ \vect r}\net_R}{N_R \paral L}
		\end{align*}
		and consider a transition $S_L \lstep{\alpha} U$.
		We proceed by induction on the structure of $\vvect{w}$:
		\begin{description}
			\item [ Case $\vvect{w} = \emptyset$] We proceed by case analysis on $\alpha$:
	          \begin{description}
	          \item[Case $\tau$] We consider the different cases listed in Lemma~\ref{lemma:structforpars}
			  for $S_L = \sys{\res{\vect{s}}\net_S}{N_S \paral L}$ and $S_L \lstep{\tau} U$:
	            \begin{description}
	              \item[Case \lrl{sync}{l}]
				  In this case, we have:
				  \begin{align*}
				  	& S = \sys{\res{\vect{s}}\net_S}{N_S} 
					     \lstep{\res{\vect{a}}\out{x}{\vect{u}}@n_\lambda} 
						 \sys{\res{\vvect{z_s}}\net_S}{N'_S} = S' \\
					& \sys{\net_S}{L} \lstep{x(\vect{u})@n_\lambda} \sys{\net_S}{L'} \\
					& \vvect{z_s} = \vect{s} \setminus \vect{a} \quad \vect{u} \cap \vect{z_s} = \emptyset \\
					& U = \sys{\res{\vect{s}}\net_S}{N'_S \paral L'}
				  \end{align*}
	  			  Since $S \wbsim R$, we have 
				  $R \lwstep{\res{\vect{a}}\out{x}{\vect{u}}@n_\lambda} \sys{\res{\vvect{w_r}}\res{\vvect{z_r}}\net'_R}{N'_R} = R'$ 
				  for some $\vect{w_r}, \net'_R, N'_R$, with
	  			   $\vvect{z_r} = \vect{r} \setminus \vect{a}$, and $R' \wbsim S'$. 
				  Now, since $R \lwstep{\tau} R_1 \lstep{\out{x}{\vect{u}}@n_\lambda} R_2 \lwstep{\tau} R'$, where
	  			  $R_1 = \sys{ \res{\vvect{w_1}} \res{\vvect{z_r}} \net^{1}_R }{N^{1}_R}$
				  for some $\net^{1}_R, N^{1}_R$, 
				  we are guaranteed that $n_\lambda$ is alive, and that
	  			  $\sys{\net^1_R}{L} \lstep{x(\vect{u})@n_\lambda} \sys{\net^1_R}{L'}$.
	  			  Also, we have $\vect{u} \cap \vvect{z_r} = \emptyset$.
	  			  By repeated applications of rule \lrl{par}{l} and by one application
	  			  of rule \lrl{sync}{l}, we obtain
	  			  $R_L \lwstep{\tau} V$, with $V = \sys{\res{\vvect{w_r}}\res{\vect{r}}\net'_R}{N'_R \paral L'}$ 
	  			  and with $\fn{L'} \cap \vvect{w_r} = \emptyset$ because of the conditions of rule \lrl{par}{l}.
				  
				  Summing up, we have:
				  \begin{align*}
				  	& U = \sys{\res{\vect{a}}\res{\vvect{z_s}}\net_S}{N'_S \paral L'} = \res{\vect{a}} S' \\
					& V \equiv  \sys{\res{\vect{a}}\res{\vvect{w_r}}\res{\vvect{z_r}}\net'_R}{N'_R \paral L'} = \res{\vect{a}} R' \\
					& S' \wbsim R' \\
					& \fn{L'} \subseteq \fn{L} \cup \vect{u}\setminus \vect{a} \quad \text{by Lemma~\ref{lemma:fnames}}\\
					& \fn{L'} \cap \vvect{w_r} \\
					& \vect{u} \cap \vvect{z_s} = \vect{u} \cap \vvect{z_r} = \emptyset
				  \end{align*}
				  Hence we have $\fn{L'} \cap \vvect{z_s} = \emptyset$ and $\fn{L'} \cap (\vvect{w_r} \cup \vvect{z_r}) = \emptyset$,
				  and $S' = \sys{\res{\vvect{z_s}}\net_S}{N'_S} \wbsim \sys{\res{\vect{w_r}}\res{\vvect{z_r}}\net'_R}{N'_R} = R'$,
				  which means that $\langle U, V \rangle \in \rel{S}$, as required.
			
	              \item[Case \lrl{sync}{r}] Similar to the case \lrl{sync}{l}, but simpler.
			
	              \item[Case \lrl{par}{l}] In this case, we have $S_L \lstep{\tau} U$, with
	  		      $S_L = \sys{\net_S}{N_S \paral L}$,
	  		      $U = \sys{\res{\vect{s}}\net'_S}{N'_S \paral L}$,
	  		      $S \lstep{\tau} \sys{\res{\vect{s}}\net'_S}{N'_S}$,
	  		      with $\fn{L} \cap \vect{s} = \emptyset$.
	  		      Since $S \wbsim R$, we have 
	  		      $R \lwstep{\tau} R'$ with $R' \wbsim S'$ and $R' = \sys{\res{\vect{r}}\net'_R}{N'_R}$.
	  		      Now applying repeatedly rule \lrl{par}{l}, we get 
	  		      $R_L \lwstep{\tau} \sys{\res{\vect{r}}\net'_R}{N'_R \paral L} = V$, with 
	  		      $\fn{L} \cap \vect{r} = \emptyset$ because of the conditions of rules \lrl{par}{l}.
	  		      Now we have $\langle U, V \rangle \in \rel{S}$, as required.
		
	  		      \item [Case \lrl{par}{r}] Similar to the case \lrl{par}{l}, but simpler.
		
	           \end{description}
		
	        \item[Case $x(\vect{u})@n_\lambda$] We consider the different cases  listed in Lemma~\ref{lemma:structforpars}
			for $S_L =  \sys{\res{\vect{s}} \net_S}{N_S \paral L}$ and $S_L \lstep{x(\vect{u})@n_\lambda} U$:
	          \begin{description}
	          	\item[Case \lrl{par}{l}] In that case, we have
	  			$S = \sys{\res{\vect{s}}\net_S}{N_S}$,
	  			$S \lstep{x(\vect{u})@n_\lambda} S'$,
	  			$S' = \sys{\res{\vect{s}}\net_S}{N'_S}$.
	  			Since $S \wbsim R$, we have 
	  			$R \lwstep{x(\vect{u})@n_\lambda} R'$, where 
	  			$R' = \sys{\res{\vect{z}}\res{\vect{r}} \net'_R}{N'_R}$.
	  			Now, by repeated application of rule \lrl{par}{l}, we get
	  			$R_L \lwstep{x(\vect{u})@n_\lambda} \sys{\res{\vect{w}}\res{\vect{r}} \net'_R}{N'_R} \paral L = V$,
	  			with $\fn{L} \cap \vect{z} = \emptyset$ because of the conditions of rule \lrl{par}{l}.
	  			Since we also have $\fn{L} \cap \vect{r} = \emptyset$ by definition, 
	  			we have $\langle U, V \rangle \in \rel{S}$, as required.

	            \item[Case \lrl{par}{r}] Similar to the case \lrl{par}{l}, but simpler.
	          \end{description}
		
	        \item[Case $\res{\vect{a}}\out{x}{\vect{u}}@n_\lambda$] We consider the different cases listed in  Lemma~\ref{lemma:structforpars}
			for $S_L =  \sys{\res{\vect{s}} \net_S}{N_S \paral L}$ and $S_L \lstep{\res{\vect{a}}\out{x}{\vect{u}}@n_\lambda} U$ :
	          \begin{description}
	          	\item[Case \lrl{par}{l}] In that case, we have
	  			$S = \sys{\res{\vect{s}}\net_S}{N_S}$,
	  			$S \lstep{\res{\vect{a}}\out{x}{\vect{u}}@n_\lambda} S'$
	  			$S' = \sys{\res{\vvect{s_a}}\net_S}{N'_S}$,
				$U = \sys{\res{\vect{s_a}}\net_S}{N'_S \paral L}$,
				$\vvect{s_a} = \vect{s} \setminus\vect{a}$.
	  			Since $S \wbsim R$, we have 
	  			$R \lwstep{\res{\vect{a}}\out{x}{\vect{u}}@n_\lambda} R'$, where 
	  			$R' \equiv \sys{\res{\vect{w}}\res{\vvect{r_a}} \net'_R}{N'_R}$,
				$\vvect{r_a} = \vect{r} \setminus\vect{a}$,
				and $S' \wbsim R'$.
	  			Now, by repeated application of rule \lrl{par}{l}, we get
	  			$R_L \lwstep{\res{\vect{a}}\out{x}{\vect{u}}@n_\lambda} \sys{\res{\vect{z}}\res{\vvect{r_a}} \net'_R}{N'_R} \paral L = V$,
	  			with $\fn{L} \cap \vect{z} = \emptyset$ because of the conditions of rule \lrl{par}{l}.
	  			Since we also have $\fn{L} \cap \vect{r} = \emptyset$ by definition, 
	  			we have $\langle U, V \rangle \in \rel{S}$, as required.

	            \item[Case \lrl{par}{r}] Similar to the case \lrl{par}{l}, but simpler.
              \end{description}
	        \item[Case $\ms{create}(n,\lambda)$] In this case, we have:
            \begin{align*}
              & S_L = \sys{\res{\vect{s}}\net_S}{N_S \paral L} \lstep{\ms{create}(n,\lambda)} \sys{\res{\vect{s}}\net'_S}{N_S \paral L} = U\\
              & S = \sys{\res{\vect{s}}\net_S}{N_S} \lstep{\ms{create}(n,\lambda)} \sys{\res{\vect{s}}\net'_S}{N_S} = S'
            \end{align*}
            Since $S\wbsim R$, we have $R \lwstep{\ms{create}(n,\lambda)} R' = \sys{\res{\vect{r}}\net'_R}{N'_R}$,
            with $S' \wbsim R'$. Now, by one application of rule \lrl{create-ext}, we
            obtain $R_L \lwstep{\ms{create}(n,\lambda)}
            \sys{\res{\vect{r}}\net'_R}{N'_R \paral L} = V$. Since we have $\fn{L}\cap
            \vect{r} = \emptyset$ by definition we have $\langle U,V \rangle\in \rel{S}$.
	        \item[Case $\ms{kill}(n,\lambda)$] Similar to above.
	        \item[Case $\oplus n \mapsto m$] Similar to above.
	        \item[Case $\ominus n \mapsto m$] Similar to above.
	        \item[Case $n \views m$] Similar to above.
	        \end{description}
		\item [Case $\vvect{w} = w,\vvect{z}$] 
		Let $\langle \res{w}\res{\vvect{z}} S, \res{w}\res{\vvect{z}} R \rangle \in \rel{S}$. 
		By construction $\langle \res{\vvect{z}} S, \res{\vvect{z}} R \rangle \in \rel{S}$.
		By induction hypothesis, if $\res{\vvect{z}} S \lstep{\alpha} U$, then $\res{\vvect{z}} R \lstep{\alpha} V$ for some
		$V$ with $\langle U, V \rangle \in \rel{S}$. Now using rule \lrl{res}{} or rule \lrl{res}{o}, we obtain
		either $S_L \lstep{\alpha} \res{w} U$ or $S_L \lstep{\res{w}\alpha} U$, which are matched respectively by
		$R_L \lstep{\alpha} \res{w} V$ or $R_L \lstep{\res{w}\alpha} V$. In both cases, we have 
		$\langle U, V \rangle \in \rel{S}$ and $\langle \res{w} U, \res{w} V \rangle \in \rel{S}$ 
		($\rel{S}$ is closed under restriction by construction), as required. \qedhere

		\end{description}
    \end{proof}
  \else
  \fi

\begin{restatable}[Soundness of weak bisimilarity]{prop}{bisimsound}\label{prop:bisimsound}
	Weak bisimilarity is sound with respect to weak barbed congruence, i.e.\ $\wbsim ~\subseteq~ \bcong$.
\end{restatable}
\begin{proof}
	Weak bisimilarity is weak barb-preserving thanks to Lemma~\ref{lemma:act-sys}.
	It is weak reduction-closed thanks to Lemma~\ref{prop:taustepsarereductions}.
	It is a system congruence thanks to Proposition~\ref{prop:bisimacong}.
	Thus $\wbsim ~\subseteq~ \bcong$ by definition of weak barbed congruence as the largest of weak barb-preserving,
	reduction-closed, system congruence.
\end{proof}

\subsection{Completeness}

In the remainder of this section, we use the following notation: if $S = \sys{\res{\vect{s}} \net}{N}$ is a closed system,
we write $S \paral M$ for the system $\sys{\res{\vect{s}} \net}{N \paral M}$ provided $\fn{M} \cap \vect{s} = \emptyset$.

\begin{restatable}[Inducing Network Changes]{lem}{netupdates}\label{lemma:netupdates}~
      \begin{itemize}
      \item Suppose $S=\sys{\res{\vect{u}}\net}{N}$ and $\net \vdash (n,\lambda):\ms{alive}$
        \begin{itemize}
        \item $S \lstep{\ms{kill}(n,\lambda)} S'$ implies $S \paral
          \proc{\skill}{n}{\lambda} \step S'$
        \item $S \paral \proc{\skill}{n}{\lambda} \step S'$, where $S'\equiv
          \sys{\res{\vect u}\net}{N}$, $\net \not\vdash (n,\lambda):
          \ms{alive}$ implies $S
          \lstep{\ms{kill}(n,\lambda)} S'$
        \end{itemize}
      \item Suppose $S=\sys{\res{\vect{u}}\net}{N}$ and $\net \vdash (n,\lambda):\ms{dead}$
        \begin{itemize}
        \item $S \lstep{\ms{create}(n,\lambda)} S'$ implies $S \paral
          \proc{\create{n}{P}}{m}{\kappa} \step S' \paral \proc{P}{n}{\lambda}$
        \item $S \paral \proc{\create{n}{P}}{m}{\kappa} \step S' \paral \proc{P}{n}{\lambda}$, where $S'\equiv
          \sys{\res{\vect u}\net}{N}$, $\net \vdash n_\lambda :
          \ms{alive}$ implies $S
          \lstep{\ms{create}(n,\lambda)} S'$
        \end{itemize}
      \item Suppose $S=\sys{\res{\vect{u}}\net}{N}$ and $\net \vdash n\connection m$
        \begin{itemize}
        \item $S \lstep{ \ominus n_\lambda \mapsto m}
          S'$, where $S' \equiv \sys{\res{\vect{u}}\net'}{N}$ implies $S \paral
          \proc{\sbreak{m}.P}{n}{\lambda} \step S' \paral \proc{P}{n}{\lambda}$
        \item $S \paral
          \proc{\sbreak{m}.P}{n}{\lambda} \step S'\paral \proc{P}{n}{}$, where $S'\equiv \sys{\res{\vect{u}}\net'}{N}$, $\net' \vdash n \not\connection m$
          implies $S \lstep{\ominus n_\lambda \mapsto m}
          S'$
        \end{itemize}
      \item Suppose $S=\sys{\res{\vect{u}}\net}{N}$ and $\net \vdash n\not\connection m$
        \begin{itemize}
        \item $S \lstep{ \oplus n_\lambda \mapsto m}
          S'$, where $S' \equiv \sys{\res{\vect{u}}\net'}{N}$ implies $S \paral
          \proc{\link{m}.P}{n}{\lambda} \step S' \paral \proc{P}{n}{\lambda}$
        \item $S \paral
          \proc{\link{m}.P}{n}{\lambda} \step S'\paral \proc{P}{n}{\lambda}$, where $S'\equiv \sys{\res{\vect{u}}\net'}{N}$, $\net' \vdash n \connection m$
          implies $S \lstep{\oplus n_\lambda \mapsto m}
          S'$
        \end{itemize}
      \end{itemize}
\end{restatable}
    \begin{proof}
      \ifwithproof
        The first clause for the action $\ms{kill}(n,\lambda)$ is proved by induction on the
        derivation of $S = \sys{\res{\vect{u}}\net}{N} \lstep{\ms{kill}(n,\lambda)}
        \sys{\res{\vect{u}}\net'}{N} = S'$. The second clause uses induction on the
        derivation of $\sys{\res{\vect{u}}\net}{N \paral \proc{\skill}{n}{\lambda}} \step
        \sys{\res{\vect{u}}\net'}{N'}$. The proof for the other clauses is similar.
      \else \fi
    \end{proof}

    Given a system $S\equiv \sys{\res{\vect{u}}\net}{N}$ and $l$ s.t.
    $l\notin fn(S)\cup bn(S)$ we define $S^l$ as $\sys{\res{\vect{u}}\net \oplus
      (l,1)}{N}$.

\begin{restatable}[]{lem}{}\label{lemma:nil-proc}
      If $S$ is a closed system and  $l,x \notin \fn{S}$ then 
	  $\res{x,l}S^l\paral \proc{\eout{x}}{l}{\lambda} \bsim S $.
\end{restatable}
\begin{proof}
	Easy, since
	$\rel{R} = \{ \langle{\res{x,l}S^l\paral \proc{\eout{x}}{l}{\lambda}, S \rangle \mid S\in \cset{S}~\text{closed}}, ~x,l \notin \fn{S} \}$ 
	is a strong bisimulation, noting that any transition $\res{x,l}S^l\paral \proc{\eout{x}}{l}{\lambda} \lstep{\alpha} T^l$
	must have been obtained by a derivation with $S^l \lstep{\alpha} U^l$ as a premise,
	with $T^l = \res{x,l}U^l$, $x,n\notin\fn{\alpha}$,
	and that for any such transition we have $S \lstep{\alpha} T$. \qedhere

\end{proof}

\begin{restatable}[]{lem}{}\label{lemma:congruence-inverted}
      If we have $S^l\paral \proc{\eout{x}}{l}{\lambda} \bcong R^l \paral
      \proc{\eout{x}}{l}{\lambda}$ with $x,l$ fresh for $S,R$ then $S
      \bcong R$.
\end{restatable}
    \begin{proof}
      If $S^l\paral \proc{\eout{x}}{l}{\lambda} \bcong R^l \paral
      \proc{\eout{x}}{l}{\lambda} $ then by system congruence we also have
      \[
        \res{x,l}S^l\paral \proc{\eout{x}}{l}{\lambda} \bcong \res{x,l}R^l \paral
        \proc{\eout{x}}{l}{\lambda}
      \]
      and then by Lemma~\ref{lemma:nil-proc}, Proposition~\ref{prop:bisimsound}, and
      transitivity of $\bcong$ we have $S\bcong R$ as required.
    \end{proof}

\begin{prop}[Completeness of $\wbsim$ w.r.t. $\bcong$]\label{prop:completeness}
      $\bcong ~ \subseteq ~ \wbsim$.
\end{prop}
    \begin{proof}
      To prove the statement, since $\bcong$ is an equivalence relation and hence symmetric, it suffices to show that $\bcong$ is a
      weak simulation up-to $\equiv$.
      Take $S \bcong R$ and suppose that $S\lstep{\alpha} S'$; we reason by case
      analysis on $\alpha$.
      \begin{description}
      \item[Case $\alpha = \tau$] thanks to Lemma~\ref{prop:reductionsaretausteps} and Lemma~\ref{prop:taustepsarereductions},
        the thesis follow by the reduction closure property.
      \item[Case $\alpha = \out{x}{\vect{u}}@n_\lambda$] consider the
        context
        \[
          L = \proc{x(\vect{y}).\ift{\vect{y} = \vect{u}}{\create{l}{(\eout{fail} \parop
                  fail.\eout{succ}})}}{n}{\lambda}
          \]
          with $l$, $succ$ and $fail$ fresh and the reduction $S\paral L \wstep
          T_1$, where $T_1 \equiv S'^l \paral \proc{\eout{succ}}{l}{}$. Now,
          since $S\bcong R$, we must
          have a transition $R \paral L \wstep T_2$ with $T_1 \bcong T_2$. Then,
          since $T_1\barb{succ@l}$, $T_1\nbarb{fail@l}$ and $T_1 \bcong T_2$ we must have
          $T_2\barb{succ@l}$ and $T_2\nbarb{fail@l}$. We remark here that we
          have $T_2\barb{succ@l}$, with a strong barb, because the only way for
          $R \paral L$ to make disappear $\barb{fail@l}$ is to have consumed it
          on the fresh location $l$, thereby showing $\barb{succ@l}$.

          The only way to obtain this derivation is if $R
          \lwstep{\out{x}{\vect{u}}@n_\lambda} R'$, hence $T_2\equiv R'^l \paral
          \proc{\eout{succ}}{l}{}$.
          Now, since $(S'^l \paral \proc{\eout{succ}}{l}{}, R'^l \paral
          \proc{\eout{succ}}{l}{}) \in \bcong$, by
          Lemma~\ref{lemma:congruence-inverted} we have $(S',R')\in
          \bcong$ as required.
        \item[Case $\alpha = \res{\vect{v}}\out{x}{\vect{u}\cdot
            \vect{v}}@n_\lambda$]

          For this case we define the following macros
          \[
            \ifte{x \not\in \vect{v}}{P}{Q} \equiv
              \ifte{x = \vect{v}_1}{Q}{\ms{if} \ldots \ms{else}~\ifte{x = \vect{v}_n}{P}{Q}}
          \]

          \[
            \ifte{\vect{u}\cap \vect{v} = \emptyset}{P}{Q} \equiv
              \ifte{\vect{u}_1 \notin \vect{v}}{Q}{\ms{if} \ldots
                  \ifte{\vect{u}_n \notin \vect{v}}{P}{Q} \ldots\ms{else}~{Q}}
          \]

          where by $\vect{v}_i$ we mean the $i$-th element of $\vect{v}$.

          In addition, for simplicity, here we assume that the private names in
          $\vect{u}$ are in the final part of the vector, i.e., $\vect{u}=
          \vect{u}'\cdot \vect{v}$. This implies a simple nesting of the
          \tms{if} primitives to test the privateness of names. In the other
          cases, when private names are freely mixed with public ones, a testing
          context can always be found by nesting the \tms{if}s following the
          order of the names. Consider the context
          \[
            L = \proc{x(\vect{y}\cdot\vect{y}').\ift{\vect{y} =
                  \vect{u}}.\ift{\vect{y}'\cap \vect{w} = \emptyset}{{\create{l}{(\eout{fail} \parop
                    fail.\eout{succ}})}}}{n}{\lambda}
            \]
            with $l$, $fail$, $succ$ fresh and $\vect{w} = fn(S,R)$.
            The reasoning then proceeds mostly as in the previous case.
			
        \item[Case $\alpha = x(\vect{u})@n_\lambda$] consider the context
          \[
            L =  \proc{\out{x}{\vect{u}}.\create{l}{(\eout{fail}\parop fail.\eout{succ})}}{n}{\lambda}
          \]
          with $l$, $fail$, $succ$ fresh and the reduction $S\paral L \wstep T_1$,
          where $T_1 \equiv S'^l \paral \proc{\eout{succ}}{l}{}$.
          Now, since $S\bcong R$, we must
          have a transition $R \paral L \wstep T_2$ s.t. $T_1 \bcong T_2$. Now,
          since $T_1\barb{succ@l}$, $T_1\nbarb{fail@l}$ and $T_1 \bcong T_2$ we must also have $T_2\barb{succ@l}$ and
          $T_2\nbarb{fail@l}$ (the barb on $succ@l$ is strong for reasons given in the second case above).
          The only way to obtain this is if $R
          \lwstep{x(\vect{u})@n_\lambda} R'$, hence $T_2\equiv R'^l \paral
          \proc{\eout{succ}}{l}{}$.
          Now, since $(S'^l \paral \proc{\eout{succ}}{l}{}, R'^l \paral
          \proc{\eout{succ}}{l}{}) \in \bcong$, by
          Proposition~\ref{lemma:congruence-inverted} we have $(S',R')\in
          \bcong$ as required.
		  
        \item[Case $\alpha = \oplus n_\lambda \mapsto m$] by
          Lemma~\ref{lemma:netupdates} we know that
          $\proc{\link{m}.P}{n}{\lambda}$ induces the desired labeled action. Consider the context
          \[
            L = \proc{\link{m}.\create{l}{(\eout{fail}\parop fail.\eout{succ})}}{n}{\lambda}
          \]
          with $l$, $fail$, $succ$ fresh and the reduction $S\paral L \wstep
          T_1$, where $T_1 \equiv S'^l \paral \proc{\eout{succ}}{l}{}$. Now,
          since $S\bcong R$ we must have a transition $R \paral L \wstep T_2$
          s.t. $T_1 \bcong T_2$. Now, since $T_1\barb{succ@l}$,
          $T_1\nbarb{fail@l}$ and $T_1 \bcong T_2$ we must also have
          $T_2\barb{succ@l}$, $T_2\nbarb{fail@l}$ (the barb on $succ@l$ is strong for reasons given in the second case above). 
		  The only way to obtain this
          is if $R \wstep R_1 \paral
          \proc{\link{m}.\create{l}{(\eout{fail}\parop
              fail.\eout{succ})}}{n}{\lambda} \step R_1 \paral
          \proc{\create{l}{(\eout{fail}\parop fail.\eout{succ})}}{n}{} \wstep
          T_2$, where $T_2\equiv R'^l \paral \proc{\eout{succ}}{l}{}$.

          Then, by Lemma~\ref{lemma:netupdates} we know that $R \lwstep{\tau} R_i
          \lstep{\ominus n_\lambda \mapsto m} R_{j} \lwstep{\tau} R'$.

          Now, since $(S'^l \paral \proc{\eout{succ}}{l}{}, R'^l \paral
          \proc{\eout{succ}}{l}{}) \in \bcong$, by
          Lemma~\ref{lemma:congruence-inverted} we have $(S',R')\in
          \bcong$ as required.
		  
        \item[Case $\alpha = \ominus n_\lambda \mapsto m$] consider the context
          \[
            L = \proc{\sbreak{m}.\create{l}{(\eout{fail} \parop fail.succ)}}{n}{\lambda}
          \]
          with $l$, $succ$ and $fail$ fresh. The reasoning is similar to the previous case.
		  
        \item[Case $\alpha = \skill (n,\lambda)$] by Lemma~\ref{lemma:netupdates}
          we know that $\proc{\skill}{n}{\lambda}$ is the process that induces the
          reduction we are looking for. Consider the context
          \[
            L = \proc{\eout{fail}}{n}{\lambda} \paral \proc{\skill}{n}{\lambda}
          \]
          with $fail$ fresh and reduction $S \paral L \wstep T_1$.

          Now, since $S\bcong R$ there must be a matching move $R\paral L \wstep
          T_2$ s.t. $T_1 \bcong T_2$. Then, since $T_1\nbarb{fail@n}$ and $T_1
          \bcong T_2$ we must have $T_2\nbarb{fail}$. Since $fail$ is fresh the
          only way to have $T_2\nbarb{fail@n}$ is to have $R \paral L \wstep R_1
          \paral \proc{\eout{fail}}{n}{\lambda} \paral \proc{\skill}{n}{\lambda}
          \step R_1 \paral \proc{\eout{fail}}{n}{\lambda} \wstep T_2$.

          By Lemma~\ref{lemma:netupdates} we know that $R \lwstep{\tau} R_1
          \lwstep{\skill(n,\lambda) } R'_1 \lwstep{\tau} R'$.

          Now, we know that $T_1 \bcong T_2$, where $T_1 \equiv S' \paral
          \proc{\eout{fail}}{n}{\lambda}$ and $T_2 \equiv R' \paral
          \proc{\eout{fail}}{n}{\lambda}$.
          We know $S' \paral
          \proc{\eout{fail}}{n}{\lambda} \bsim S'$ since $n_\lambda$ is not alive
          in $S'$ hence $S \paral
          \proc{\eout{fail}}{n}{\lambda} \bcong S$ and by transitivity of $\bcong$ we get
          $S \bcong R \paral \proc{\eout{fail}}{n}{\lambda}$. By using a similar
          reasoning we get $(S',R') \in \bcong$ as required.

        \item[Case $\alpha = \ms{create}(n,\lambda)$] consider the context
          \[
            L = \proc{\create{n}{(\create{l}{(\eout{fail} \parop fail.\eout{succ})})}}{\godloc}{}
          \]
          with $l$, $fail$ and $succ$ fresh; the reasoning is similar to above.

        \item[Case $\alpha = n_\lambda \views m$] in this case the context is
          slightly more involved than the other ones. This because the label
          $n_\lambda \views m$ only informs us about the correct view of
          location $n_\lambda$ toward location name $m$, but does not tell us
          anything about the aliveness of the link between the two. Hence, we
          need to build a context that accounts for the case in which the two are
          already connected and for the case in which a connection between the
          two must be established.
          Consider the context
          \[
            L = \left[
              \res{x}\left( \eout{x} \parop
                \begin{array}{l}
                  x.\link{m}.(\spawn{m}{(\sbreak{n}.\create{l}{(\eout{fail} \parop fail.\eout{succ_1}})}) \parop\\
                  x.\spawn{m}({\create{l}{(\eout{fail} \parop fail.\eout{succ_2})}})
                \end{array}
            \right)
            \right]^n_\lambda
          \]
          with $l$, $fail$, $succ_1$ and $succ_2$ fresh. Now, since $S\bcong R$
          then the public network of the two systems must agree on aliveness of locations,
          links, and views. We consider now the two cases.
          \begin{description}
          \item[Case $n\not\connection m$] consider the transition
            \[
              S \paral L \wstep T_1 \equiv \res{x}S^l
              \paral
              \left[
                \begin{array}{l}
                  x.\spawn{m}{\\
                  \create{l}{(\eout{fail} \parop
                    fail.\eout{succ})}}
                \end{array}
              \right]^n_\lambda \paral
              \proc{\eout{succ_1}}{l}{}
              \]
            where $\eout{x}$ synchronizes with the first process (and the second one remains inactive).  
              
            Now, since $S\bcong R$ there must be a
            matching move $R \paral L \wstep T_2$ such that $T_1 \bcong T_2$.
            Then, since
            $T_1\barb{succ_1@l}$ and $T_1\nbarb{fail@l}$ and $T_1 \bcong T_2$ we
            also have $T_2\barb{succ_1@l}$ and
            $T_2\nbarb{fail@l}$.
            Hence, we have $T_2 \equiv \res{x}R'^l \paral
            \proc{x.\spawn{m}{\create{l}{(\eout{fail} \parop
                  fail.\eout{succ})}}}{n}{\lambda} \paral
            \proc{\eout{succ_1}}{l}{}$ ($x$ is fresh in $R$). This is possible only if
            $R\lwstep{n_\lambda \views m}R'$.
            Now, $\res{x}S^l \paral \proc{x.\spawn{m}{\create{l}{(\eout{fail} \parop
                  fail.\eout{succ})}}}{n}{\lambda} \paral
            \proc{\eout{succ_1}}{l}{} \bsim S^l \paral
            \proc{\eout{succ_1}}{l}{}$, because $x\notin fn(S)$.

            By using a similar reasoning for $T_2$ and transitivity of $\bcong$
            we have  $(S'^l \paral \proc{\eout{succ_1}}{l}{\lambda}, R'^l \paral
            \proc{\eout{succ_1}}{l}{\lambda}) \in \bcong$. Finally by
            Lemma~\ref{lemma:congruence-inverted} we have $(S',R')\in
            \bcong$ as required.

          \item[Case $n\connection m$] consider the transition
            \[
              S \paral L \wstep T_1 \equiv S^l \paral
              \left[ 
                \begin{array}{l}
                  x.\link{m}.(\spawn{m}{(\sbreak{n}.\\
                  \create{l}{(\eout{fail}
                    \parop fail.\eout{succ_1}})})
                \end{array}
              \right]^n_\lambda \paral
              \proc{\eout{succ_2}}{l}{}
              \]
              where $\eout{x}$ synchronizes with the second process (and the first one remains inactive).
              
            Now, since $S\bcong R$ we have a
            transition $R \paral L \wstep T_2$ such that $T_1 \bcong T_2$. By
            $T_1\barb{succ_2@l}$, $T_1\nbarb{fail@l}$ and $T_1 \bcong T_2$ we
            have $T_2\barb{succ_2@l}$ and
            $T\nbarb{fail@l}$. Hence we have
            \[
              T_2 \equiv R'^l \paral \left[
                \begin{array}{l}
                  x.\link{m}.(\spawn{m}{(\sbreak{n}.\\
                  \create{l}{(\eout{fail}
                    \parop fail.\eout{succ_1}})})
                \end{array}
              \right]^n_\lambda
              \paral
              \proc{\eout{succ_2}}{l}{}
            \]
            This is possible only if
            $R\lwstep{n_\lambda \views m}R'$.

            Now,
            \[\res{x}S^l \paral
              \proc{\eout{succ_2}}{l}{} \paral
              \left[
                \begin{array}{l}
                  x.\link{m}.(\spawn{m}{(\sbreak{n}.\\
                \create{l}{(\eout{fail}
                  \parop fail.\eout{succ_1}})})
              \end{array}
              \right]^n_\lambda \bsim S^l \paral
              \proc{\eout{succ_2}}{l}{}
              \] because $x\notin fn(S)$.

            By using a similar reasoning for $T_2$ and transitivity of $\bcong$
            we have  $(S'^l \paral \proc{\eout{succ_2}}{l}{}, R'^l \paral
            \proc{\eout{succ_2}}{l}{}) \in \bcong$. Finally by
            Lemma~\ref{lemma:congruence-inverted} we have $(S',R')\in
            \bcong$ as required. \qedhere
          \end{description}
        \end{description}
      \end{proof}

%%% Local Variables:
%%% mode: latex
%%% TeX-master: "main-distributed"
%%% End:

\section{Bisimulation Of Running Example}\label{app:bisimulation}

We now show the proof of Example~\ref{ex:running-ex}

We recall briefly the two systems.

\begin{align*}
  & \textbf{servD} =
  \begin{array}{ll}
    \sys{\res{n_r,n_b,r_1,r_2,b}\net}{
    \proc{ I  }{n_i}{}
    \paral \proc{ R }{n_r}{}
    \paral \proc{ B }{n_b}{}
    }
  \end{array}
  \\
  &\textbf{servDFR} =
  \sys{\res{n_r,n_b,\green{n_c},r_1,r_2,b,c,\green{retry}} \; \net'}{
      \proc{ K }{n_i}{} \paral
      \proc{ R }{n_r}{} \paral
      \red{\proc{ \skill }{n_r}{}} \paral
      \proc{ !B }{n_b}{} \paral
      \green{\proc{ C }{n_c}{}}
  }
\end{align*}
where:
\begin{align*}
	& I = req(y,z).\spawn{n_r}{\out{r_1}{y,z}} \\
	& R = (r_1(y,z).\spawn{n_b}{\out{b}{y,z}}) \parop (r_2(y,z).\spawn{n_i}{\out{z}{y}}) \\
	& B = b(y,z).\spawn{n_r}{\out{r_2}{z,w_y}}\\
  & K = req(y,z).((\spawn{n_r}{\out{r_1}{y,c}}) \parop
    c(w).\out{z}{w} \parop
    retry.\spawn{n_r}{\out{r_1}{y,c}}) \\
	  & C = \res{c}\eout{c}\; \parop \; !c.(\eout c  \; \parop \create{n_r}{(R \parop \spawn{n_i}{\eout{retry}})})
\end{align*}
and where the networks $\net$ and $\net'$ are defined as follows:
\begin{align*}
	& \net_{\mc{A}} = \{ n_i \mapsto 1, n_r \mapsto 1, n_b \mapsto 1 \}   \\
	& \net_{\mc{L}} = \{ n_i \connection n_r, n_r \connection n_b \} &  \\
	& \net_{\mc{V}} = \{ n_i \mapsto \{ n_r \mapsto 1 \}, n_b \mapsto \{ n_r \mapsto 1 \}, n_r \mapsto \{ n_i \mapsto 1, n_b \mapsto 1 \} \} \\
	& \\
	& \net'_{\mc{A}} = \{ n_i \mapsto 1, n_r \mapsto 1, n_b \mapsto 1, n_c \mapsto 1 \} \\
	& \net'_{\mc{L}} = \{ n_i \connection n_r, n_r \connection n_b \} \\
	&  \net'_{\mc{V}} = \{ n_i \mapsto \{ n_r \mapsto 1 \}, n_b \mapsto \{ n_r \mapsto 1 \}, n_r \mapsto \{ n_i \mapsto 1, n_b \mapsto 1 \}, n_c \mapsto \fnull \} 
\end{align*}

\begin{prop}
  $\textbf{servD} \wbsim \textbf{servDFR}$
\end{prop}

\begin{proof}
  In the following, for the sake of legibility, we will omit in the action label the
  incarnation number of the location doing the action.

  Consider relation $\rel{R} = \{(\textbf{servD},\textbf{servDFR})\} \cup \rel{S}_0 \cup
  \rel{S}_1 \cup \rel{S}_2$
  where
  \begin{align*}
    & \rel{S}_0 = \{(\textbf{servD}, R_0) \mid \textbf{servDFR} \lwstep{\tau}R_0\}\\
    & \rel{S}_1 = \{(S_1,R_1) \mid (S_0,R_0)\in \rel{S}_0, S_0\lwstep{req(x,y)@n_i} S_1\wedge R_0\lwstep{req(x,y)@n_i} R_1\}\\
    & \rel{S}_2 = \{(S_2,R_2) \mid (S_1,R_1)\in \rel{S}_1, S_1\lwstep{\out{z}{w}@n_1} S_1\wedge R_1 \lwstep{\out{z}{w}@n_1} R_2\}
  \end{align*}

  Note that $(\textbf{servD},\textbf{servDFR}) \in \rel{S}_0$. Now we analyze the moves of the various pairs to show that indeed $\rel{R}$ is
  a bisimulation.
  \begin{itemize}
  \item Pair $(\textbf{servD},\textbf{servDFR})$. Now we proceed by case analysis on the possible transitions.
    \begin{itemize}
    \item Case $req(y,z)@n_i$.
      \begin{itemize}
      \item Case $\textbf{servD} \lstep{req(y,z)@n_i} T_1$,\\
        where
        \[
           T_1 \equiv
          \sys{\res{n_r,n_b,r_1,r_2,b}\net}{ \proc{ \spawn{n_r}{\out{r_1}{y,z}}
            }{n_i}{} \paral \proc{ R }{n_r}{} \paral \proc{ B }{n_b}{} }
        \]
        The move can then be matched by
        \begin{align*}
          &\textbf{servDFR} \lwstep{req(y,z)@n_i}\\
          & T_2 \equiv
            \res{n_r,n_b,\green{n_c},r_1,r_2,b,c,\green{retry}} \; \net'
            \vartriangleright \\
          & \left(
            \begin{array}{l}
              \proc{(\spawn{n_r}{\out{r_1}{y,c}}) \parop
              c(w).\out{z}{w} \parop
              retry.\spawn{n_r}{\out{r_1}{y,c}}}{n_c}{}
              \\ \paral
              \proc{ R }{n_r}{} \paral
              \red{\proc{ \skill }{n_r}{}} \paral
              \proc{ !B }{n_b}{} \paral
              \green{\proc{C}{n_c}{}}
            \end{array}
            \right)
        \end{align*}
        and $(T_1, T_2)\in \rel{S}_1 \subseteq \rel{R}$.
      \item Case $\textbf{servDFR} \lstep{req(y,z)@n_i} T_2$. Similar to above.
      \end{itemize}
    \item Case $\tau$.
      \begin{itemize}
      \item Consider the transition
        \begin{align*}
          &\textbf{servDFR} \lstep{\tau} \\
          & T_2 \equiv
            \res{n_b,n_r,b,\purple{s},\green{retry,n_c,c}}\net \vartriangleright\\
          & \proc{K}{n_i}{}
            \paral \proc{R}{n_r}{} \paral \red{\proc{\skill}{n_r}{}} \paral
            \proc{ !B }{n_b}{}  \paral\\
          & \green{\proc{\eout{c}\parop !c.(\create{n_r}{(R \parop \spawn{n_i}{\eout{retry}})} \parop \eout c)}{n_c}{}}
        \end{align*}

        Then, the move can be matched by the other system by doing nothing,
        indeed $(\textbf{servD}, T_2) \in \rel{S}_1 \subseteq \rel{R}$

        \item Consider the transition
          \begin{align*}
             \textbf{servDFR} \lstep{\tau} T_2 \equiv
             \res{n_r,n_b,a,b,\purple{s},\green{retry,n_c}}\net \ominus
            n_r
            \vartriangleright \\
             \proc{K}{n_i}{} \paral \proc{R}{n_r}{} \paral \proc{!B}{n_b}{} \paral
            \green{\proc{C}{n_c}{}}
          \end{align*}
          Then, the move can be matched by the other system by doing nothing,
          indeed $(\textbf{servD}, T_2) \in \rel{S}_1 \subseteq  \rel{R}$
      \end{itemize}
    \end{itemize}
	
  \item Pairs $(\textbf{servD},R_0)\in\rel{S}_0$ are handled similarly as in the case above.
  
  \item Pairs $(S_1,R_1)\in\rel{S}_1$
    \begin{description}
    \item[Case $\tau$] Here, any possible $\tau$ transition can be matched by
      the other system by doing nothing.
    \item[Case $\out{z}{w}@n_i$]
      Here there is only one possible system $S$ that can perform
      the step $\out{z}{w}@n_i$, that is
      \[
        \sys{\res{\vect{u}}\net}{\proc{\out{z}{w}}{n_1}{}}
      \]

      Now, there are an unbounded number of different system $R_1$ that can
      match the move, according to the state of the recovery. We only show the
      following example.
      Consider system
      \begin{align*}
        & R_1 \equiv
            \res{n_r,n_b,\green{n_c},r_1,r_2,b,c,\green{retry}}  \net''\vartriangleright \\
        &    \left(
          \begin{array}{l}
              \begin{array}{l}
                \proc{c(w).\out{z}{w}}{n_i}{} \paral
                \proc{retry.\spawn{n_r}{\out{r_1}{y,c}}}{n_i}{} \paral\\
                \proc{ \eout{retry} \parop R }{n_r}{} \paral
                \red{\proc{ \skill }{n_r}{}} \paral
                \proc{ !B }{n_b}{} \paral
                \green{\proc{ C }{n_c}{}}
              \end{array}
            \end{array}
            \right)
      \end{align*}
      In $R_1$ the failure has occurred (we omit dead locations) and the recovery process has started
      already. By doing the following steps $R$ can match the move.
      \begin{align*}
        R_1 \lstep{\tau}\\
        \sys{\res{\vect{u}}\net''}{
        \begin{array}{l}
            \proc{c(w).\out{z}{w}}{n_i}{} \paral
            \proc{retry.\spawn{n_r}{\out{r_1}{y,c}}}{n_i}{} \paral
            \proc{\eout{retry}}{n_r}{} \paral
            \proc{ R }{n_r}{} \paral\\
            \red{\proc{ \skill }{n_r}{}} \paral
            \proc{ !B }{n_b}{} \paral
            \green{\proc{ C }{n_c}{}}
        \end{array}
        } \lstep{\tau}
        \\
        \sys{\res{\vect{u}}\net''}{
        \proc{c(w).\out{z}{w}}{n_i}{} \paral
        \proc{\spawn{n_r}{\out{r_1}{y,c}}}{n_i}{} \paral
        \proc{ R }{n_r}{} \paral
        \red{\proc{ \skill }{n_r}{}} \paral
        \proc{ !B }{n_b}{} \paral
        \green{\proc{ C }{n_c}{}}
        } \lstep{\tau}
        \\
        \sys{\res{\vect{u}}\net''}{
        \proc{c(w).\out{z}{w}}{n_i}{} \paral
        \proc{\out{r_i}{y,c}}{n_r}{} \paral
        \proc{ R }{n_r}{} \paral
        \red{\proc{ \skill }{n_r}{}} \paral
        \proc{ !B }{n_b}{} \paral
        \green{\proc{ C }{n_c}{}}
        }\lstep{\tau}
        \\
        \res{\vect{u}}\net'' \vartriangle
        \left(
        \begin{array}{l}
        \proc{c(w).\out{z}{w}}{n_i}{} \paral
        \proc{\out{r_1}{y,c}}{n_r}{} \paral
        \proc{r_1(y,z).\spawn{n_b}{\out{b}{y,z}} }{n_r}{} \paral\\
        \proc{r_2(y,z).\spawn{n_i}{\out{z}{y}}}{n_r}{} \paral
        \red{\proc{ \skill }{n_r}{}} \paral
        \proc{ !B }{n_b}{} \paral
        \green{\proc{ C }{n_c}{}}
        \end{array}
        \right)
        \lstep{\tau}
        \\
        \res{\vect{u}}\net'' \vartriangleright
        \left(
        \begin{array}{l}
          \proc{c(w).\out{z}{w}}{n_i}{} \paral
          \proc{\spawn{n_b}{\out{b}{y,c}} }{n_r}{} \paral
          \proc{r_2(y,z).\spawn{n_i}{\out{z}{y}}}{n_r}{} \paral\\
          \red{\proc{ \skill }{n_r}{}} \paral
          \proc{ !B }{n_b}{} \paral
          \green{\proc{ C }{n_c}{}}
        \end{array}
        \right)
        \lstep{\tau}
        \\
        \res{\vect{u}}\net'' \vartriangleright
        \left(
        \begin{array}{l}
          \proc{c(w).\out{z}{w}}{n_i}{} \paral
          \proc{{\out{b}{y,c}} }{n_b}{} \paral
          \proc{r_2(y,z).\spawn{n_i}{\out{z}{y}}}{n_r}{} \paral\\
          \red{\proc{ \skill }{n_r}{}} \paral
          \proc{ !B }{n_b}{} \paral
          \green{\proc{ C }{n_c}{}}
        \end{array}
        \right)
        \lstep{\tau}
        \\
        \res{\vect{u}}\net'' \vartriangleright
        \left(
        \begin{array}{l}
          \proc{c(w).\out{z}{w}}{n_i}{} \paral
          \proc{{\out{b}{y,c}} }{n_b}{} \paral
          \proc{r_2(y,z).\spawn{n_i}{\out{z}{y}}}{n_r}{} \paral\\
          \red{\proc{ \skill }{n_r}{}} \paral
          \proc{ b(y,z).(!B \parop \spawn{n_r}{\out{r_2}{z,w_y}}) }{n_b}{} \paral
          \green{\proc{ C }{n_c}{}}
        \end{array}
        \right)
        \lstep{\tau}
        \\
        \res{\vect{u}}\net'' \vartriangleright
        \left(
        \begin{array}{l}
          \proc{c(w).\out{z}{w}}{n_i}{} \paral
          \proc{r_2(y,z).\spawn{n_i}{\out{z}{y}}}{n_r}{} \paral
          \red{\proc{ \skill }{n_r}{}} \paral\\
          \proc{!B \parop \spawn{n_r}{\out{r_2}{c,w_y}}) }{n_b}{} \paral
          \green{\proc{ C }{n_c}{}}
        \end{array}
        \right)
        \lstep{\tau}
        \\
        \res{\vect{u}}\net'' \vartriangleright
        \left(
        \begin{array}{l}
          \proc{c(w).\out{z}{w}}{n_i}{} \paral
          \proc{r_2(y,z).\spawn{n_i}{\out{z}{y}}}{n_r}{} \paral
          \red{\proc{ \skill }{n_r}{}} \paral\\
          \proc{!B}{n_b}{} \paral
          \proc{\spawn{n_r}{\out{r_2}{c,w_y}}}{n_b}{} \paral
          \green{\proc{ C }{n_c}{}}
        \end{array}
        \right)
        \lstep{\tau}
        \\
        \res{\vect{u}}\net'' \vartriangleright
        \left(
        \begin{array}{l}
          \proc{c(w).\out{z}{w}}{n_i}{} \paral
          \proc{r_2(y,z).\spawn{n_i}{\out{z}{y}}}{n_r}{} \paral
          \red{\proc{ \skill }{n_r}{}} \paral\\
          \proc{!B}{n_b}{} \paral
          \proc{\out{r_2}{c,w_y}}{n_r}{} \paral
          \green{\proc{ C }{n_c}{}}
        \end{array}
        \right)
        \lstep{\tau}
        \\
        \res{\vect{u}}\net'' \vartriangleright
        \left(
          \proc{c(w).\out{z}{w}}{n_i}{} \paral
          \proc{\spawn{n_i}{\out{c}{w_y}}}{n_r}{} \paral
          \red{\proc{ \skill }{n_r}{}} \paral
          \proc{!B}{n_b}{} \paral
          \green{\proc{ C }{n_c}{}}
        \right)
        \lstep{\tau}
        \\
        \res{\vect{u}}\net'' \vartriangleright
        \left(
          \proc{c(w).\out{z}{w}}{n_i}{} \paral
          \proc{\out{c}{w_y}}{n_i}{} \paral
          \red{\proc{ \skill }{n_r}{}} \paral
          \proc{!B}{n_b}{} \paral
          \green{\proc{ C }{n_c}{}}
        \right)
        \lstep{\tau}
        \\
        \res{\vect{u}}\net'' \vartriangleright
        \left(
        \proc{\out{z}{w_y}}{n_i}{} \paral
        \red{\proc{ \skill }{n_r}{}} \paral
        \proc{!B}{n_b}{} \paral
        \green{\proc{ C }{n_c}{}}
        \right)
        \lstep{\out{z}{w_y}@n_i}
        \\
        T_2 \equiv
        \res{\vect{u}}\net'' \vartriangleright
        \left(
        \red{\proc{ \skill }{n_r}{}} \paral
        \proc{!B}{n_b}{} \paral
        \green{\proc{ C }{n_c}{}}
        \right)
      \end{align*}
      with $(T_1, T_2)\in \rel{R}$
    \end{description}
  \item Pairs $S_2 \times R_2$
    \begin{description}
    \item[Case $\tau$] Here there exist infinite $R_2$ that can perform a $\tau$
     (the controller still attempting to recreate the location), anyway all the
     moves can matched by $S_2$ by
      doing nothing.
    \end{description}
  \end{itemize}
\end{proof}
%%% Local Variables:
%%% mode: latex
%%% TeX-master: "main-distributed"
%%% End:

\end{document}